\documentclass[aps,prl,notitlepage,twocolumn,amsmath,amssymb,10pt]{revtex4-1}
\usepackage{times}

\usepackage[usenames,dvipsnames]{xcolor}
\usepackage{amsthm}
\usepackage{tikz}  
\usetikzlibrary{arrows,shapes,positioning,shadows,backgrounds,fit}

\newtheorem*{observation*}{Observation}

\newcommand{\GGrev}[1]{\textcolor{black}{#1}}

\usepackage{epstopdf} 

\usepackage{tikz}  
\usetikzlibrary{arrows,shapes,positioning,shadows,svg.path,backgrounds,fit}

\usepackage[english]{babel}
\usepackage{graphicx}
\usepackage{dcolumn}
\usepackage{bm}
\usepackage{color}
\usepackage{amsmath}
\usepackage{relsize,amsmath,dsfont,mathrsfs,empheq,verbatim,upgreek}
\usepackage{etoolbox}
\usepackage[caption = false]{subfig}

\usepackage[colorlinks=true,linkcolor=blue,urlcolor=blue,citecolor=blue]{hyperref}
\usepackage{graphics}
\usepackage{bm}
\usepackage{color}
\usepackage{amscd}
\usepackage{amsfonts}
\usepackage{amssymb}
\usepackage{graphicx}
\usepackage{tabularx}

\newcommand{\bra}[1]{\langle #1 |}
\newcommand{\ket}[1]{| #1 \rangle}

\newcommand{\mean}[1]{\langle #1 \rangle}
\providecommand{\bgreek}[1]{\mbox{\boldmath$#1$}}
\newcommand{\vett}[1]{\mathbf{#1}}

\newcommand{\Ham}{\mathcal{H}}
\newcommand{\norm}[1]{\| #1 \|}
\newcommand{\Id}{\mathds{1}}

\newcommand{\HILB}{\mathscr{H}}

\begin{document}


\newdimen\origiwspc%
\newdimen\origiwstr%
\preprint{}

\title{Steady-state coherences by composite system-bath interactions}
\author{Giacomo Guarnieri$^{1,2}$, Michal Kol\'{a}\v{r}$^{1}$, Radim Filip$^{1}$}
\email{gguarnieri88@gmail.com}

\affiliation{$^1$\mbox{Department of Optics, Palack\'{y} University, 17. listopadu 1192/12, 771 46 Olomouc, Czech Republic}\\
$^2$\mbox{Department of Physics, Trinity College Dublin, Dublin 2, Ireland}}

\begin{abstract}
We identify sufficient conditions on the structure of the interaction Hamiltonian between a two-level quantum system and a thermal bath which, without any external drive or coherent measurement, guarantee the generation of steady-state coherences (SSC). The SSC this way obtained remarkably turn out to be independent on the initial state of the system, which therefore could be even taken initially incoherent. We characterize in detail this phenomenon first analytically in the weak coupling regime for two paradigmatic models, and then numerically in more complex systems without any assumption on the coupling strength. In all these cases we find that SSC become increasingly significant as the bath is cooled down. These results can be therefore directly verified in many experimental platforms.  
\end{abstract}
\date{\today}
\maketitle

Standard textbook quantum mechanics deals with closed systems and their coherent unitary evolution.
However, every realistic quantum system has to be considered as open in light of its unavoidable interaction with its surroundings. The resulting reduced non-unitary system dynamics takes into account for irreversible processes, such as decoherence and dissipation \cite{Alicki2007, Breuer2002}. Especially the former still withstands among the major obstacles to all the countless applications relying on the maintenance and exploitation of quantum coherences, ranging from quantum metrology \cite{Smirne2016a}, state engineering \cite{Wineland1998}, to even relatively far fields such as quantum thermodynamics \cite{Goold2016a} and quantum biology \cite{Huelga2013a, chin2013nature}. 
Due to this prominence, lots of efforts have thus been devoted to conceive strategies to oppose or even neutralize the detrimental effects of environmental couplings, e.g. error-correction schemes \cite{Shor1995}, dynamical decoupling \cite{Viola1999} and quantum feedback control \cite{Rabitz2005}, just to mention a few. 

Nowadays the theory of coherence represents a wide research field, encompassing theoretical developments of a resource theory of coherences \cite{Chitambar2016}, characterization of suitable quantifiers and measures of coherence \cite{Streltsov2015}, investigation of coherence dynamics \cite{Bromley2015} and experimental applications \cite{Napoli2016}.
Recently it was also shown that, under specific conditions on the parameters determining the dynamics, a spin undergoing a pure-dephasing evolution may, in the long-time dynamics, \textit{retain} some of its initial coherences in the energy eigenbasis \cite{Roszak2006, Addis2014, Radim2015SciRep}. In particular, \textit{Addis et al.} provided a clear-cut connection between this phenomenon, which they named \textit{coherence trapping}, and the properties of the environmental spectrum \cite{Addis2014}: while its temperature and low-frequency part determines the partial survival or complete erasure of the initial coherences, its high-frequency band dictates their maximum attainable amount. 

The main limitations of such result are however twofold. The first technical one is that it relies on the specific model and interaction considered, i.e. a pure-dephasing spin-boson \cite{Breuer2002}.
When a more general dissipative spin-boson dynamics is in fact taken into account \cite{Leggett1987}, i.e. when the interaction does not commute anymore with the system's bare Hamiltonian, all the coherences inevitably vanish in the long-time limit, irrespectively of the type of environment considered. 
Individual attempts at generalizing this result and extending it to other models have been pursued in \cite{Zhang2015}, where qubit states initially correlated with the environment were considered, and in \cite{Bhattacharya2016a}, where a slow down of the coherences decay was characterized for two qubits interacting with a harmonic oscillator. In both cases, however, additional resources to achieve \textit{steady-state coherences} (SSC), such as initial correlations or a mediating system, were employed.
The second fundamental limitation is its \textit{initial-state dependence}: coherence trapping dictates in fact a way to preserve, by means of clever engineering of the environment, a fraction of initial coherences which, if not present, are not thus generated by this process. This process thus requires state preparation of coherent superposition.

\begin{figure}[!htbp]
\includegraphics[width=0.9\columnwidth]{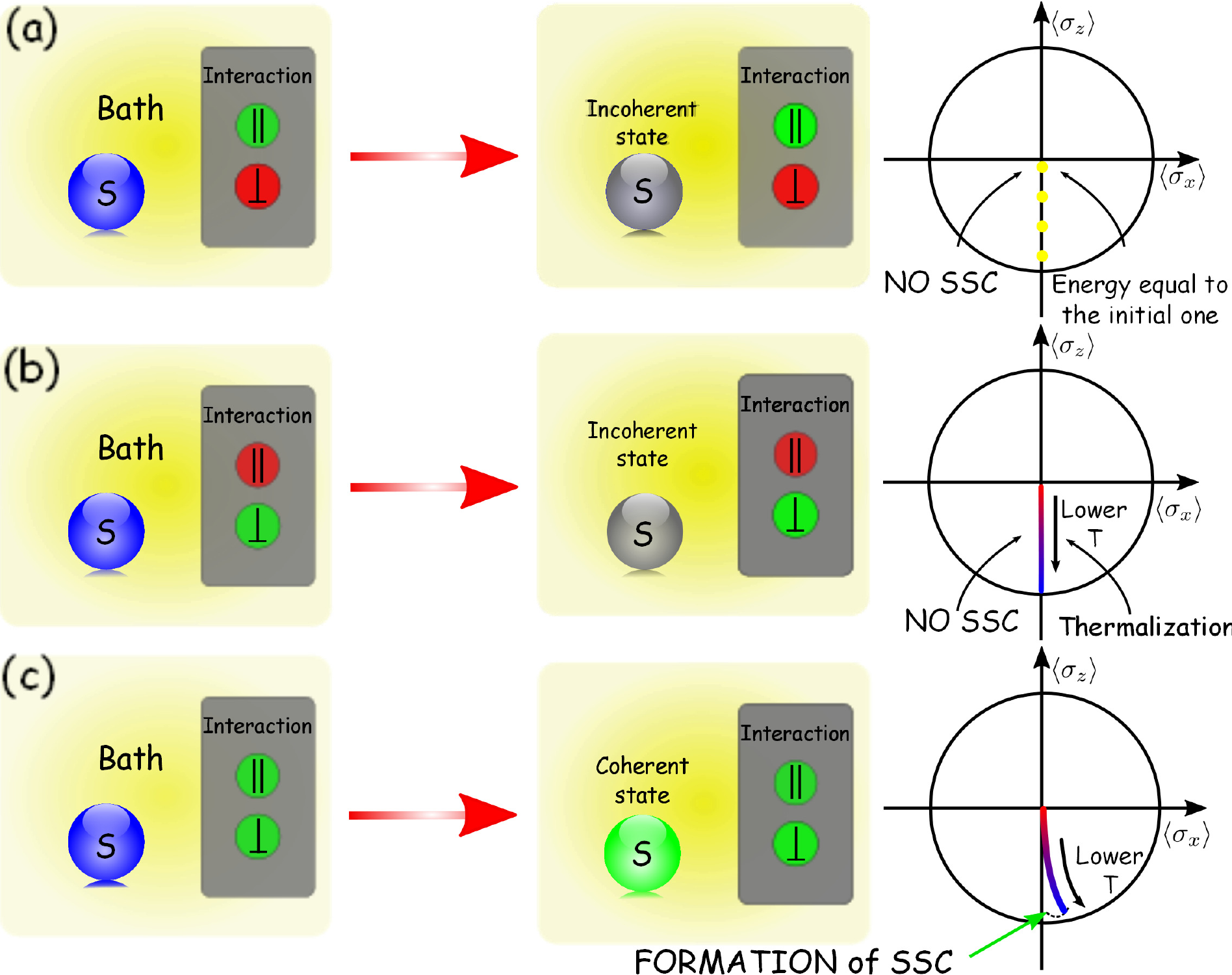}
\caption{A \GGrev{two-level} system is put in contact with a thermal bath until the steady state is reached. 
The steady-state solution is shown in the right column using a $xz-$planar section of the Bloch sphere, where for illustrative purposes we have assumed the coherences to be real.
In the case of a interaction \textit{parallel} to $\Ham_S$ {\bf{(a)}}, the steady-state solution is an incoherent state depicted on the $z-$axis (yellow points), relative to a final energy which is equal to the initial one. In the case of a interaction \textit{orthogonal} to $\Ham_S$ {\bf{(b)}}, the system will thermalize with the bath and SSC will not be present, its steady state thus again be given by a point on the $z-$axis determined by the Boltzmann factor relative to the temperature $T$. However, when the interaction is a linear combination of \textit{parallel} and \textit{orthogonal} interactions {\bf{(c)}} the steady state solution acquires a deviation from the thermal state, due to the \textit{formation of coherences}, which is in general increasingly pronounced as long as the bath is cooled down.}
\label{figSchem}
\vspace*{-0.5cm}
\end{figure}

In this Letter we aim to remove these two significant constraints and provide instead sufficient conditions concerning the structure of the system-bath interaction Hamiltonian which guarantee the \textit{formation} (and not mere trapping) of SSC \GGrev{in a generic two-level system \textit{independently} on its initial state}. In this case no additional systems, state preparations or measurement procedures are required. It is in fact well known that quantum coherence can be induced by means of external classical driving \cite{Bloch1946} or by coherent measurement \cite{Cable2005PRA, Radim2014PRA}.  The only resource we employ here is instead a composite unitary system-bath dynamics achievable in many experimental platforms. 
In particular, we will give conclusive evidences of the following statement (see also the schematics in Fig.~\ref{figSchem}):
\begin{observation*}
Consider a two-level system interacting with a single thermal environment such that the total Hamiltonian is $\Ham = \Ham_S + \Ham_E + \Ham_{SE}$. The interaction Hamiltonian $\Ham_{SE}$ is considered to be an hermitian operator of the form
$\Ham_{SE} = \sum_j O_{S,j}\otimes b_E + h.c. $, with $O_{S,j}$ denoting system's operators and $b_E = \sum_k g_k b_k$ denoting the multimode environmental annihilation operator.

Let $\Ham_{SE}^{\parallel \Ham_S} \equiv h_S^{-1} \mathrm{Tr}_{S} [\Ham_S\! \allowbreak\sum_j O_{S,j} ] \allowbreak \Ham_S\!\otimes b_E + h.c. $ (with $h_S \equiv \mathrm{Tr}_S\left[\Ham_S^2\right]$) denote the projection of the interaction Hamiltonian parallel to $\Ham_S$ (according to the Hilbert-Schmidt scalar product \cite{Breuer2002}), and let $\Ham_{SE}^{\perp \Ham_S} \equiv \Ham_{SE} - \Ham_{SE}^{\parallel \Ham_S}$ denote its orthogonal complement. 

If the interaction Hamiltonian has \textbf{both} non-zero projections over the parallel and orthogonal components with respect to $\Ham_S$, i.e. if $\Ham_{SE} = \Ham_{SE}^{\parallel \Ham_S} + \Ham_{SE}^{\perp \Ham_S} $, then the two-level system will show SSC with respect to the eigenbasis of $\Ham_S$ independently on the initial state.
\end{observation*}
Moreover, we will show that the SSC this way generated can be generically enhanced simply when the bath is cooled down and that they decay slowly with increasing temperature, thus ensuring their possible observation even for non-zero temperature in experimental setups. These results apply to a wide class of systems and interaction Hamiltonians and here we will explicitly provide a paradigmatic analysis for experimentally relevant examples of them. 
The implications of this result pave the way for optimization of the system-bath interaction in experimental platforms in order to achieve autonomous SSC.  

Let us start by considering a two-level system coupled to a thermal bosonic bath such that the total Hamiltonian is given by
\vspace*{-0.2cm}
\begin{equation}\label{eq:Ham}
\Ham = \frac{\omega_0}{2}\sigma_z +\sum_k \omega_k b^{\dagger}_k b_k + \left( f_1\sigma_z + f_2\sigma_x\right)\otimes B_E,
\vspace*{-0.2cm}
\end{equation}
with $\sigma_{x,y,z}$ being the usual Pauli matrices 
, $B_E=\sum_k g_k (b_k + b^{\dagger}_k )$ representing the multimode quadrature operator with $\lbrace b^{\dagger}_k, b_k\rbrace_k$ denoting the family of bosonic creation and annihilation operators and $f_{1,2}$ being two generic coupling constants independent on the bath modes $k$.
\GGrev{We stress again that the crucial point is that the system-bath interaction Hamilonian satisfies the condition stated in the Proposition. The parallel projection $\Ham_{SE}^{\parallel \Ham_S} = f_1\sigma_z\otimes B_E$ induces a pure dephasing dynamics on the reduced system while the orthogonal projection $\Ham_{SE}^{\perp \Ham_S} = f_2\sigma_x\otimes B_E$ will generate a dynamics involving both populations and coherences of the two-level system \cite{Breuer2002, Leggett1987}.}
In the following calculations, we will assume the system and the environment to be \textit{weakly coupled} and starting in a product state $\rho_{SE}(0)=\rho_S(0)\otimes\rho_{\beta} $, with $\rho_{\beta}\equiv Z^{-1}e^{-\beta\Ham_E}$ being the Gibbs state at inverse temperature $\beta=\left( k_B T\right)^{-1}$ ($Z=\mathrm{Tr}_E\left[e^{-\beta\Ham_E}\right]$, $\Ham_E=\sum_k\omega_k b^{\dagger}_k b_k$).
When $f_2\to 0$, we retrieve the pure dephasing dynamics \cite{Breuer2002} while for $f_1\to 0$ the evolution retraces the usual decoherent dynamics of a dissipative spin-boson \cite{Breuer2002, Clos2012, Leggett1987}.
By means of standard techniques (i.e. time-convolutionless expansion of the dynamical generator up to second order in the coupling constant \cite{Breuer2002}), we derived a time-dependent non-Markovian master equation for the system's density operator and then, from the latter, obtained the equations of motion for the three components of the Bloch vector $\vett{v}(t)$, defined through the relation $\rho_S(t)=\frac{1}{2}\left(\Id_2+\vett{v}(t)\cdot\bgreek{\sigma}\right)$, with $v_{x,y,z}(t) =\mathrm{Tr}_S\left[\sigma_{x,y,z}\rho_S(t)\right]$.
In particular, $v_3(t)$ gives the population imbalance of the qubit while $v_{1,2}(t)$ denote the real and imaginary parts of coherences in the $\sigma_z$ basis, respectively. 
The interested reader is referred to the Supplementary Material \cite{SM} (SM, Section A) for detailed derivation and additional comments concerning the dynamics of such system. 

Since for many practical purposes it is easiest to exploit the steady-state properties of a system, we have focused on the long-time limit. The resulting steady-state solutions for $v_{1,2}$ characterize the SSC and turn out to be given by (see SM, Eq.~(42)) 
\begin{equation}\label{eq:SSsolution}
\overline{v}_1 = \frac{f_1f_2 \left[ \Delta_1\tanh\left(\frac{\omega_0}{2T}\right) +4\lambda\Omega\Gamma(s) + \Delta_2 \right]}{\omega_0 + f^2_2 \Delta_1},\quad
\overline{v}_2 = 0,
\end{equation}
where $\Gamma(x)$ is the Euler's function and where we have defined (omitting for brevity their parameter dependence)
\begin{align}	\label{eq:Deltas}
&\Delta_1(T,\Omega) = -2\!\!\int_0^{+\infty}\!\!\!\!\!\!\!d\omega \, J_{\mathrm{eff}}(\Omega,T)\!\!\left[\mathcal{P} \frac{1}{\omega+\omega_0}  -  \mathcal{P} \frac{1}{\omega-\omega_0}\!\right], \notag\\
&\Delta_2(\Omega) = -2 \int_0^{+\infty}\!\!\!\!\!\!\!d\omega\, J(\omega) \left[\mathcal{P} \frac{1}{\omega+\omega_0} + \mathcal{P} \frac{1}{\omega-\omega_0}\right],
\end{align}
with $\mathcal{P}$ standing for Cauchy principal value, $J(\omega)$ denoting the spectral density of the bath and $J_{\mathrm{eff}}(\omega,T) \equiv J(\omega) \coth\left(\frac{\omega_0}{2T}\right)$ being the so-called effective spectral density \cite{Clos2012}. The former has been moreover taken to be of the general Ohmic-dependent form $J(\omega) =\lambda\frac{\omega^s}{\Omega^{s-1}}e^{-\omega/\Omega}$, with $\lambda$ being a coupling constant, $\Omega$ the cut-off frequency and $s$ the Ohmicity parameter ruling over the low-frequency behavior. The latter is known to lead to a sub-Ohmic spectrum for $s<1$, to an Ohmic one for $s=1$ and finally to a super-Ohmic spectrum for $s>1$ \cite{Leggett1987}.
As can be immediately seen from Eq.~\eqref{eq:SSsolution}, if either one of the two couplings $f_{1,2}$ goes to zero, $\overline{v}_1$ vanishes. It is then the simultaneous presence of both terms in the interaction Hamiltonian \eqref{eq:Ham} which guarantees the occurrence of SSC.
It is moreover worth pointing out that the reason why only $\overline{v}_1 $, has survived in the steady-state while $\overline{v}_2 = 0$ is only due to the specific choice of the structure of Eq.~\eqref{eq:Ham} whose orthogonal projection $\Ham_{SE}^{\perp \Ham_S}$ was proportional to $\sigma_x$. An exchange $\sigma_x\to\sigma_y$ in such interaction Hamiltonian produces in fact a corresponding non-zero value of $\overline{v}_2$ and $\overline{v}_1 = 0$.

It is finally central to notice that the result \eqref{eq:SSsolution} is remarkably \textit{independent on the initial state of the system}. This draws a neat line of distinction with the previous phenomenon of coherence trapping studied, e.g., in \cite{Addis2014}. While in their case an opportunely engineered environment and interaction was exploited in order to make a fraction of the initial coherences survive the dephasing process (due to the damping coefficient going to zero in a finite time interval), in our case non-zero SSC have been built up even in the case the system starts in an incoherent state. We emphasize that such SSC generation stems purely from the system-bath interaction and is thus autonomous, in the sense that no coherent driving or measurement is introduced in the scheme.    

To further discuss the above result and in light of comparisons in the subsequent models, we will employ the $l_1-$ norm of coherence $\mathcal{C}=\sqrt{\overline{v}_1^2+\overline{v}_2^2}$, first introduced in \cite{Baumgratz2014}, which can be shown to satisfy all the properties to be considered as a valid coherence measure \cite{Adesso2017RMP}.
\GGrev{First of all, one can immediately notice that  $\overline{v}_1$ is linear in $f_1$, i.e. the strength determining the dephasing, and so $\mathcal{C}$ will be given in units of it. It is important to keep in mind however that the range of $f_1$ remains firmly limited by the weak-coupling condition according to which $\lambda f_{1,2} \ll \omega_0$ \citep{Breuer2002}. }
The maximum of $\mathcal{C}$ with respect to $f_2$ can be instead analytically calculated 
and thus one obtains
\begin{equation}\label{eq:maxCoherences}
\max_{f_2} \mathcal{C}/f_1 = \left|\frac{\left[ \Delta_1\tanh\left(\frac{\omega_0}{2T}\right) +4\lambda\Omega\Gamma(s) + \Delta_2 \right]}{2\sqrt{\omega_0\Delta_1}}\right|.
\end{equation}

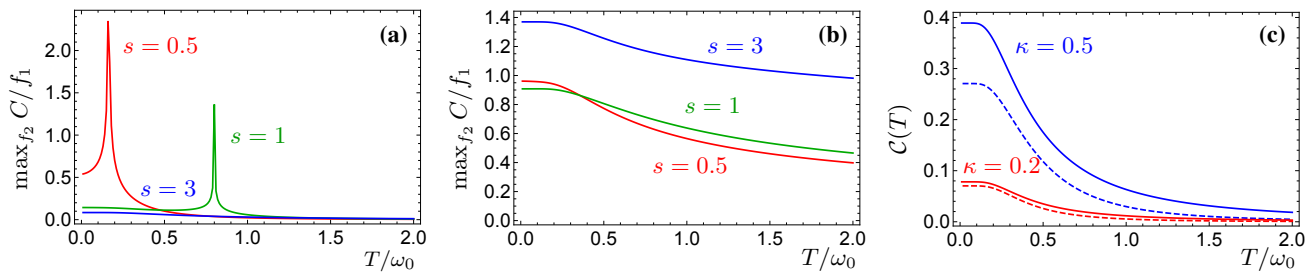
\begin{figure*}[htbp!]
\begin{center}
\begin{tikzpicture} 
  \node (img1)  {\includegraphics[scale=0.4]{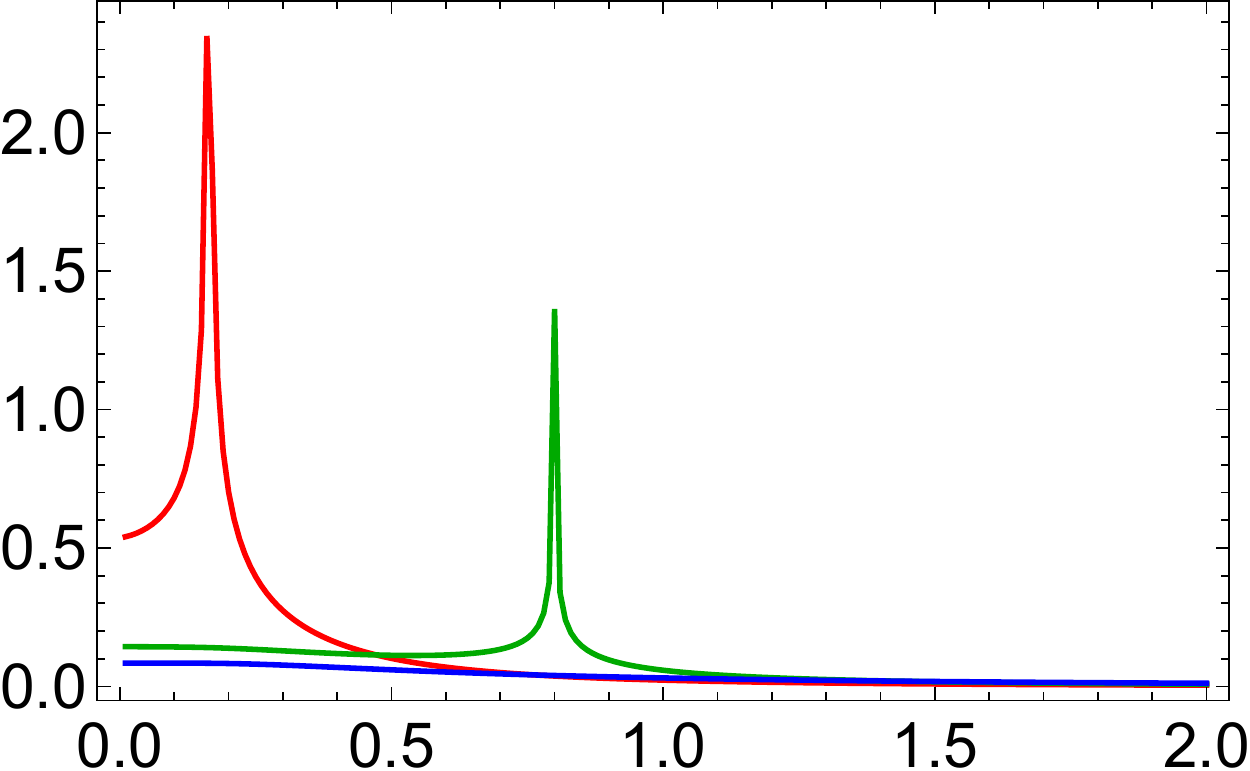}};
    \node[above=of img1, node distance=0cm, yshift=-1.8cm,xshift=-1cm] {{\color{red}$s=0.5$}};
    \node[above=of img1, node distance=0cm, yshift=-1.7cm,xshift=2.1cm] {{\color{black}{\bf{(a)}}}};
     \node[above=of img1, node distance=0cm, yshift=-3cm,xshift=.3cm] {{\color{black!40!green}$s=1$}};
   \node[above=of img1, node distance=0cm, yshift=-3.7cm,xshift=-0.9cm] {{\color{blue}$s=3$}};
      \node[left=of img1, node distance=0cm, rotate=90, anchor=center, yshift=-.8cm,xshift=0cm] {{$\max_{f_2}{C/f_1}$}};
       \node[above=of img1, node distance=0cm, yshift=-4.7cm,xshift=2.05cm] {{\fontsize{8}{10}$T/\omega_0$}};
\end{tikzpicture}
\begin{tikzpicture} 
  \node (img1)  {\includegraphics[scale=0.4]{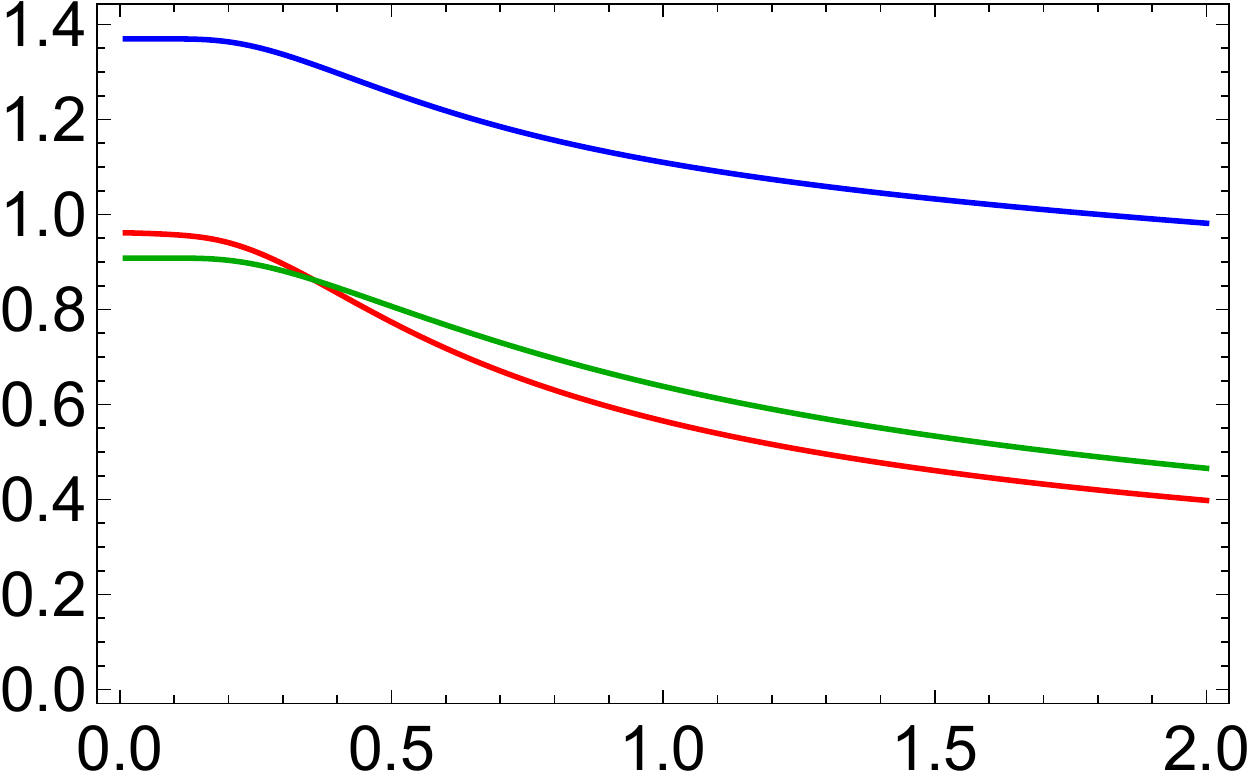}};
      \node[above=of img1, node distance=0cm, yshift=-1.7cm,xshift=2.1cm] {{\color{black}{\bf{(b)}}}};
    \node[above=of img1, node distance=0cm, yshift=-3.4cm,xshift=0.2cm] {{\color{red}$s=0.5$}};
     \node[above=of img1, node distance=0cm, yshift=-2.6cm,xshift=0.5cm] {{\color{black!40!green}$s=1$}};
   \node[above=of img1, node distance=0cm, yshift=-1.8cm,xshift=0.8cm] {{\color{blue}$s=3$}};
      \node[left=of img1, node distance=0cm, rotate=90, anchor=center, yshift=-.8cm,xshift=0cm] {{$\max_{f_2}{C/f_1}$}};
       \node[above=of img1, node distance=0cm, yshift=-4.7cm,xshift=2.05cm] {{$T/\omega_0$}};
\end{tikzpicture}
\begin{tikzpicture} 
  \node (img1)  {\includegraphics[scale=0.4]{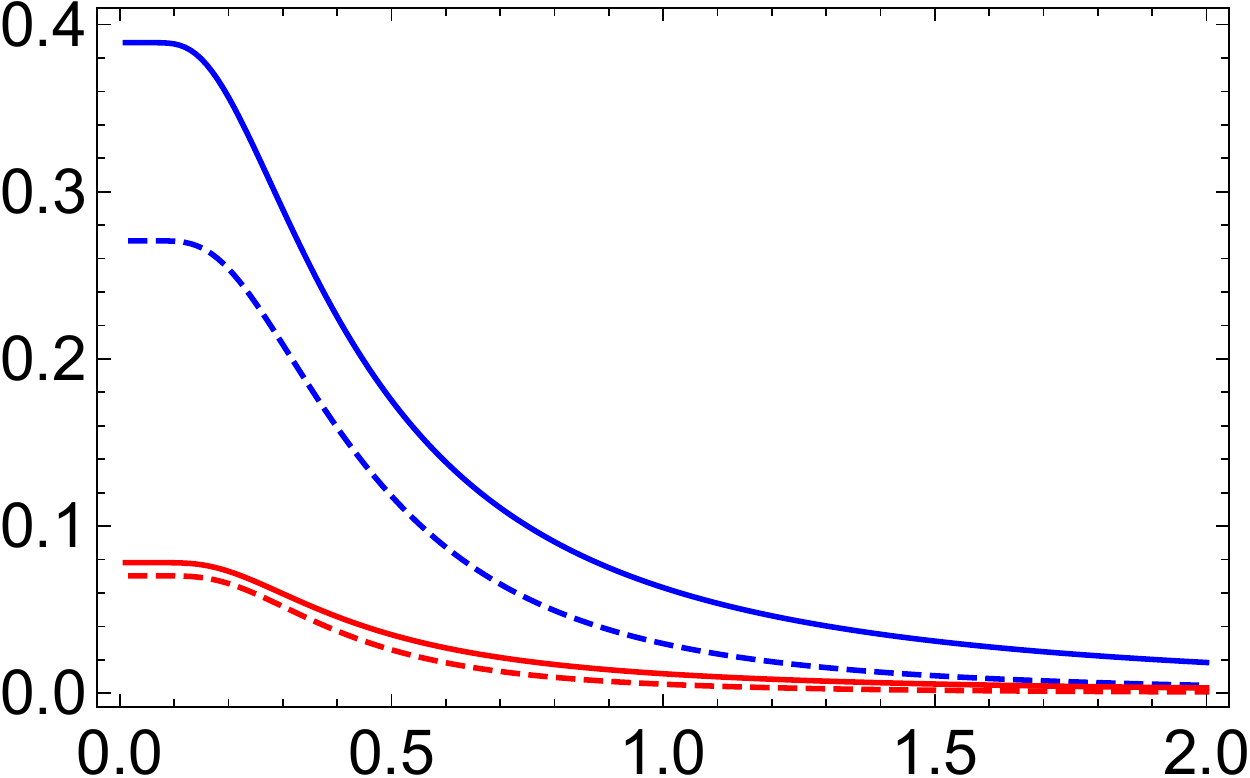}};
      \node[above=of img1, node distance=0cm, yshift=-1.7cm,xshift=2.1cm] {{\color{black}{\bf{(c)}}}};
  \node[above=of img1, node distance=0cm, yshift=-3.4cm,xshift=-1.5cm] {{\color{red}$\kappa=0.2$}};
   \node[above=of img1, node distance=0cm, yshift=-1.8cm,xshift=-0.8cm] {{\color{blue}$\kappa=0.5$}};
      \node[left=of img1, node distance=0cm, rotate=90, anchor=center, yshift=-0.8cm,xshift=-0.0cm] {{${\mathcal{C}}(T)$}};
       \node[above=of img1, node distance=0cm, yshift=-4.7cm,xshift=2.05cm] {{$T/\omega_0$}};
\end{tikzpicture}
\vspace*{-0.2cm}
\caption{Temperature dependence of SSC:  {\bf{(a)}} Eq.~\eqref{eq:maxCoherences} for $\lambda=10^{-2}\omega_0$, and cut-off frequency $\Omega=5\omega_0$; {\bf{(b)}} Eq.~(78) of SM for the same choice of parameters; {\bf{(c)}} the coherence measure $\mathcal{C}$ for different couplings $\kappa$ between the qubit and its effective bath, namely an harmonic mode in turn coupled to a thermal bath (solid lines) and a spin-chain in turn coupled to a thermal bath (dashed lines), see Section E of SM.} \vspace*{-0.5cm}
\label{fig2}
\end{center}
\end{figure*}

Two important general features of Eq.~\eqref{eq:maxCoherences} with respect to $\Omega$ and $T$ can be seen to be generally valid irrespectively of the Ohmicity parameter $s$. First of all, for any fixed value of the temperature, the coherence measure turns out to be a non-decreasing function of the cutoff frequency $\Omega$. This is because a larger $\Omega$ reflects in this model in a higher value of $J(\omega_0)$.
More interestingly, it turns out that the SSC are progressively enhanced as the bath is cooled down, reaching its maximum (as a function of $T$ for any fixed $\Omega$) for $T \to 0$.
On the other side, they are found to vanish in the high-temperature limit, in consistency with the intuition that a more 'classical' hot bath prevents the observation of such phenomenon. \GGrev{However,  remarkably in view of experimental applications, the decay of the generated SSC with increasing $T$ is slow, thus allowing for their observation even at non-zero bath temperatures.}
Alongside these general properties, the remaining parameter $s$ ruling over the low-frequency shape of the spectrum, induces a different behavior of Eq.~\eqref{eq:maxCoherences} whether $s\leq 1$ or $s > 1$. For further details, see Section A.4 of SM.

\paragraph{Ohmic and sub-Ohmic case.---}Performing a first-order Taylor expansion of the integrands of \eqref{eq:Deltas} around $\omega=\omega_0$, allows to realize that in these there is a pole of first order along the so-called \textit{resonance curve} \cite{Clos2012}, implicitly defined by the condition
$\partial_\omega J_{\mathrm{eff}}(\omega,T)|_{\omega=\omega_0}=0$.
The resonance curve physically indicates a match between the system's frequency $\omega_0$ and the frequency $\omega_{\mathrm{max}}$ for which the spectrum reaches its maximum, i.e. such that $\partial_\omega J_{\mathrm{eff}} = 0$. When such dominant environmental frequency $\omega_{\mathrm{max}}$ coincides with $\omega_0$, i.e. is resonant with the system, then the system practically interacts with a locally flat spectrum, the latter notoriously leading to a Markovian dynamics \cite{Clos2012}.
The enhancement of the SSC along the resonance curve can be seen in Fig.~\ref{fig2} {\bf{(a)}}, where the 'spike' of \eqref{eq:maxCoherences}, plotted as function of the bath temperature $T$ for a fixed $\Omega=5\omega_0$, is located on a point of that curve. The resonance curve thus allows a relatively higher coherence for the same $f_1$ and optimized $f_2$. 
The sub-Ohmic case does not qualitatively differ from the Ohmic case, as the resonance condition above highlighted still plays the same role here; see the red curve in Fig.~\ref{fig2} {\bf{(a)}}.  A more thorough discussion can be found in Section A.4 of SM.

\paragraph{Super Ohmic case.---}The singular behavior of the integrand of $\Delta_{1,2}$ in Eq.~\eqref{eq:Deltas} disappears when $s\geq 2$. 
The coherence measure consequently shows a more regular behavior, as shown by the absence of resonance peaks of the blue line in Fig.~\ref{fig2} {\bf{(a)}} which refers to the case $s=3$. It is worth mentioning that such a spectral density is of prominent importance in the context of polarons, when defects or electron tunnelling in a solid coupled to a 3D phononic bath is considered \cite{Leggett1987}.

Since many physical systems, especially in quantum optics, are described by interaction Hamiltonians in the so-called \textit{Rotating Wave Approximation} (RWA), we now 
re-consider the same two-level system and bosonic bath as above, but this time coupled according to
\begin{equation}\label{eq:Ham2}
\Ham_{SE} = f_1\sigma_z\otimes B_E + f_2\left(\sigma_+\otimes b_E + \sigma_-\otimes b^{\dagger}_E\right),
\end{equation}
with $B_E=b_E+b^{\dagger}_E$ and $b_E= \sum_k g_k b_k $. We emphasize that interaction Eq.~\eqref{eq:Ham2}, despite being in RWA, still satisfies the Observation.
Performing the same master-equation based analysis, we obtain also in this case non-zero SSC which remarkably show the same qualitative features and behavior as in the previous model (see Section B of SM for quantitative results and discussions).
The behavior of the maximum coherence measure as a function of the bath temperature $T$ is shown in Fig.~\ref{fig2} {\bf{(b)}}, for the same values of $\Omega$ and $s$ used for the previous model Fig.~\ref{fig2} {\bf{(a)}}. A comparison between the two plots immediately shows that the general trend of SSC to increase when $T$ decreases is found also in this model. At variance with the previous case however, all the singular behavior of the steady-state complenents $\overline{v}_{1,2}$ at the resonance frequency $\omega_0$ is removed for every value of $s$, as detailed and discussed in Section B.4 of SM. This reflects in the absence of enhancement peaks of the SSC even for Ohmic or sub-Ohmic spectral densities.

To provide a complete picture, we have further investigated what happens if we split the two projections $\Ham_{SE}^{\parallel \Ham_S , \perp \Ham_S}$ of an interaction Hamiltonian of the form Eq.~\eqref{eq:Ham} and attribute them to two separate independent thermal baths attached to the system (their temperature being arbitrary and eventually different), neither of which will therefore satisfy the Observation. It turns out that in this case all the SSC vanish (see Section C of SM), this clearly indicating that the SSC obtained above cannot be equivalently generated through alternate sequences of interactions with independent baths each one not generating SSC.
On the other hand, remarkably, we have checked that the generation of SSC by means of an interaction Hamiltonian of the suitable structure evidenced, e.g. \eqref{eq:Ham} and \eqref{eq:Ham2}, are robust even in the presence of an additional dephasing channel on top of it (see Section D of SM). This represents an important support to the feasibility of an eventual experimental test of such theoretically predicted phenomenon, as in many physical situations there often is an unwanted secondary environment.

\paragraph{Equilibration picture- } 
An alternative approach with respect to the master-equation-based one can also be pursued through the equilibration theory. The latter is based on the strong suggestion, widely assumed in the community of closed quantum many-body systems, that quantum systems coupled to a large thermal bath should equilibrate with it, so that the stationary state is given by the local reduced state of the global Gibbs state, i.e. $\rho_S =\mathrm{Tr}_E [\,Z^{-1} e^{-\beta\Ham} ]$, with $Z=\mathrm{Tr}_{SE} [e^{-\beta\Ham} ]$ \cite{Subasi2012,FordPRL1985}. The global Gibbs state differs from the local Gibbs state due to the presence of the interaction Hamiltonian. While being particularly enhanced in the strong coupling regime, even in the weak-coupling regime the corrections to the thermal state $ e^{-\beta\Ham_S}/\mathrm{Tr}_S [e^{-\beta\Ham_S}]$ can become significant \cite{SeifertPRL2016,kolar}. This approach however allows to characterize only steady-state properties and moreover can work only when the system is coupled to a single thermal bath inducing equilibration. Nevertheless, it proves extremely useful in order to characterize SSC even in strongly coupled systems as well as more complex many-body systems.
First of all, we have then employed a perturbative expansion up to second order in the coupling strength of the local reduced state of the global Gibbs state, in the same spirit as done in \cite{Mori2008a}, for the model described by Eq.~\eqref{eq:Ham} (see SM Section E). The result obtained through this different approach have notably confirmed all the above conclusions.

Finally, we employed the equilibration method to access the SSC in different models.  In particular we have firstly considered a qubit, the subsystem of interest, coupled to an harmonic oscillator through an interaction having the crucial composite structure put in evidence in this work and equilibrated by means of an interaction with a thermal reservoir.
Subsequently we also studied the case where the role of the harmonic oscillator is taken by another two-level system; the reader is referred to Section E of SM for all the details.
In both cases a fully numerical approach has been pursued, and thus no weak coupling assumption has been invoked.
In Fig.~\ref{fig2} {\bf{(c)}} we show the temperature dependence of the coherence measure $\mathcal{C}$ for different values of the coupling strength $\kappa$. Solid lines refer to the first model (with Hamiltonian given by Eq.~(107) of SM) while dashed lines refer to the second one (with Hamiltonian given by Eq.~(108) of SM).
A comparison between the curves in Fig.~\ref{fig2} {\bf{(a,b)}} and {\bf{(c)}} shows a great consistency in the behavior of the SSC with respect to $T$, namely its enhancing for a cold bath, thus supporting the feasibility of the SSC  formation for different experimental situations. Finally, this trend of the SSC is remarkably left unchanged significantly even outside the weak coupling regime, as highlighted in Fig.~\ref{fig2} {\bf{(c)}} by the choice $\kappa = 0.5 \omega_0$.

In conclusion, we have provided sufficient prescriptions concerning the structure of the interaction Hamiltonian which allow the formation of \textit{steady-state coherences} in a \GGrev{generic two-level quantum system coupled to a generalized} thermal bath. The SSC this way obtained are remarkably independent on the initial state of the system, so that this scheme can be used to obtain coherences even from an initially purely incoherent system state, and are generally enhanced as the bath temperature is lowered. Interesting outlooks range from theoretical to experimental. 
On the experimental side, the results presented in this work spurs the immediate possibility to generate and observe autonomous SSC, for the first time, in many platforms such as trapped ions \cite{lo2015Nature, Kienzler2016PRL} or superconducting circuits \cite{Lisenfeld2015NatComm,YaleGuys, chiorescu2004Nature, Andre2006NatPhys, yoshihara2017NatPhys}. 
On the other hand, it will be interesting to investigate \GGrev{the generalization to higher dimensional systems and} the situation where two thermal baths at different temperatures are attached to the system through interaction Hamiltonians all of the composite form as, e.g., in Eq.~\eqref{eq:Ham} or \eqref{eq:Ham2}, thus leading to a non-equilibrium steady-state solution (NESS).

{\bf Acknowledgements --}  The Authors acknowledge the support of the Czech Science Foundation (GACR) (grant No. GB14-36681G). R.F. also acknowledges national funding from the MEYS and from European Union's Horizon 2020 (2014-2020) research and innovation framework program under grant agreement No. 731473.

\bibliographystyle{apsrev4-1}

\vspace*{-0.5cm}

\bibliography{SSC}

\widetext
\clearpage
\begin{center}
\textbf{\large Supplementary Material: Steady-state coherences by composite system-bath interactions}\\
\end{center}
\setcounter{equation}{0}
\setcounter{figure}{0}
\setcounter{table}{0}
\setcounter{page}{1}
\makeatletter
\renewcommand{\theequation}{S\arabic{equation}}
\renewcommand{\thefigure}{S\arabic{figure}}

In this Supplementary Material we provide all the detailed calculations of the results presented in the main text, as well as some additional considerations concerning the models discussed.
Note that throughout the paper we set $\hbar = 1$ and $k_B=1$.
The organization is as follows: In Section A we will present the full analysis for the spin-boson undergoing an evolution generated by the particular Hamiltonian considered in Eq. (1) of the main text. In particular in Subsection A.1 the derivation of the second-order master equation will be illustrated, while in Subsection A.2 we will derive the corresponding differential equations for the Bloch vector components. In Subsection A.3 we will perform the long-time limit on the time-dependent coefficients of the above-mentioned master equation, the solution of which will give the steady-state Bloch vector. Finally, in Subsection A.4 additional information and considerations will be given concerning the measure of coherences employed and its behavior in the model under consideration.
Section B will be then devoted to analyze the other model governed by the Hamiltonian Eq. (6) of the main text and the structure of this Section will retrace the previous one.
In Section C the discussion will be focused on proving that the splitting of the two terms composing the interaction Hamiltonian Eq.(1) to two separate independents heat baths makes any steady-state coherences (SSC for brevity throughout the text) disappear, while in Section D we will show the robustness of our results by proving that the addition of another dephasing channel to a bath possessing an interaction of the crucial form as in Eq.(1) will not hinder the formation of SSC.
Finally, in Section E we will discuss the alternative approach based on equilibration theory, explicitly presenting the calculations and results of a second-order expansion of the local reduced state of the global Gibbs state. 

\section{A --- The First Model}
\label{Sec:FirstModel}

We briefly remind that the model under consideration consists of a two-level system interacting with a bosonic bath such that the total Hamiltonian of the composite system is given by
\begin{equation}\label{eq:Ham1}
\Ham = \Ham_S + \Ham_E + \Ham_{SE} = \frac{\omega_0}{2}\sigma_z\otimes\Id_E + \Id_S\otimes\sum_k \omega_k b^{\dagger}_k b_k + \left( f_1\sigma_z + f_2\sigma_x\right)\otimes B_E,
\end{equation}
with $B_E = \sum_k g_k \left(b_k + b^{\dagger}_k\right)$ and $f_{1,2}$ being two completely generic coupling constants.
In what follows we will assume that the system starts in a product state of the form
\begin{equation}\label{eq:factcond}
\rho_{SE}(0) = \rho_S(0)\otimes\rho_{\beta},
\end{equation}
with $\rho_{\beta} \equiv Z^{-1}e^{-\beta\Ham_E}$ ($Z = \mathrm{Tr}_E\left[e^{-\beta\Ham_E}\right]$ with $\beta$ being the inverse temperature) and where $\rho_S(0)$ is left completely generic.

\subsection{A.1 --- The second-order time-local master equation}
\label{Subsec:ME1}

We now proceed to show the detailed derivation of the time-local master equation describing the dynamics of the reduced system within a perturbative approach up in the coupling strength.
The whole procedure relies on the well-known projection-operator technique and time-convolutionless expansion of the dynamical generator up to second-order, to which we refer the interested reader to the extensive exposition e.g. in \cite{Breuer2002}.

Starting from the overall unitary evolution, the evolution of the open system is given by
\begin{equation}\label{eq:2oTCL}
\frac{d}{dt}\rho_S(t) = -i \left[ \Ham_S,\rho_S(t) \right] - \int_0^t d\tau \mathrm{Tr}_E \left\{ \left[ \tilde{\Ham}_{SE},\, \left[ \tilde{\Ham}_{SE}(-\tau),\,\rho_S(t) \right]\right] \right\}.
\end{equation}
where 
\begin{equation}\label{eq:interactionHamevol}
\tilde{\Ham}_{SE} (t) = \left[f_1 \sigma_z + f_2 \left( \sigma_+e^{i\omega_0 t} + \sigma_-e^{-i\omega_0 t} \right)\right]\otimes \sum_k g_k \left( b_k e^{-i\omega_k t} +  b^{\dagger}_k e^{i\omega_k t}\right),
\end{equation}
is the system-bath interaction Hamiltonian evolved in interaction picture, i.e. $\tilde{\Ham}_{SE} (t) = e^{i\left(\Ham_S+\Ham_E\right) t} \Ham_{SE} e^{-i\left(\Ham_S+\Ham_E\right) t}$.
Eq. \eqref{eq:2oTCL} can be recast in the familiar time-dependent Gorini-Kossakowski-Sudashan-Lindblad (GKSL) form \cite{Gorini1976}
\begin{equation}\label{eq:MEgeneral}
\frac{d}{dt}\rho_S(t)\!=\!\mathcal{K}_{TCL}(t)\rho_S(t)\!=\!-i \left[ \mathbf{\Ham}^{LS}(t),\,\rho_S(t)\right] + \sum_{\alpha\beta=1}^{3} \mathit{A}_{\alpha\beta}(t) \left( \sigma_{\alpha}\rho_S(t)\sigma^{\dagger}_{\beta} - \frac{1}{2}\lbrace\sigma^{\dagger}_{\beta}\sigma_{\alpha},\rho_S(t)\rbrace \right),
\end{equation}
where the Lamb-Shift Hamiltonian $\mathbf{\Ham}^{LS}(t)$, of coefficients $\mathit{\Ham}^{LS}_{\alpha\beta}(t)$, and the Kossakowski matrix $\mathbf{A}(t)$, of coefficients $\mathit{A}_{\alpha\beta}(t)$, are $3\times 3$ matrices.

In order to accomplish the task at hand, we first fix the orthonormal Hilbert-Schmidt (HS) basis of $\mathcal{B}(\HILB)$ (i.e. the set of bounded operators acting on $\HILB$) to be $\lbrace \Id/\sqrt{2} , \sigma_+, \allowbreak \sigma_-, \sigma_z /\sqrt{2} \rbrace$, such that it meets the requirements of \cite{Gorini1976}. We stress that the subsequent results are obviously independent of the particular choice of HS basis.
It is worth stressing that Eq. \eqref{eq:MEgeneral} does not rely on the Born-Markov or secular approximation, thus being suitable for describing non-Markovian dynamics even at short timescale.

The explicit expansion of the double commutator of the interaction Hamiltonian entering Eq. \eqref{eq:2oTCL} consists of the sum of the following four contributions:
\begin{itemize}
\item 
\begin{align*}
&- f^2_1 \int_0^t d\tau \left[ \sigma_z \sigma_z \rho_S(t) \mean{B_E B_E(-\tau)}_{\beta} - \sigma_z\rho_S(t)\sigma_z \mean{B_E B_E(-\tau)}_{\beta}\right. \\
&\left.\qquad\qquad\qquad\qquad- \sigma_z\rho_S(t)\sigma_z \mean{B(-\tau)B}_{\beta} + \rho_S(t) \sigma_z\sigma_z \mean{B_E(-\tau) B_E}_{\beta}\right]\\
&= - f^2_1 \int_0^t d\tau \left[ \rho_S(t) \left( \mean{B_E B_E(-\tau)}_{\beta} + \mean{B_E(-\tau) B_E}_{\beta}\right)  \right.\\
&\left.\qquad\qquad\qquad\qquad- \sigma_z\rho_S(t)\sigma_z \left( \mean{B_E B_E(-\tau)}_{\beta} + \mean{B_E(-\tau) B_E}_{\beta}\right)\right]\\
&= \left( \sigma_z\rho_S(t)\sigma_z - \rho_S(t)\right) f^2_1  \int_0^t d\tau \left( \mean{B_E B_E(-\tau)}_{\beta} + \mean{B_E(-\tau) B_E}_{\beta}\right).
\end{align*}
Note that we have used the fact that $\sigma_z^2 = \Id_S$.
\item
\begin{align*}
&-f_1f_2 \int_0^t d\tau \left[ \sigma_z \sigma_x(-\tau) \rho_S(t) \mean{B_E B_E(-\tau)}_{\beta} - \sigma_z\rho_S(t)\sigma_x(-\tau) \mean{B_E B_E(-\tau)}_{\beta}\right. \\
&\left.\qquad\qquad\qquad\qquad- \sigma_x(-\tau)\rho_S(t)\sigma_z \mean{B(-\tau)B}_{\beta} + \rho_S(t) \sigma_x(-\tau)\sigma_z \mean{B_E(-\tau) B_E}_{\beta}\right]\\
&= -f_1f_2 \int_0^t d\tau  \mean{B_E B_E(-\tau)}_{\beta} \left[ \sigma_z\sigma_+\rho_S(t) e^{-i\omega_0\tau} + \sigma_z\sigma_-\rho_S(t) e ^{i\omega_0\tau} \right.\\
&\left.\qquad\qquad\qquad\qquad\qquad\qquad\qquad\qquad\qquad- \sigma_z\rho_S(t)\sigma_+e^{-i\omega_0\tau} - \sigma_z\rho_S(t)\sigma_-e^{i\omega_0\tau}\right] \\
& - f_1f_2 \int_0^t d\tau  \mean{B (-\tau) B}_{\beta} \left[ \rho_S(t) \sigma_+\sigma_z e^{-i\omega_0\tau} + \rho_S(t)\sigma_-\sigma_z e^{i\omega_0\tau} \right.\\
&\left.\qquad\qquad\qquad\qquad\qquad\qquad\qquad\qquad\qquad- \sigma_+\rho_S(t)\sigma_ze^{-i\omega_0\tau} - \sigma_-\rho_S(t)\sigma_ze^{i\omega_0\tau}\right] 
\end{align*}
\item
\begin{align*}
&- f_1f_2 \int_0^t d\tau \left[ \sigma_x \sigma_z \rho_S(t) \mean{B_E B_E(-\tau)}_{\beta} - \sigma_x\rho_S(t)\sigma_z \mean{B_E B_E(-\tau)}_{\beta}\right. \\
&\left.\qquad\qquad\qquad\qquad- \sigma_z\rho_S(t)\sigma_x \mean{B(-\tau)B}_{\beta} + \rho_S(t) \sigma_z\sigma_x \mean{B_E(-\tau) B_E}_{\beta}\right]\\
&= - f_1f_2 \int_0^t d\tau \mean{B_E B_E(-\tau)}_{\beta}\left[ \sigma_+\sigma_z\rho_S(t) + \sigma_-\sigma_z\rho_S(t) - \sigma_+\rho_S(t)\sigma_z - \sigma_-\rho_S(t)\sigma_z \right]\\
& - f_1f_2 \int_0^t d\tau \mean{B_E(-\tau) B_E}_{\beta}\left[ \rho_S(t)\sigma_z\sigma_+ + \rho_S(t)\sigma_z\sigma_- - \sigma_z\rho_S(t)\sigma_+ - \sigma_z\rho_S(t)\sigma_- \right]
\end{align*}
\item 
\begin{align*}
& - f^2_2 \int_0^t d\tau \left[ \sigma_x \sigma_x(-\tau) \rho_S(t) \mean{B_E B_E(-\tau)}_{\beta} - \sigma_x\rho_S(t)\sigma_x(-\tau) \mean{B_E B_E(-\tau)}_{\beta}\right. \\
&\left.\qquad\qquad\qquad\qquad- \sigma_x(-\tau)\rho_S(t)\sigma_x \mean{B(-\tau)B}_{\beta} + \rho_S(t) \sigma_x(-\tau)\sigma_x \mean{B_E(-\tau) B_E}_{\beta}\right]\\
&= - f^2_2  \int_0^t d\tau \left[ \left( \sigma_++\sigma_- \right)\left(\sigma_+e^{-i\omega_0 \tau}+\sigma_-e^{i\omega_0 \tau}\right) \rho_S(t) \mean{B_E B_E(-\tau)}\right.\\
&\qquad\qquad\qquad\qquad\qquad\qquad \left. - \left( \sigma_++\sigma_- \right)\rho_S(t) \left(\sigma_+e^{-i\omega_0 \tau}+\sigma_-e^{i\omega_0 \tau}\right) \mean{B_E B_E(-\tau)}\right.\\
&\qquad\qquad\qquad\qquad\qquad\qquad \left. - \left(\sigma_+e^{-i\omega_0 \tau}+\sigma_-e^{i\omega_0 \tau}\right)\rho_S(t)\left( \sigma_++\sigma_- \right)\mean{B_E(-\tau) B_E }\right.\\
&\qquad\qquad\qquad\qquad\qquad\qquad \left. + \rho_S(t)\left(\sigma_+e^{-i\omega_0 \tau}+\sigma_-e^{i\omega_0 \tau}\right)\left( \sigma_++\sigma_- \right)\mean{B_E(-\tau) B_E }\right]\\
&= - f^2_2 \int_0^t d\tau \mean{B_E B_E(-\tau)}_{\beta} \left[ \sigma_+\sigma_-\rho_S(t) e^{i\omega_0 \tau} + \sigma_-\sigma_+\rho_S(t) e^{-i\omega_0 \tau} \right.\notag\\
&\left.\qquad\qquad - \sigma_+\rho_S(t)\sigma_+ e^{-i\omega_0 \tau} - \sigma_+\rho_S(t)\sigma_- e^{i\omega_0 \tau} - \sigma_-\rho_S(t)\sigma_+ e^{-i\omega_0 \tau} - \sigma_-\rho_S(t)\sigma_- e^{i\omega_0 \tau}\right]\\
& \quad- f^2_2 \int_0^t d\tau \mean{B_E(-\tau) B_E}_{\beta} \left[ \rho_S(t) \sigma_+\sigma_-e^{-i\omega_0 \tau} + \rho_S(t) \sigma_-\sigma_+e^{i\omega_0 \tau}\right.\notag\\
&\left.\qquad\qquad - \sigma_+\rho_S(t)\sigma_+ e^{-i\omega_0 \tau} - \sigma_+\rho_S(t)\sigma_- e^{-i\omega_0 \tau} - \sigma_-\rho_S(t)\sigma_+ e^{i\omega_0 \tau} - \sigma_-\rho_S(t)\sigma_- e^{i\omega_0 \tau}\right]
\end{align*}
\end{itemize}
We explicitly give here the derivation by comparison of the coefficients of the time-local master equation. By making use of the relation $ \left(\mean{BB(-\tau)}_{\beta}\right)^* = \mean{B(-\tau)B}_{\beta}$, it is now straightforward to derive by comparison the time-dependent coefficients of the time-local master equation \eqref{eq:MEgeneral}. In particular, the Lamb-Shift Hamiltonian in Eq. \eqref{eq:MEgeneral} is given by the matrix
\begin{equation}
\mathbf{\Ham}^{LS}(t) = \begin{pmatrix}
h_{11}(t) & 0 & \frac{h_{13}(t)}{\sqrt{2}} \\ 
0 & h_{22}(t) & \frac{h_{23}(t)}{\sqrt{2}} \\
\frac{h^*_{13}(t)}{\sqrt{2}} &\frac{h^*_{23}(t)}{\sqrt{2}} & 0
\end{pmatrix},
\end{equation}
where the non-zero entries are given by
\begin{align*}
&h_{11}(t) = f^2_2\, \int_0^t d\tau\, \mathrm{Im}\left[\mean{B B(-\tau)}_{\beta} e^{i\omega_0\tau}\right]\\
&h_{22}(t) =  f^2_2\int_0^t d\tau\, \mathrm{Im}\left[\mean{B B(-\tau)}_{\beta} e^{-i\omega_0\tau}\right]\\
&h_{13}(t) = \sqrt{2}\left(f_1 f_2\right) \int_0^t d\tau\, e^{-i\frac{\omega_0\tau}{2}} \,\,\mathrm{Im}\left[ \mean{B B(-\tau)}_{\beta} e^{i\frac{\omega_0\tau}{2}} \right] \\
&h_{23}(t) = \sqrt{2}\left(f_1 f_2\right) \int_0^t d\tau\, e^{i\frac{\omega_0\tau}{2}} \,\,\mathrm{Im}\left[ \mean{B B(-\tau)}_{\beta} e^{-i\frac{\omega_0\tau}{2}} \right].
\end{align*}
Note that, for brevity, we employed the notation $\mean{\cdot}_\beta$ to indicate $\mathrm{Tr}_E\left[\cdot\rho_{\beta}\right]$.
Similarly, the Kossakowski matrix responsible for dissipation 
\begin{equation}
\mathbf{A}(t) = \begin{pmatrix}
a_{11}(t) & a_{12}(t) & \frac{a_{13}(t)}{\sqrt{2}} \\ 
a^*_{12}(t) & a_{22}(t) & \frac{a_{23}(t)}{\sqrt{2}} \\
\frac{a^*_{13}(t)}{\sqrt{2}} & \frac{a^*_{23}(t)}{\sqrt{2}} & \frac{a_{33}(t)}{2}
\end{pmatrix},
\end{equation}
has the following entries
\begin{align*}
&a_{11}(t) = f^2_2 \int_0^t d\tau\, 2\mathrm{Re}\left[\mean{B B(-\tau)}_{\beta} e^{-i\omega_0\tau}\right]\\
&a_{22}(t) =f^2_2 \int_0^t d\tau\, 2\mathrm{Re}\left[\mean{B B(-\tau)}_{\beta} e^{i\omega_0\tau}\right]\\
&a_{33}(t) = f^2_1  \int_0^t d\tau \,\,4\mathrm{Re}\left[\mean{B B(-\tau)}_{\beta}\right]\\
&a_{12}(t) =  f^2_2 \int_0^t d\tau \,e^{-i\omega_0\tau}\,\,2\mathrm{Re}\left[ \mean{B B(-\tau)}_{\beta}\right]\\
&a_{13}(t) = \sqrt{2}\left(f_1 f_2\right) \int_0^t d\tau\, e^{-i\frac{\omega_0\tau}{2}} \,\,2\mathrm{Re}\left[ \mean{B B(-\tau)}_{\beta} e^{-i\frac{\omega_0\tau}{2}} \right]\\
&a_{23}(t) = \sqrt{2}\left(f_1 f_2\right) \int_0^t d\tau\, e^{i\frac{\omega_0\tau}{2}} \,\,2\mathrm{Re}\left[ \mean{B B(-\tau)}_{\beta} e^{i\frac{\omega_0\tau}{2}} \right].
\end{align*}
All the information determining the dynamics of the reduced system are thus encoded in the time-dependent functions obtained above, all of which are weighted linear combinations of the \textit{environmental correlation function} 
\begin{align}\label{eq:envcorrgeneric}
\mean{B_E B_E(-\tau)}_{\beta} = \sum_k |g_k|^2 \left[ \mean{n_k}e^{i\omega_k\tau} + \left(1+\mean{n_k}\right)e^{-i\omega_k\tau}\right].
\end{align}
Taking the continuum limit of bath modes, the spectral density \cite{Breuer2002} becomes $J(\omega) = 4 f(\omega) |g(\omega)|^2$ (we will assume without loss of generality that $f(\omega)$, which is the modes distribution, is constant and equal to $1$) and therefore Eq. \eqref{eq:envcorrgeneric} becomes
\begin{equation}\label{eq:envcorrD1D2}
\mean{B_E B_E(-\tau)}_{\beta} = \int_0^{+\infty} d\omega\, J(\omega)\left[\coth\left(\frac{\omega}{2T}\right)\cos(\omega \tau) - i\sin(\omega \tau)\right] \equiv \frac{1}{2}\left[D_1(\tau) - i D_2(\tau)\right],
\end{equation}
where
\begin{align}\label{eq:D1D2}
&D_1(\tau) = 2\int_0^{+\infty} d\omega J_{\mathrm{eff}}(\omega,\Omega,T_E) \cos(\omega\tau) \quad\quad \left( \text{with} \,\, J_{\mathrm{eff}}(\omega,\Omega,T_E) \equiv J(\omega) \coth\left(\frac{\omega}{2T_E}\right)\right)\\
&D_2(\tau) = 2\int_0^{+\infty} d\omega J(\omega) \sin(\omega\tau)
\end{align}
are known in the literature as \textit{noise} and \textit{dissipation} kernels, respectively \cite{Breuer2002}.

In what follows we will choose the spectral density $J(\omega)$ to be of following general form \cite{Leggett1987}
\begin{equation}\label{eq:SD}
J(\omega) = \lambda\frac{\omega^s}{\Omega^{s-1}}e^{-\frac{\omega}{\Omega}},
\end{equation}
where $\lambda$ is the coupling strength, $\Omega$ is a cut-off frequency and $s$ is the Ohmicity parameter. The latter is known to lead to a sub-Ohmic spectrum for $s<1$, to an Ohmic one for $s=1$ and finally to a super-Ohmic spectrum for $s>1$.
Closed analytic expressions for the noise and dissipation kernels are known in the cases of either Ohmic spectral density and generic temperature $T$ of the environment, or for generic $s$ and zero-temperature environment.
Here, a closed analytic expression also for generic $s$, cutoff frequency $\Omega$ and bath temperature $T$, is given and reads
\begin{align}
&D_1(t) = \lambda  T \Omega  \left(-\frac{T}{\Omega }\right)^s \left[ \psi^{(s)}\left(\frac{T}{\Omega } (1-i \Omega t)\right)+\psi ^{(s)}\left(\frac{T}{\Omega }(1+i \Omega t)\right)\right.\notag\\
&\qquad\qquad\qquad\qquad\qquad\qquad\qquad\qquad\left.+\psi^{(s)}\left(1+\frac{T}{\Omega }(1-i \Omega t)\right)+\psi ^{(s)}\left(1+\frac{T}{\Omega }(1+i \Omega t)\right)\right]\\
&D_2(t) = \frac{2 \lambda  \Omega^2 \Gamma (s+1)}{\left[1+(\Omega t)^2 \right]^{s+1}} \mathrm{Im}\left[(1+i \Omega t)^{s+1}\right],
\end{align}
where $\Gamma (z)$ denotes the Euler's gamma function and $\psi^{(n)}(z)$ is the $n-$ derivative of the Euler's digamma function $\psi(z)\equiv \Gamma'(z)/\Gamma(z)$.
Re-expressing the coefficients of the Lamb-Shift Hamiltonian and of the Kossakowski matrix derived above in terms of the noise and dissipation kernels finally leads to
\begin{align*}
&h_{11}(t) = f^2_2\int_0^t d\tau\,\, \frac{1}{2}\left[ D_1(\tau)\sin(\omega_0\tau) - D_2(\tau) \cos(\omega_0\tau) \right]\\
&h_{22}(t) = - f^2_2\int_0^t d\tau\,\, \frac{1}{2}\left[ D_1(\tau)\sin(\omega_0\tau) + D_2(\tau) \cos(\omega_0\tau) \right]\\
&h_{13}(t) = \left( h_{31}(t) \right)^* = \sqrt{2}\left(f_1 f_2\right) \int_0^t d\tau\, e^{-i\frac{\omega_0\tau}{2}} \,\,\frac{1}{2}\left[ D_1(\tau)\sin\left(\frac{\omega_0\tau}{2}\right) - D_2(\tau) \cos\left(\frac{\omega_0\tau}{2}\right) \right]
\\
&h_{23}(t) = \left( h_{32}(t) \right)^* = \sqrt{2}\left(f_1 f_2\right) \int_0^t d\tau\, e^{i\frac{\omega_0\tau}{2}} \,\,\frac{1}{2}\left[ D_1(\tau)\sin\left(\frac{\omega_0\tau}{2}\right) + D_2(\tau) \cos\left(\frac{\omega_0\tau}{2}\right) \right],
\end{align*}
and
\begin{align*}
&a_{11}(t) = f^2_2 \int_0^t d\tau\,\, \left[D_1(\tau)\cos(\omega_0\tau) - D_2(\tau) \sin(\omega_0\tau)\right]\\
&a_{22}(t) =   f^2_2 \int_0^t d\tau\,\, \left[D_1(\tau)\cos(\omega_0\tau) + D_2(\tau) \sin(\omega_0\tau)\right]\\
&a_{33}(t) =  f^2_1\int_0^t d\tau\, 2 D_1(\tau)\\
&a_{12}(t) =  \left(a_{21}(t)\right)^* = f^2_2 \int_0^t d\tau \, D_1(\tau) \left[\cos(\omega_0\tau) - i\sin(\omega_0\tau)\right]\\
&a_{13}(t) =  \left(a_{31}(t)\right)^* = \sqrt{2}\left(f_1 f_2\right) \int_0^t d\tau\, e^{-i\frac{\omega_0\tau}{2}}\left[D_1(\tau)\cos\left(\frac{\omega_0\tau}{2}\right) - D_2(\tau) \sin\left(\frac{\omega_0\tau}{2}\right)\right]\\
&a_{23}(t) = \left(a_{32}(t)\right)^* = \sqrt{2}\left(f_1 f_2\right) \int_0^t d\tau\, e^{i\frac{\omega_0\tau}{2}}\left[D_1(\tau)\cos\left(\frac{\omega_0\tau}{2}\right) + D_2(\tau) \sin\left(\frac{\omega_0\tau}{2}\right)\right].
\end{align*}

\subsection{A.2 --- From the master equation to the Bloch differential equations}
\label{Subsec:Bloch1}

From Eq. \eqref{eq:2oTCL}, one can derive the equations of motion for the Bloch vector $\vett{v}(t)$, implicitly defined through the relation
\begin{equation}
\rho_S(t) = \frac{1}{2}\left(\Id_2 + \vett{v}(t)\cdot\bgreek{\sigma}\right), \,\, v_{x,y,z}(t) = \mathrm{Tr}_S\left[\sigma_{x,y,z} \, \rho_S(t)\right],
\end{equation}
The three components of the Bloch vector are explicitly given by
\begin{align*}
&v_1(t) = \rho_{10}(t) + \rho_{01}(t) = 2 \mathrm{Re}\left[\rho_{01}(t)\right]\\
&v_2(t) = i \left( \rho_{10}(t) - \rho_{01}(t)\right) = 2 \mathrm{Im}\left[\rho_{01}(t)\right]\\
&v_3(t) = \rho_{11}(t) - \rho_{00}(t) = 2 \rho_{11}(t) - 1.
\end{align*}
The third component of the Bloch vector gives the population imbalance of the qubit while the first two components relate with the real and imaginary part of coherences, respectively. 

\subsubsection{Equation of motion for $v_3(t)$}

Let us then start with the differential equation ruling over the evolution of $v_3(t)$. A simple calculation yields
\begin{align}\label{v3.1}
\frac{d}{dt}v_3(t) &= -\left[a_{11}(t) + a_{22}(t)\right] v_3(t) + \left[ a_{11}(t) - a_{22}(t)\right] + \frac{1}{\sqrt{2}} v_1(t) \mathrm{Re}\left[ f(t) \right] -  \frac{1}{\sqrt{2}} v_2(t) \mathrm{Im}\left[ f(t) \right]\notag\\
&= \sum_{j=1}^3 M_{3j}(t) v_j(t) + b_3(t)
\end{align}
where
\begin{equation}
f(t) \equiv a_{13}(t) + a_{32}(t) + 2i \left[ h_{13}(t) - h_{32}(t) \right].
\end{equation}
In order to gain some insight on this expression, we firstly substitute the expressions of the entries of the Lamb-Shift Hamiltonian and of the Kossakowski matrix derived above.
Moreover, we will exploit the following relation, valid $\forall z \in \mathbb{C}$:
\begin{equation}\label{ReIm}
\mathrm{Re}\left[i z\right] = - \mathrm{Im}\left[z\right],\qquad \qquad \mathrm{Im}\left[i z\right] = \mathrm{Re}\left[z\right].
\end{equation}
Simple calculations lead to
\begin{align*}
M_{13}(t) \equiv \frac{1}{\sqrt{2}} \mathrm{Re}\left[ f(t) \right] &= f_1f_2\int_0^t d\tau\,\, 2 D_1(\tau) \notag\\
&= 2 f_1f_2\int_0^{+\infty}\,d\omega J_{\mathrm{eff}}(\omega,T) \int_0^t\,d\tau\cos(\omega\tau)
\end{align*}
with $J_{\mathrm{eff}}(\omega,T) \equiv J(\omega) \coth\left(\frac{\omega}{2T}\right)$ being the effective spectral density.
Similarly we get
\begin{equation}
M_{23}(t) \equiv \frac{1}{\sqrt{2}} \mathrm{Im}\left[ f(t) \right] = 0.
\end{equation}
as well as
\begin{align*}
& M_{33} (t) \equiv -\left[ a_{11}(t) + a_{22}(t)\right] = -f^2_2 \int_0^t d\tau\,\,  2 D_1(\tau)\cos(\omega_0\tau) \\
& b_3(t) \equiv a_{11}(t) - a_{22}(t) = - f^2_2 \int_0^t d\tau\,\, 2 D_2(\tau)\sin(\omega_0\tau).
\end{align*}

\subsubsection{Equation of motion for $v_1(t)$}

We now consider the differential equation that governs the evolution of $v_1(t)$.
While the third component rules over the population (inversion) dynamics, the first and second component takes into account for the coherence dynamics. 
We have that
\begin{align}\label{v1.1}
\frac{d}{dt}v_1(t) &= -\frac{1}{2}\mathrm{Re}\left[l(t)\right] v_1(t) + \frac{1}{2}\mathrm{Im}\left[l(t)\right] v_2(t) + \frac{1}{\sqrt{2}} m(t) v_3(t) + \frac{2}{\sqrt{2}} \left( \mathrm{Re}\left[a_{23}(t)\right]  - \mathrm{Re}\left[a_{13}(t)\right]  \right)\notag\\
&= \sum_{j=1}^3 M_{1j}(t) v_j(t) + b_1(t)
\end{align}
with
\begin{align*}
&l(t) \equiv a_{11}(t) + a_{22}(t) + 2a_{33}(t) -2 a_{12}(t) - 2i \left( h_{11}(t) - h_{22}(t) + \omega_0 \right)\\
&m(t) \equiv \mathrm{Re}\left[a_{13}(t)\right]  + \mathrm{Re}\left[a_{23}(t)\right] + 2\mathrm{Im}\left[h_{13}(t)\right] + 2\mathrm{Im}\left[h_{23}(t)\right].
\end{align*}
Making use of \eqref{ReIm}, of the fact that $a_{ii}(t), h_{ii}(t) \in \mathbb{R}, \,\,\forall i=1,2,3$, and finally that 
\begin{align}\label{noteq1}
& a_{11}(t) + a_{22}(t) - 2 \mathrm{Re}\left[a_{12}(t)\right] = 0\\
& h_{11}(t) - h_{22}(t) + \mathrm{Im}\left[a_{12}(t) \right] = 0,
\end{align}
we find
\begin{align}
&M_{11}(t) \equiv -\frac{1}{2}\mathrm{Re}\left[l(t)\right] = - a_{33}(t) = - f^2_1\int_0^t d\tau\, 2 D_1(\tau)\\
&M_{12}(t) \equiv \frac{1}{2}\mathrm{Im}\left[l(t)\right] = 0\\
&M_{13}(t) \equiv \frac{1}{\sqrt{2}} m(t) = f_1f_2 \int_0^t d\tau\,\, 2D_1(\tau)\cos\left(\omega_0\tau\right)
\end{align}
and
\begin{align}
b_1(t) &\equiv \frac{2}{\sqrt{2}} \left( \mathrm{Re}\left[a_{23}(t)\right]  - \mathrm{Re}\left[a_{13}(t)\right]  \right) \notag\\
&= f_1f_2 \int_0^t d\tau\, \, D_2(\tau)\sin\left(\omega_0\tau\right).
\end{align}

\subsubsection{Equation of motion for $v_2(t)$}

We finally derive the proper differential equation for $v_2(t)$. We have that
\begin{align}\label{v2.1}
&\frac{d}{dt}v_2(t) = \frac{1}{2}\mathrm{Re}\left[r(t)\right] v_1(t) - \frac{1}{2}\mathrm{Im}\left[r(t)\right] v_2(t) + \frac{1}{\sqrt{2}}  n(t) v_3(t) + \frac{2}{\sqrt{2}}  \mathrm{Im}\left[a_{23}(t)\right]  + \frac{2}{\sqrt{2}}  \mathrm{Im}\left[a_{13}(t)\right]\notag\\
&= \sum_{j=1}^3 M_{2j}(t) v_j(t) + b_2(t),
\end{align}
with
\begin{align*}
&r(t) \equiv i a_{11}(t) + i a_{22}(t) + 2i a_{33}(t) +2 i a_{12}(t) + 2 \left( h_{11}(t) - h_{22}(t) + \omega_0 \right)\\
&n(t) \equiv \mathrm{Im}\left[a_{23}(t)\right]  - \mathrm{Im}\left[a_{13}(t)\right] + 2\mathrm{Re}\left[h_{13}(t)\right] - 2\mathrm{Re}\left[h_{23}(t)\right].
\end{align*}
Making use of \eqref{ReIm} and of \eqref{noteq1}, one finally arrives to
\begin{align}
&M_{21}(t) \equiv  \frac{1}{2}\mathrm{Re}\left[r(t)\right] = 2\left( h_{11}(t) - h_{22}(t) + \omega_0 \right) =\omega_0 + f^2_2 \int_0^t d\tau\,\,  2 D_1(\tau)\sin(\omega_0\tau)\\
&M_{22}(t) \equiv - \frac{1}{2}\mathrm{Im}\left[r(t)\right] =  a_{11}(t) + a_{22}(t) + 2a_{33}(t) = f^2_2 \int_0^t d\tau\,\,  2 D_1(\tau)\cos(\omega_0\tau) + f^2_1\int_0^t d\tau\, 2 D_1(\tau)\\
&M_{23}(t) \equiv \frac{1}{\sqrt{2}} n(t) = f_1f_2 \int_0^t d\tau\,\,2 D_1(\tau)\sin\left(\omega_0\tau\right)
\end{align}
and finally
\begin{align}
b_2(t) &\equiv \frac{2}{\sqrt{2}}  \left(\mathrm{Im}\left[a_{23}(t)\right]  +   \mathrm{Im}\left[a_{13}(t)\right]\right) \notag\\
&= f_1f_2 \int_0^t d\tau\, \, 2 D_2(\tau)\left[ 1 - \cos\left(\omega_0\tau\right)\right]
\end{align}

\subsubsection{The Bloch Equations}

In light of the above calculations, we obtain the following system of linear differential equations
\begin{equation}\label{eq:BVdiffEqs}
\frac{d}{dt}\vett{v}(t) = \mathbf{M}(t)\vett{v}(t) + \mathbf{b}(t),
\end{equation}
where
\begin{align*}
&\mathbf{M}(t) = \begin{pmatrix}
-f^2_1 \gamma_1(t) & -\omega_0 &  f_1f_2\gamma_1^c(t) \\
\omega_0 + f^2_2 \gamma_1^s(t) & -f^2_1\gamma_1(t) - f_2^2 \gamma_1^c(t) & f_1f_2\gamma_1^s(t) \\
f_1f_2\gamma_1(t) & 0 & -f^2_2\gamma_1^c(t) \\
\end{pmatrix},\\
&\mathbf{b}(t) = \begin{pmatrix}
f_1f_2\gamma_2^s(t) \\
f_1f_2 \left[\gamma_2(t)-\gamma_2^c(t)\right]\\
-f^2_2\gamma_2^s(t)\\
\end{pmatrix}
\end{align*}
with 
\begin{align}\label{def:gammas}
&\gamma_{1,2}(t) = \int_0^t d\tau\,\,2 D_{1,2}(\tau),\,\, \gamma_{1,2}^{c} = \int_0^t d\tau\,\,2 D_{1,2}(\tau)\cos\left(\omega_0\tau\right),\notag\\
&\gamma_{1,2}^{s} = \int_0^t d\tau\,\,2 D_{1,2}(\tau)\sin\left(\omega_0\tau\right).
\end{align} 
The structure of the differential equations for the Bloch vector components is very informative. 
First of all we notice that if we set $f_2 \to 0$, Eq. \eqref{eq:BVdiffEqs} describe a pure-dephasing dynamics, i.e. $\mathbf{b}(t)|_{f_2=0} = \mathbf{0}$ and
\begin{align*}
&\mathbf{M}(t)|_{f_2=0} = \begin{pmatrix}
-f^2_1 \gamma_1(t) & -\omega_0 & 0 \\
\omega_0 + f^2_1 \gamma_1^s(t) & -f^2_1\gamma_1(t) & 0 \\
0 & 0 & 0 \\
\end{pmatrix}.
\end{align*}
When instead we set $f_1 \to 0$ we retrieve the dynamics of a spin-boson \cite{Breuer2002, Clos2012}, i.e. 
\begin{align*}\label{eq:ClosDyn}
&\mathbf{M}(t)|_{f_1=0} = \begin{pmatrix}
0 & -\omega_0 &  0 \\
\omega_0 + f^2_2 \gamma_1^s(t) & - f_2^2 \gamma_1^c(t) & 0 \\
0 & 0 & -f^2_2\gamma_1^c(t) \\
\end{pmatrix},\\
&\mathbf{b}(t)|_{f_1=0} = \begin{pmatrix}
0 \\
0\\
-f^2_2\gamma_2^s(t)\\
\end{pmatrix}.
\end{align*}
The important thing is then to notice is that the elements $\mathit{M}_{13}(t)$ and $\mathit{M}_{23}(t)$ (as well as their conjugate transpose) depend on the product $f_1f_2$ of the two coupling strengths; same observation holds for the first two components of the affine vector $\mathbf{b}(t)$. As a consequence of this, all such terms vanish whenever the interaction Hamiltonian is \textit{not} given by the crucial structure evidenced in the main text, e.g. Eq. (1). 

Note that the above Bloch equations have been obtained with the only assumption of weak coupling between system and environment (so without any secular or Born-Markov approximation) and thus they provide a faithful description of the dynamics also at the short time-scale, where non-Markovian effects can take place \cite{Breuer2002}.
By letting the simulation time to become large enough, one can use directly such equations of motion and obtain the steady state solutions. This way to proceed is however computationally expensive and moreover does not allow to get a real insight on the resulting values.
Since we are interested in the steady-state properties of the system, we can obtain the analytic results into such regime by taking the long-time limit $t\to\infty$ in the upper integration limit of the time-dependent coefficients, making them as a result, time-independent. The resulting equations of motion for the Bloch vector components will suitably describe the dynamics at large time-scales and the values have been cross-checked with the above-mentioned 'brute-force' method, see Fig. \ref{fig1SM} below.

\subsection{A.3 --- The steady-state solution}
\label{Subsec:LTL1}

In the present Subsection we derive the long-time limit version of Eq. \eqref{eq:BVdiffEqs}. Despite the latter being also suitable, as already stressed, to describe the short timescale, we are interested in the steady-state dynamics and so we will set $t\to\infty$ in the upper integration limit of the time-dependent coefficients, making them as a result, time-independent.
The crucial relation which will be used is the well-known Sokhotski-Plemelj formulae
\begin{equation}\label{eq:PSrelation}
\int_0^{+\infty} \, d\tau e^{\pm i \omega\tau} = \pi\delta(\omega) \mp i \mathcal{P}\frac{1}{\omega},
\end{equation}
with $\mathcal{P}$ denoting the Cauchy Principal value integral.

\subsubsection{The long-time limit version of the coefficients}

Here we will show the explicit evaluation of the long-time limit version of the time-dependent coefficients of the Bloch equations.
\begin{itemize}
\item 
\begin{align*}
\gamma_1(+\infty) &= \int_0^{+\infty} \,d\tau 2D_1(\tau) = 4\int_0^{+\infty}\,d\tau \int_0^{+\infty}\,d\omega J_{\mathrm{eff}}(\omega,T) \cos\left(\omega \tau\right) \\
&= 4\int_0^{+\infty}\,d\omega J_{\mathrm{eff}}(\omega,T) \mathrm{Re}\left[ \int_0^{+\infty}\,d\tau e^{i\omega\tau}\right] = 4\pi\int_0^{+\infty}\,d\omega J_{\mathrm{eff}}(\omega,T)\delta(\omega) = 4\pi J_{\mathrm{eff}}(0,T) = 0
\end{align*}
\item 
\begin{align*}
\gamma_1^c(+\infty) &= \int_0^{+\infty} \,d\tau 2D_1(\tau)\cos\left(\omega_0\tau\right) = 4\int_0^{+\infty}\,d\omega J_{\mathrm{eff}}(\omega,T) \int_0^{+\infty}\,d\tau \mathrm{Re}\left[e^{i\omega\tau}\right] \mathrm{Re}\left[e^{i\omega_0\tau}\right] \\
&= \int_0^{+\infty}\,d\omega J_{\mathrm{eff}}(\omega,T) \int_0^{+\infty}\,d\tau \left( e^{i\omega\tau} + e^{-i\omega\tau} \right) \left( e^{i\omega_0\tau} + e^{-i\omega_0\tau} \right)\\
&= \int_0^{+\infty}\,d\omega J_{\mathrm{eff}}(\omega,T) \int_0^{+\infty}\,d\tau \left[\left( e^{i(\omega+\omega_0)\tau} + e^{i(\omega-\omega_0)\tau} \right) + h.c.\right]\\
&= 2 \int_0^{+\infty}\,d\omega J_{\mathrm{eff}}(\omega,T) \mathrm{Re}\left[ \int_0^{+\infty}\,d\tau \left( e^{i(\omega+\omega_0)\tau} + e^{i(\omega-\omega_0)\tau} \right) \right]\\
&= 2\pi \int_0^{+\infty}\,d\omega J_{\mathrm{eff}}(\omega,T) \left[ \delta(\omega+\omega_0)+\delta(\omega-\omega_0\right] = 2\pi J_{\mathrm{eff}}(\omega_0,T)
\end{align*}
where $h.c.$ stands for \textit{hermitan conjugate} and where we have used the fact that the first Dirac's delta does not give a contribution because the integration of the frequencies is only on $\mathbb{R}^+$ (or, more formally, the spectral density can be prolonged to the negative frequencies with a constant zero function, so that the integral over the positive frequencies becomes extended to the whole real axis and the action of the Dirac's delta is the same as before).
\item 
\begin{align*}
\gamma_1^s(+\infty) &= \int_0^{+\infty} \,d\tau 2D_1(\tau)\sin\left(\omega_0\tau\right) = 4\int_0^{+\infty}\,d\omega J_{\mathrm{eff}}(\omega,T) \int_0^{+\infty}\,d\tau \mathrm{Re}\left[e^{i\omega\tau}\right] \mathrm{Im}\left[e^{i\omega_0\tau}\right] \\
&= -i \int_0^{+\infty}\,d\omega J_{\mathrm{eff}}(\omega,T) \int_0^{+\infty}\,d\tau \left[\left( e^{i(\omega+\omega_0)\tau} + e^{-i(\omega-\omega_0)\tau} \right) - h.c.\right]\\
&= 2 \int_0^{+\infty}\,d\omega J_{\mathrm{eff}}(\omega,T) \mathrm{Im}\left[ \int_0^{+\infty}\,d\tau \left( e^{i(\omega+\omega_0)\tau} + e^{-i(\omega-\omega_0)\tau} \right) \right]\\
&= -2 \int_0^{+\infty}\,d\omega J_{\mathrm{eff}}(\omega,T) \left[\mathcal{P} \frac{1}{\omega+\omega_0} - \mathcal{P} \frac{1}{\omega-\omega_0}\right] \equiv \Delta_1(\Omega,T)
\end{align*}
\item 
\begin{align*}
\gamma_2(+\infty) &= \int_0^{+\infty} \,d\tau 2D_2(\tau) = 4\int_0^{+\infty}\,d\tau \int_0^{+\infty}\,d\omega J(\omega) \sin\left(\omega \tau\right) \\
&= 4\int_0^{+\infty}\,d\omega J(\omega) \mathrm{Im}\left[ \int_0^{+\infty}\,d\tau e^{i\omega\tau}\right] = - 4 \mathcal{P} \int_0^{+\infty}\,d\omega \frac{J(\omega)}{\omega}\\
&= - 4 \mathcal{P} \int_0^{+\infty}\,d\omega \lambda\left(\frac{\omega}{\Omega}\right)^{s-1} e^{-\omega/\Omega} = -4\lambda\Omega\Gamma(s)
\end{align*}
where we have performed the change of variables $\omega \to \omega' \equiv \omega/\Omega$ and used the definition of the Euler's Gamma function.
\item 
\begin{align*}
\gamma_2^c(+\infty) &= \int_0^{+\infty} \,d\tau 2D_2(\tau)\cos\left(\omega_0\tau\right) = 4\int_0^{+\infty}\,d\omega J(\omega) \int_0^{+\infty}\,d\tau \mathrm{Im}\left[e^{i\omega\tau}\right] \mathrm{Re}\left[e^{i\omega_0\tau}\right] \\
&= -i \int_0^{+\infty}\,d\omega J(\omega) \int_0^{+\infty}\,d\tau \left[\left( e^{i(\omega+\omega_0)\tau} + e^{i(\omega-\omega_0)\tau} \right) - h.c.\right]\\
&= -2 \int_0^{+\infty}\,d\omega J(\omega) \mathrm{Im}\left[ \int_0^{+\infty}\,d\tau \left( e^{i(\omega+\omega_0)\tau} + e^{i(\omega-\omega_0)\tau} \right) \right]\\
&= -2 \int_0^{+\infty}\,d\omega J(\omega) \left[\mathcal{P} \frac{1}{\omega+\omega_0} + \mathcal{P} \frac{1}{\omega-\omega_0}\right] \equiv \Delta_2(\Omega)
\end{align*}
\item 
\begin{align*}
\gamma_2^s(+\infty) &= \int_0^{+\infty} \,d\tau 2D_2(\tau)\sin\left(\omega_0\tau\right) = 4\int_0^{+\infty}\,d\omega J(\omega) \int_0^{+\infty}\,d\tau \mathrm{Im}\left[e^{i\omega\tau}\right] \mathrm{Im}\left[e^{i\omega_0\tau}\right] \\
&= - \int_0^{+\infty}\,d\omega J(\omega) \int_0^{+\infty}\,d\tau \left[\left( e^{i(\omega+\omega_0)\tau} - e^{i(\omega-\omega_0)\tau} \right) + h.c.\right]\\
&= -2 \int_0^{+\infty}\,d\omega J(\omega) \mathrm{Re}\left[ \int_0^{+\infty}\,d\tau \left( e^{i(\omega+\omega_0)\tau} - e^{i(\omega-\omega_0)\tau} \right) \right]\\
&= -2\pi \int_0^{+\infty}\,d\omega J(\omega) \left[\delta(\omega+\omega_0) - \delta(\omega-\omega_0)\right] = 2\pi J(\omega_0)
\end{align*}
\end{itemize}

\subsubsection{The steady-state Bloch equations and their solution}
\label{Subsubsec:SSol1}

The resulting differential equations, now with constant coefficients, are 
\begin{align}\label{eq:LTBVeq}
\frac{d v_1(t)}{dt} &= -\omega_0 v_2(t) + 2\pi f_1f_2 J(\omega_0) \left[\coth\left(\frac{\omega_0}{2T}\right) v_3(t) +1 \right] \\
\frac{d v_2(t)}{dt} &= \left[ \omega_0 + f^2_2 \Delta_1 \right] v_1(t) - 2\pi f^2_2 J(\omega_0)\coth\left(\frac{\omega_0}{2T}\right) v_2(t) + f_1f_2 \Delta_1 v_3(t) - f_1f_2 \left[ 4\lambda\Omega\Gamma(s) + \Delta_2 \right] \\
\frac{d v_3(t)}{dt} &= -2\pi f^2_2 J(\omega_0) \left[ \coth\left(\frac{\omega_0}{2T}\right) v_3(t) +1 \right],
\end{align}
where we have used the shortcut notation
\begin{align*}
\Delta_1 &\equiv D_1(\Omega,T) = -2 \int_0^{+\infty}\,d\omega J_{\mathrm{eff}}(\omega,T) \left[\mathcal{P} \frac{1}{\omega+\omega_0} - \mathcal{P} \frac{1}{\omega-\omega_0}\right], \\
\Delta_2 &\equiv D_2(\Omega) = -2 \int_0^{+\infty}\,d\omega J(\omega) \left[\mathcal{P} \frac{1}{\omega+\omega_0} + \mathcal{P} \frac{1}{\omega-\omega_0}\right].
\end{align*}
It is interesting to briefly compare the structure of set of Bloch equations with that of nuclear magnetization \cite{Bloch1946}. The main difference lies in the presence, in Eq. \eqref{eq:LTBVeq}, of the inhomogeneous terms in the differential equations for $v_{1,2}(t)$. The latter are in fact the main responsible, as we will see short after, of the presence of SSC in our model and directly stem from the particular form of interaction Hamiltonian considered.

First of all, we can immediately see that, in the long time limit, the equation of motion for the population imbalance $v_3(t)$ is not affected by the values of the coherences, i.e. $v_{1,2}(t)$, but the viceversa is not true, namely the former contributes in the determination of the long-time limit solution of the latter. 
We have in fact that the solution of the equation of motion for the third component of the Bloch vector reads
\begin{equation}
v_3(t) = v_3(0) \exp\left[-2\pi f^2_2 J(\omega_0)\coth\left(\frac{\omega_0}{2T}\right) t \right] - \tanh\left(\frac{\omega_0}{2T}\right),
\end{equation}
which, in turn, leads to the following steady-state solution
\begin{equation}\label{eq:v3s}
\overline{v}_3 \equiv \lim_{t\to +\infty} v_3(t) = - \tanh\left(\frac{\omega_0}{2T}\right).
\end{equation}
The latter represents the well-known Maxwell-Boltzmann thermal distribution, guaranteeing detailed balance. Further important considerations on this solution is given below.
Upon substituting Eq. \eqref{eq:v3s} into the Eqs. \eqref{eq:LTBVeq} for the first two components $v_{1,2}$, we get
\begin{align}
\frac{dv_1(t)}{dt} &= -\omega_0 v_2(t)\\
\frac{dv_2(t)}{dt} &= \left[ \omega_0 + f^2_2 \Delta_1 \right] v_1(t) - 2\pi f^2_2 J(\omega_0)\coth\left(\frac{\omega_0}{2T}\right) v_2(t) \notag\\
&\qquad\qquad- f_1f_2 \left[ \Delta_1\tanh\left(\frac{\omega_0}{2T}\right) +4\lambda\Omega\Gamma(s) + \Delta_2 \right].
\end{align}
As done for the third component $v_3$, in order to find the steady-state solutions of this coupled set of differential equations we set $\frac{dv_{1,2}(t)}{dt} = 0$ and get the following result
\begin{equation}\label{eq:MainRes}
\overline{v}_1 = \frac{f_1f_2 \left[ \Delta_1\tanh\left(\frac{\omega_0}{2T}\right) +4\lambda\Omega\Gamma(s) + \Delta_2 \right]}{\omega_0 + f^2_2 \Delta_1},\,\, \overline{v}_2 = 0.
\end{equation}
The latter represents the first main result of the present work as it states that the real part of the coherences reaches a non-zero steady-state value which crucially depends on the couplings constants $f_1$ and $f_2$ (more specifically on the simultaneous presence of both this couplings) and on the specific form of the spectral density (entering in the definitions of $\Delta_{1,2}$).

It is very important now to notice that the solution for the third component of the Bloch vector, i.e. $\overline{v}_3$ given by Eq. \eqref{eq:v3s}, turns out to be  independent on both $f_1$ and $f_2$ while the stationary value for $\overline{v}_1$ as a functional dependence on $f_{1,2}$, through $f_1f_2 \cdot g(f_2,\Omega,T,s,\lambda)$ with $g(f_2,\Omega,T,s,\lambda)$ being given by comparison with Eq. \eqref{eq:MainRes}. 
Furthermore, as already highlighted before, the solution for $v_3(t)$ as given by Eq. \eqref{eq:LTBVeq} did not depend on $v_{1,2}(t)$ which means automatically that no corrections due to the SSC are encompassed within the second-order approximation.
An eventual expansion of the dynamical generator up to the next non-zero order, i.e. the fourth one, would therefore inevitably be needed in order to have access to the latter.
Since our interest is however mainly focused on the characterization of SSC and the result Eq. \eqref{eq:MainRes} is correct up to $o\left(f_1^2f_2^2\right)$, we will not go further with this approach. More importantly, one can rely on the alternative approach based on equilibration theory which would give the correct value also for $\overline{v}_3$ already at second order, as it will be detailed in Section E of this SM.

\begin{figure}[htbp!]
\begin{tikzpicture} 
  \node (img1)  {\includegraphics[width=0.8\columnwidth]{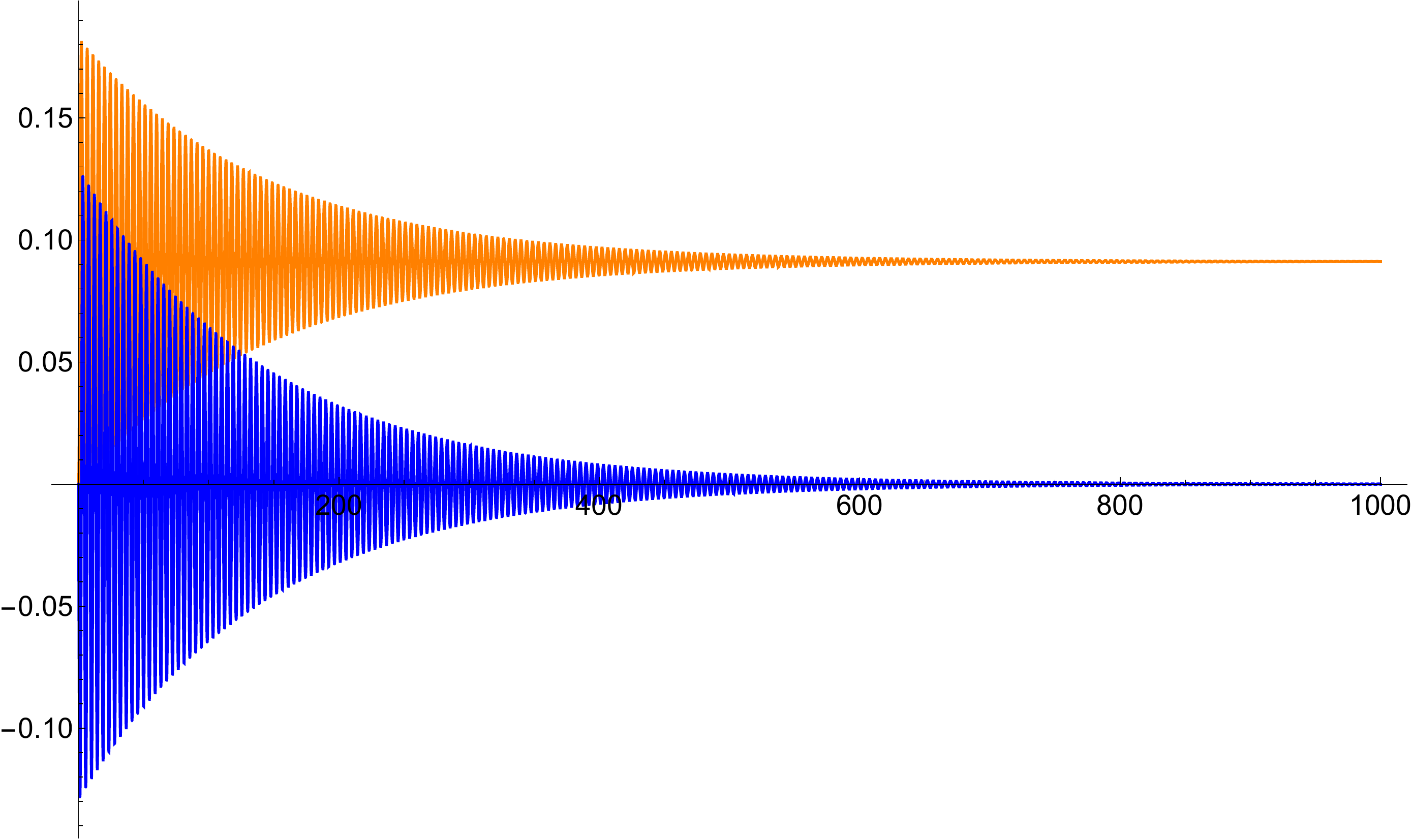}};
  \node[above=of img1, node distance=0cm, yshift=-5cm,xshift=-2.4cm] {{\color{blue}\fontsize{14}{116}$v_2(t)$}};
   \node[above=of img1, node distance=0cm, yshift=-3cm,xshift=-2.4cm] {{\color{black!40!orange}\fontsize{14}{116}$v_1(t)$}};
       \node[above=of img1, node distance=0cm, yshift=-7cm,xshift=7.5cm] {{\fontsize{14}{116}$\omega_0 t$}};
\end{tikzpicture}
\caption{(Color online) Solution of the three components of the Bloch vector in the long-time limit dynamics, where the latter is generated by the set of differential equations \eqref{eq:LTBVeq}. The orange curve represents the time-behavior of $v_1(t) \equiv \mean{\sigma_x(t)}$, while the blue one displays the behavior of $v_2(t)  \equiv \mean{\sigma_y(t)}$. The parameters have been chosen to be $\Omega = 10\omega_0$, $s=3$, $\lambda=10^{-2}\omega_0$ and $T=0.1\omega_0$. Due to the independence of the steady-state solution with respect to the initial condition, the latter has been chosen randomly within the Bloch sphere.}
\label{fig1SM}
\end{figure}

In Fig.\ref{fig1SM} we show the result of a simulation of the dynamics of the first two components of the Bloch vector as given by the set of differential equations \eqref{eq:LTBVeq} for a generic set of values of the parameters, namely $s=3$, $T=0.1\omega_0$, $\lambda=10^{-2}\omega_0$, $\Omega = 10\omega_0 $ and finally $f_2 = \sqrt{\omega_0/\Delta_1}$. From this plot one can immediately see that the imaginary part of the coherences, accounted for by the behavior of $v_2(t)$ (blue curve) correctly vanishes in the steady-state, while the real part, i.e. $v_1(t)$, approaches a non-zero steady-state value.

We finally stress that the fact that only the real part of the coherences, i.e. $\overline{v}_1 \equiv \mean{\sigma_x}_s$, survive in the steady-state is due to the fact that the orthogonal projection of the interaction Hamiltonian $\Ham_{SE}^{\perp}$ considered was proportional to $\sigma_x$. 
We have in fact checked that an exchange $\sigma_x\to\sigma_y$ in Eq.~\eqref{eq:Ham1} would produce a correspondent non-zero value of $\overline{v}_2$ instead, i.e. non-zero imaginary steady-state coherences.

\subsection{A.4 --- Additional details on the steady-state coherences and their measure}
\label{Subsec:SSC1}

This Subsection is devoted to additional details and information concerning the coherence measure employed and its steady-state properties in the model above-considered described by the Hamiltonian of the form Eq.~\eqref{eq:Ham1}.

Let us start by briefly recalling the most recent theoretical results and criteria regarding the definition and construction of a proper measure of coherence.
Let $\mathcal{I}\subset\mathcal{B}(\HILB)$ denote the set of all \textit{incoherent state}. This notion, as the notion of coherences itself, is basis-dependent, since it refers to a particular choice of basis in $\HILB$. Let then $\lbrace\ket{n}\rbrace$ an orthonormal basis for the Hilbert space of the system of interest; then $\mathcal{I} \lbrace\rho\in\mathcal{B}(\HILB) | \rho = \sum_n p_n\ket{n}\bra{n}\rbrace$. According to the axiomatic approach first put forward in \cite{Baumgratz2014}, any suitable quantifier of coherences $\mathcal{C}(\rho)$ should meet the following conditions
\begin{itemize}
\item[(C1)] \textit{Non-negativity}: $\mathcal{C}(\rho) \geq 0$ for any $\rho$, with the equality holding only on the set of incoherent states.
\item[(C2)] \textit{Monotonicity}: $\mathcal{C}(\Lambda[\rho]) \geq \mathcal{C}(\rho)$ for every incoherent operation $\Lambda$, i.e. a map from the set of incoherent states to itself.
\item[(C3)] \textit{Strong monotonicity}: $\sum_i q_i\mathcal{C}(\rho_i) \leq \mathcal{C}(\rho)$ where $\rho_i = K_i\rho K^{\dagger}_i/q_i$ are the post-measurement states with $q_i = \mathrm{Tr}\left[K_i\rho K^{\dagger}_i\right]$ being the corresponding probability and $K_i$ being an incoherent Kraus operator.
\item[(C4)] \textit{Convexity}: $\sum_i p_i \mathcal{C}(\rho_i) \geq \mathcal{C}\left(\sum_i p_i \rho_i \right)$.
\end{itemize}
In the main text we chose to employ the $l_1-$norm of coherences, which generically reads
\begin{equation}\label{def:cl1}
\mathcal{C}(\rho) = \min_{\sigma\in\mathcal{I}} \norm{\rho-\sigma}_1 = \sum_{i\neq j} |\rho_{ij}|.
\end{equation}
The latter, first introduced in \cite{Baumgratz2014}, can be proven to fulfill all the criteria (C1) - (C4) illustrated above.
For the model at hand, Eq. \eqref{def:cl1} reduces to
\begin{equation}\label{eq:CohMeas}
\mathcal{C} = \sqrt{\overline{v}^2_1 + \overline{v}^2_2} = \left|\overline{v}_1\right|,
\end{equation}
where we have omitted for simplicity the operator in the argument and where with $\overline{v}_1$ being given by Eq. \eqref{eq:MainRes}.

First of all, let's proceed with the idea that the coupling constants $f_1$ and $f_2$ could somehow be tuned by an experimenter and therefore let's perform a maximization over these two parameters and study the resulting behavior with respect to the remaining variables determining the environmental spectrum, i.e. the Ohmicity parameter $s$, the cutoff frequency $\Omega$ and the temperature $T$.

\begin{figure}[!htbp]
\includegraphics[width=.33\columnwidth]{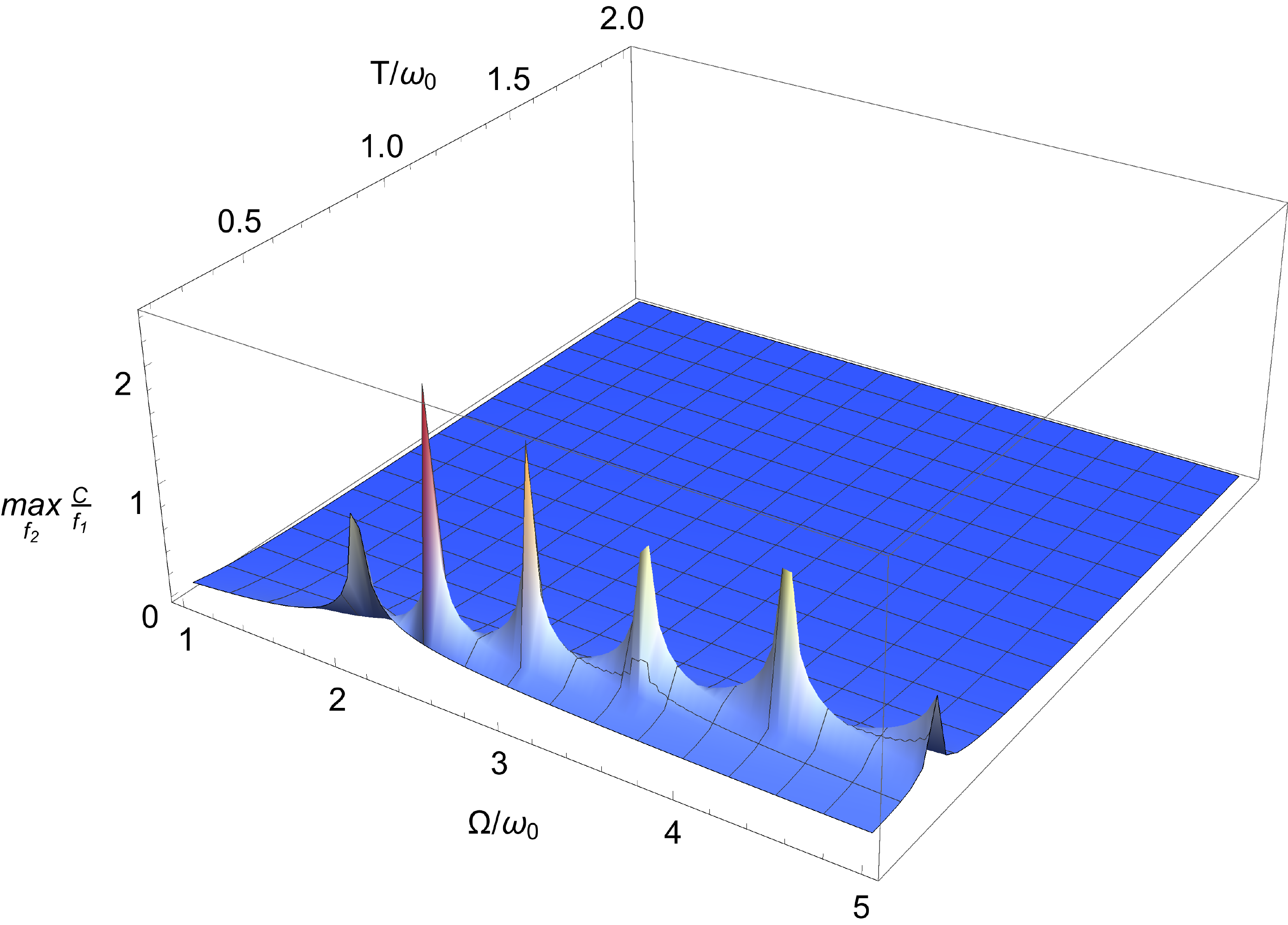}
\includegraphics[width=.33\columnwidth]{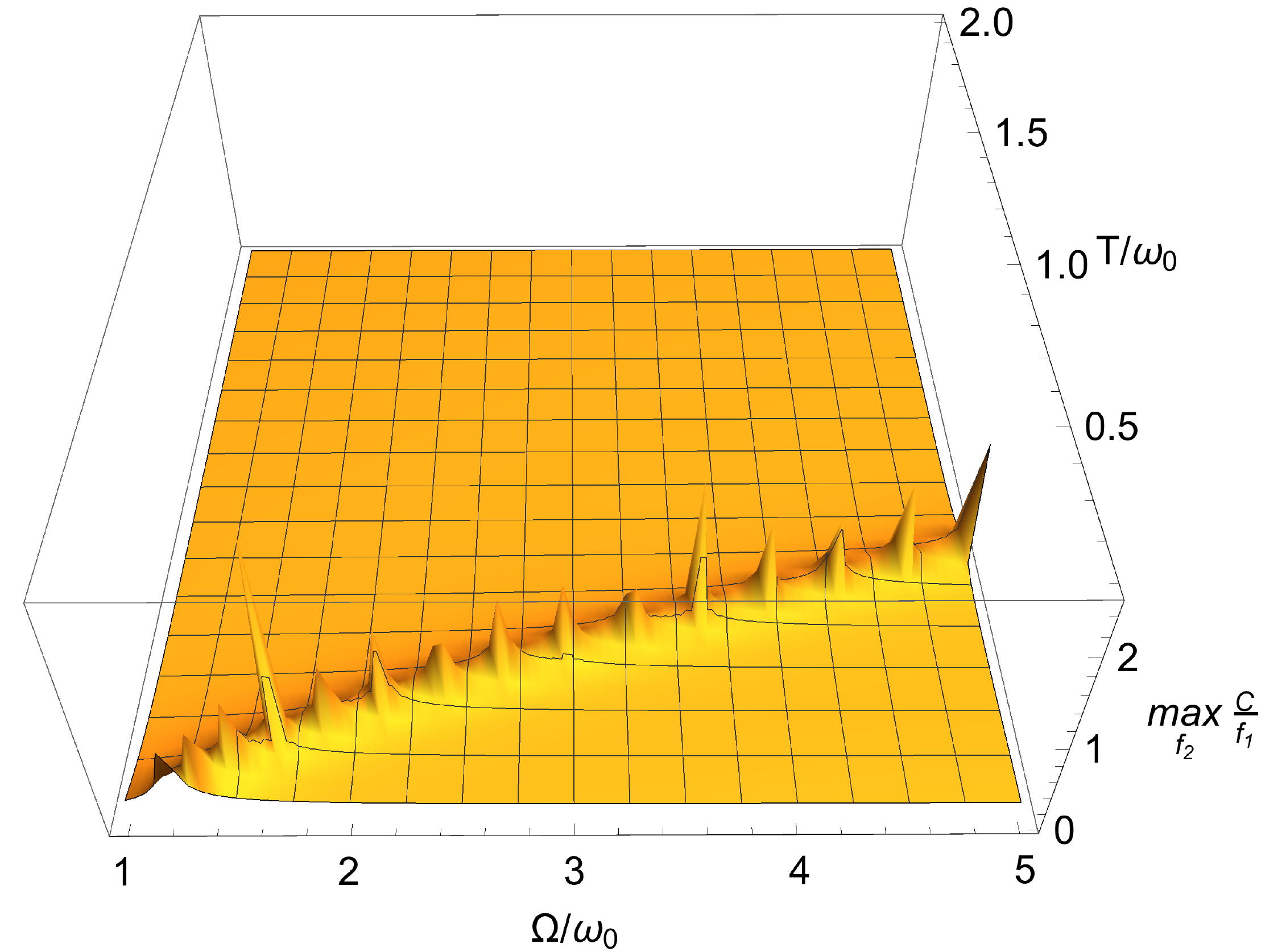}
\includegraphics[width=.33\columnwidth]{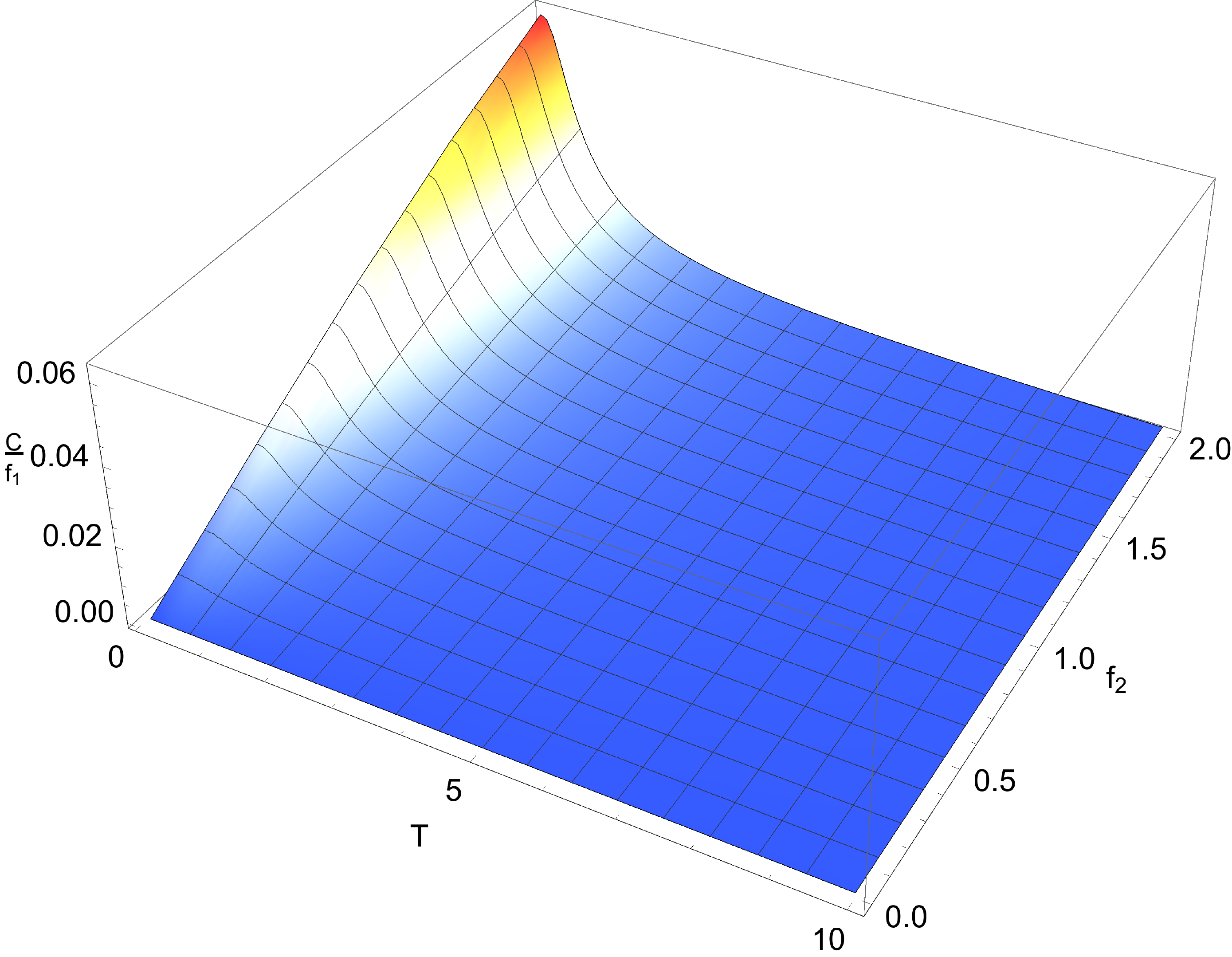}\\
{\bf{(a)}}\hspace*{6cm}{\bf{(b)}}\hspace*{6cm}{\bf{(c)}}
\caption{(Color online) Plot of Eq. \eqref{eq:maxCoherences} as a function of the bath temperature $T$ and of the cutoff frequency $\Omega$ for {\bf{(a)}} a sub-Ohmic ($s=0.5$), {\bf{(b)}} an Ohmic ($s=1$) and {\bf{(c)}} a super-Ohmic ($s=3$) environment. The coupling constant $\lambda$ has been fixed to $10^{-2}\omega_0$.}
\label{fig1SMold}
\end{figure}

First of all, since we are in the weak-coupling approximation, both the coupling strengths $\lambda f_1$ and $\lambda f_2$ must be taken much smaller than the system's free Hamiltonian $\omega_0$ (we have set $\hbar = 1$ throughout this work). In compliance with this constraint, we will fix $\lambda = 10^{-2}\omega_0$ and $\omega_0 = 1$, while $f_1$ and $f_2$ will be of the order of $10^{-1} \omega_0$. 
As highlighted also in the main text, it is immediate to notice that the $\mathcal{C} \equiv |\overline{v}_1|$ has just a simple linear dependence on $f_1$, i.e. the intensity of the channel ruling over the pure dephasing dynamics.
The coherence measure will be then from now on calculated in units of $f_1$, keeping in mind that the maximum amount of SSC will be achieved by taking the greatest value of $f_1$ allowed by the experimental conditions at hand.
The maximum of $\mathcal{C}$ with respect to $f_2$ can be instead analytically calculated and turns out to be given by
\begin{equation}
f_2 = \sqrt{\omega_0 / \Delta_1},
\end{equation}
for which one gets
\begin{equation}
\max_{f_2} \mathcal{C}/f_1 = \left|\frac{\left[ \Delta_1\tanh\left(\frac{\omega_0}{2T}\right) +4\lambda\Omega\Gamma(s) + \Delta_2 \right]}{2\sqrt{\omega_0\Delta_1}}\right|.
\end{equation}
The behavior of this quantity with respect to the remaining parameters, i.e. $\Omega, T$ and $s$ has been discussed in the main text. 
However, we complete here the discussion by adding additional information and plots concerning such behavior.

Eq. \eqref{eq:maxCoherences} is shown in Fig. \ref{fig1SMold} for $\lambda=10^{-2}\omega_0$ as a function of the cutoff frequency $\Omega$ and of the bath temperature $T$ in the sub-Ohmic case (panel {\bf{(a)}}, $s=0.8$), Ohmic case (panel {\bf{(b)}}, $s=1$) and finally super-Ohmic case (panel {\bf{(c)}}, $s=3$).
For $s\leq 1$, i.e. for Ohmic and sub-Ohmic spectral densities, the integrand of both $\Delta_{1,2}$ show a singular behavior at $\omega = \omega_0$. This results in the presence of spikes in the coherence measure located along the \textit{resonance curve}. The latter, as explained in the main text, is identified by the condition
\begin{equation}
\frac{\partial J_{\mathrm{eff}}(\omega,T)}{\partial \omega}|_{\omega=\omega_0} = 0,
\end{equation}
which, in the Ohmic case, implicitly identifies the following curve in the $\left(\Omega, T\right)-$plane 
\begin{equation}\label{rescurveSM}
\Omega_{\mathrm{res}}(T) = \frac{T}{T/\omega_0-\mathrm{Cosech}\left(\omega_0/T\right)}.
\end{equation}
The resonance curve physically means a match between the system's frequency $\omega_0$ and the frequency $\omega_{\mathrm{max}}$ . The latter is identified by the constraint $\partial_\omega J_{\mathrm{eff}} = 0$ and denotes the dominant environmental mode, i.e. the one which by definition maximizes the spectral density. When such frequency $\omega_{\mathrm{max}}$ matches $\omega_0$, i.e. is resonant with the system, then the system mainly interacts with a locally flat spectrum, the latter notoriously leading to a Markovian dynamics \cite{Clos2012}.

Moreover, in the case of cold environment, one can exploit the relations $\coth\left(\omega/(2T)\right) = \tanh\left(\omega/(2T)\right) = 1 $ and perform the analytical integration of $\Delta_{1,2} $. Fixing for example $s=3$, one obtains
\begin{equation}\label{eq:partialresult}
\left.\max_{f_2} \mathcal{C}/f_1 \right|_{T\to 0, s=3} = -\frac{2 \sqrt{2} \lambda  \Omega  \left(3 e^{\frac{1}{\Omega }}
   Ei\left(\frac{1}{\Omega }\right)-1\right)}{\sqrt{\lambda 
   \left(2-\frac{e^{\frac{1}{\Omega }} \text{Ei}\left(-\frac{1}{\Omega
   }\right)+e^{-1/\Omega } \text{Ei}\left(\frac{1}{\Omega }\right)}{\Omega
   ^2}\right)}},
\end{equation}
where $Ei(x)$ is exponential integral function of $x$. The quantity in Eq. \eqref{eq:partialresult} is a monotonic increasing function of $\Omega$ which reaches its maximum for $\Omega\to+\infty$, where the above quantity simply reduces to $\sqrt{\lambda}$. 

\begin{figure}[htbp!]
\begin{tikzpicture} 
  \node (img1)  {\includegraphics[width=0.5\columnwidth]{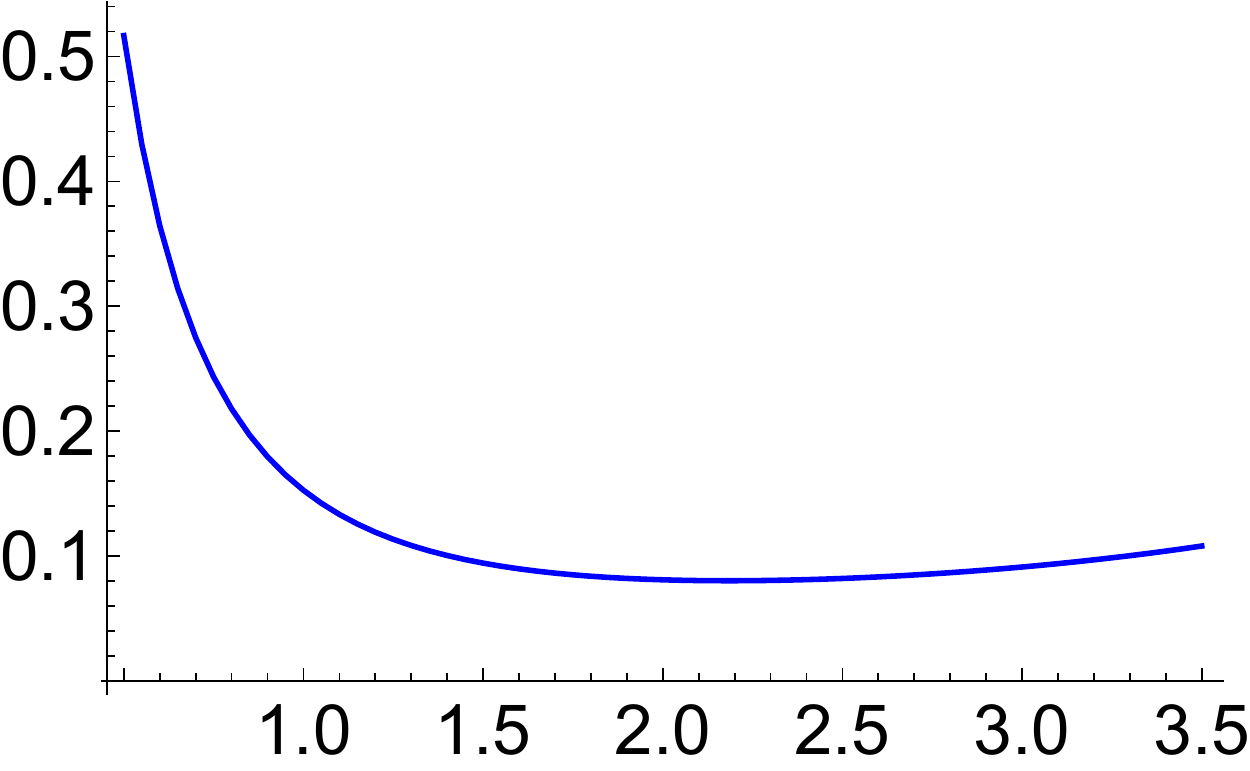}};
  \node[left=of img1, node distance=0cm, rotate=90, anchor=center, yshift=-.4cm,xshift=.3cm] {{\fontsize{16}{116}$\max_{f_2}{C/f_1}$}};
       \node[above=of img1, node distance=0cm, yshift=-7cm,xshift=4.5cm] {{\fontsize{14}{116}$s$}};
\end{tikzpicture}
\caption{Plot of the coherence measure Eq. \eqref{eq:maxCoherences} as function of the Ohmicity parameter $s$ for fixed values of the remaining parameters, i.e. $\Omega=10\omega_0$, $T=10^{-3}\omega_0$ and $\lambda=10^{-2}\omega_0$.}
\label{figCohVSsMod1}
\end{figure}

We close this Section by showing in Fig. \ref{figCohVSsMod1} the behavior of the coherence measure $\mathcal{C}$, in units of $f_1$ and opportunely maximized over $f_2$, for large cutoff frequency $\Omega = 10\omega_0$ and very small bath temperature $T = 10^{-3}\omega_0$ as a function of the Ohmicity parameter $s$. Note that the range of $s$ considered in this plot is arbitrarily chosen and one can go even beyond $s=3.5$. From this plot it is evident how the coherences are present for every type of low frequency profile of spectral densities belonging to the class \eqref{eq:SD}.

\section{B --- The second model}
\label{Sec:SecondModel}

We now present the detailed analysis of the second model discussed in the main body. The latter still consists of the same actors, namely a two-level system and a bosonic bath, but the interaction between them is different, namely given by
\begin{equation}\label{eq:Ham2SM}
\Ham_{SE,2} = f_1\sigma_z\otimes B_E + f_2\left(\sigma_+\otimes b_E + \sigma_-\otimes b^{\dagger}_E\right),
\end{equation}
with $b_E = \sum_k g_k b_k $ and $B_E = b_E + b^{\dagger}_E$. As already mentioned in the main paper, the latter can also be seen as the approximated version of \eqref{eq:Ham1} after the so-called \textit{Rotating Wave Approximation} (RWA) is performed, thus serving the purpose of highlighting the fact that the SSC were not just an artefact created by the presence of the counter-rotating terms in the Hamiltonian \eqref{eq:Ham1} but only depend on the structure of the Hamiltonian evidenced in the paper. Also in this case, in fact, the interaction Hamiltonian has the crucial property of having non-zero projections on both the parallel and orthogonal components with respect to the free Hamiltonian $\Ham_S$.

\subsection{B.1 --- The second-order time local master equation}
\label{Subsec:ME2}

We hereby proceed along the lines of Subsection A.1 and show the detailed derivation of the second-order time-local master equation.
Following the same steps as in the previous model, one obtains that the explicit expansion of the double commutator entering Eq. \eqref{eq:2oTCL} consists of the following four terms
\begin{itemize}
\item 
\begin{align*}
& - f^2_1 \int_0^t d\tau \left[ \sigma_z \sigma_z \rho_S(t) \mean{B_E B_E(-\tau)}_{\beta} - \sigma_z\rho_S(t)\sigma_z \mean{B_E B_E(-\tau)}_{\beta}\right. \\
&\left.\qquad\qquad\qquad\qquad- \sigma_z\rho_S(t)\sigma_z \mean{B(-\tau)B}_{\beta} + \rho_S(t) \sigma_z\sigma_z \mean{B_E(-\tau) B_E}_{\beta}\right]\\
&= \left( \sigma_z\rho_S(t)\sigma_z - \rho_S(t)\right) f^2_1  \int_0^t d\tau \left( \mean{B_E B_E(-\tau)}_{\beta} + \mean{B_E(-\tau) B_E}_{\beta}\right),
\end{align*}
which is the channel responsible for pure dephasing and it is equal to the corresponding term in the previous model.
\item
\begin{align*}
& -f_1f_2 \int_0^t d\tau  \sigma_z\sigma_+(-\tau)\rho_S(t) \mean{B_E b_E(-\tau)}_{\beta} + \sigma_z\sigma_-(-\tau)\rho_S(t) \mean{B_E b^{\dagger}_E(-\tau)}_{\beta}\notag\\
&\qquad\qquad-\sigma_z\rho_S(t)\sigma_+(-\tau) \mean{B_E b_E(-\tau)}_{\beta} -\sigma_z\rho_S(t)\sigma_-(-\tau) \mean{B_E b^{\dagger}_E(-\tau)}_{\beta}\notag\\
&\qquad\qquad-\sigma_+(-\tau)\rho_S(t)\sigma_z \mean{b_E(-\tau) B_E}_{\beta} -\sigma_-(-\tau)\rho_S(t)\sigma_z \mean{b^{\dagger}_E(-\tau) B_E}_{\beta}\notag\\
&\qquad\qquad+\rho_S(t)\sigma_+(-\tau)\sigma_z \mean{b_E(-\tau) B_E}_{\beta} -\rho_S(t)\sigma_-(-\tau)\sigma_z \mean{b^{\dagger}_E(-\tau) B_E}_{\beta}\notag\\
&= -f_1f_2 \int_0^t d\tau \mean{B_E b_E(-\tau)}_{\beta} e^{-i\omega_0\tau}\left[\sigma_z\sigma_+\rho_S(t) - \sigma_z\rho_S(t)\sigma_+\right]\notag\\
&-f_1f_2 \int_0^t d\tau \mean{B_E b^{\dagger}_E(-\tau)}_{\beta} e^{+i\omega_0\tau}\left[\sigma_z\sigma_-\rho_S(t) - \sigma_z\rho_S(t)\sigma_-\right]\notag\\
&-f_1f_2 \int_0^t d\tau \mean{b_E(-\tau) B_E}_{\beta} e^{-i\omega_0\tau}\left[\rho_S(t)\sigma_+\sigma_z - \sigma_+\rho_S(t)\sigma_z\right]\notag\\
&-f_1f_2 \int_0^t d\tau \mean{b^{\dagger}_E(-\tau) B_E}_{\beta} e^{i\omega_0\tau}\left[\rho_S(t)\sigma_-\sigma_z - \sigma_-\rho_S(t)\sigma_z\right].
\end{align*}
Using the relations, valid for a thermal bath, 
\begin{equation}\label{relationOK}
\mean{b_E b_E}_{\beta} = \mean{b^{\dagger}_E b^{\dagger}_E}_{\beta} = 0,
\end{equation}
the expressions above reduce respectively to
\begin{align*}
& -f_1f_2 \int_0^t d\tau \mean{b^{\dagger}_E b_E(-\tau)}_{\beta} e^{-i\omega_0\tau}\left[\sigma_z\sigma_+\rho_S(t) - \sigma_z\rho_S(t)\sigma_+\right]\notag\\
&-f_1f_2 \int_0^t d\tau \mean{b_E b^{\dagger}_E(-\tau)}_{\beta} e^{+i\omega_0\tau}\left[\sigma_z\sigma_-\rho_S(t) - \sigma_z\rho_S(t)\sigma_-\right]\notag\\
&-f_1f_2 \int_0^t d\tau \mean{b_E(-\tau) b^{\dagger}_E}_{\beta} e^{-i\omega_0\tau}\left[\rho_S(t)\sigma_+\sigma_z - \sigma_+\rho_S(t)\sigma_z\right]\notag\\
&-f_1f_2 \int_0^t d\tau \mean{b^{\dagger}_E(-\tau) b_E}_{\beta} e^{i\omega_0\tau}\left[\rho_S(t)\sigma_-\sigma_z - \sigma_-\rho_S(t)\sigma_z\right].
\end{align*}
\item
\begin{align*}
& -f_1f_2 \int_0^t d\tau  \sigma_+\sigma_z\rho_S(t) \mean{b_E B_E(-\tau)}_{\beta} + \sigma_-\sigma_z\rho_S(t) \mean{b^{\dagger}_E B_E(-\tau)}_{\beta}\notag\\
&\qquad\qquad-\sigma_+\rho_S(t)\sigma_z \mean{b_E B_E(-\tau)}_{\beta} -\sigma_-\rho_S(t)\sigma_z \mean{b^{\dagger}_EB_E(-\tau)}_{\beta}\notag\\
&\qquad\qquad-\sigma_z\rho_S(t)\sigma_+ \mean{B_E(-\tau) b_E}_{\beta} -\sigma_z\rho_S(t)\sigma_- \mean{B_E(-\tau) b^{\dagger}_E}_{\beta}\notag\\
&\qquad\qquad+\rho_S(t)\sigma_z\sigma_+ \mean{B_E(-\tau)b_E}_{\beta} -\rho_S(t)\sigma_z\sigma_- \mean{B_E(-\tau)b^{\dagger}_E}_{\beta}\notag\\
&= -f_1f_2 \int_0^t d\tau \mean{B_E(-\tau) b_E}_{\beta} \left[\rho_S(t)\sigma_z\sigma_+ - \sigma_z\rho_S(t)\sigma_+\right]\notag\\
&-f_1f_2 \int_0^t d\tau \mean{B_E(-\tau) b^{\dagger}_E}_{\beta} \left[\rho_S(t)\sigma_z\sigma_- - \sigma_z\rho_S(t)\sigma_-\right]\notag\\
&-f_1f_2 \int_0^t d\tau \mean{b_EB_E(-\tau)}_{\beta} \left[\sigma_+\sigma_z \rho_S(t) - \sigma_+\rho_S(t)\sigma_z\right]\notag\\
&-f_1f_2 \int_0^t d\tau \mean{b^{\dagger}_E B_E(-\tau)}_{\beta} \left[\sigma_-\sigma_z\rho_S(t) - \sigma_-\rho_S(t)\sigma_z\right]\notag\\
&= -f_1f_2 \int_0^t d\tau \mean{b^{\dagger}_E(-\tau) b_E}_{\beta} \left[\rho_S(t)\sigma_z\sigma_+ - \sigma_z\rho_S(t)\sigma_+\right]\notag\\
&-f_1f_2 \int_0^t d\tau \mean{b_E(-\tau) b^{\dagger}_E}_{\beta} \left[\rho_S(t)\sigma_z\sigma_- - \sigma_z\rho_S(t)\sigma_-\right]\notag\\
&-f_1f_2 \int_0^t d\tau \mean{b_Eb^{\dagger}_E(-\tau)}_{\beta} \left[\sigma_+\sigma_z \rho_S(t) - \sigma_+\rho_S(t)\sigma_z\right]\notag\\
&-f_1f_2 \int_0^t d\tau \mean{b^{\dagger}_E b_E(-\tau)}_{\beta} \left[\sigma_-\sigma_z\rho_S(t) - \sigma_-\rho_S(t)\sigma_z\right]
\end{align*}
where Eq. \eqref{relationOK} has been used in the last passage.
\item 
Making use of Eq. \eqref{relationOK}, we also have that
\begin{align*}
& - f^2_2  \int_0^t d\tau \sigma_+\sigma_-(-\tau)\rho_S(t) \mean{b_E b^{\dagger}_E(-\tau)}_{\beta} +  \sigma_-\sigma_+(-\tau)\rho_S(t) \mean{b^{\dagger}_E b_E (-\tau)}_{\beta} \notag\\
&\qquad\qquad -\sigma_+\rho_S(t)\sigma_-(-\tau)\mean{b_E b^{\dagger}_E(-\tau)}_{\beta} -\sigma_-\rho_S(t)\sigma_+(-\tau)\mean{b^{\dagger}_E b_E(-\tau)}_{\beta}\notag\\
&\qquad\qquad +\rho_S(t)\sigma_+(-\tau)\sigma_-\mean{b_E(-\tau) b^{\dagger}_E}_{\beta}+\rho_S(t)\sigma_-(-\tau)\sigma_+\mean{b^{\dagger}_E(-\tau) b_E}_{\beta}\notag\\
&= - f^2_2  \int_0^t d\tau \mean{b_E(-\tau) b^{\dagger}_E}_{\beta}e^{-i\omega_0\tau} \left[\rho_S(t)\sigma_+\sigma_- - \sigma_+\rho_S(t)\sigma_-\right]\notag\\
& - f^2_2  \int_0^t d\tau \mean{b_E b^{\dagger}_E(-\tau)}_{\beta}e^{i\omega_0\tau} \left[\sigma_+\sigma_-\rho_S(t) - \sigma_+\rho_S(t)\sigma_-\right]\notag\\
& - f^2_2  \int_0^t d\tau \mean{b^{\dagger}_E(-\tau) b_E}_{\beta}e^{i\omega_0\tau} \left[\rho_S(t)\sigma_-\sigma_+ - \sigma_-\rho_S(t)\sigma_+\right]\notag\\
& - f^2_2  \int_0^t d\tau \mean{b^{\dagger}_E b_E(-\tau)}_{\beta}e^{-i\omega_0\tau} \left[\sigma_-\sigma_+\rho_S(t) - \sigma_-\rho_S(t)\sigma_+\right]
\end{align*}
\end{itemize}
Let us now define the following two environmental correlation functions
\begin{align}
\mean{b^{\dagger}_E b_E(-\tau)}_{\beta} \equiv \sum_k |g_k|^2\overline{n}_k e^{i\omega_k\tau} \to \int_0^{\infty} d\omega\, J(\omega) \overline{n}(\omega)e^{i\omega\tau} \equiv \frac{1}{2}\left[d_1(\tau) + i d_2(\tau)\right]\\
\mean{b_E(-\tau) b^{\dagger}_E}_{\beta} \equiv \sum_k |g_k|^2\left(1+\overline{n}_k\right) e^{i\omega_k\tau} \to \int_0^{\infty} d\omega\, J(\omega) \left(1+\overline{n}(\omega)\right)e^{i\omega\tau} \equiv \frac{1}{2}\left[\tilde{d}_1(\tau) + i \tilde{d}_2(\tau)\right]
\end{align}
with $\overline{n}(\omega) = \left[\exp\left(\beta\omega\right)-1\right]^{-1}$.
The structure of the master equation is then of the form Eq. \eqref{eq:MEgeneral}, where the non-zero entries of the Lamb-Shift Hamiltonian $\mathbf{\Ham}^{LS}(t)$ and of the Kossakowski matrix $\mathbf{A}(t)$, after few straightforward calculations, are found to be respectively given by
\begin{align*}
&h_{11}(t) = f^2_2\int_0^t d\tau\,\, \frac{1}{2}\left[ \tilde{d}_1(\tau)\sin(\omega_0\tau) - \tilde{d}_2(\tau) \cos(\omega_0\tau) \right]\\
&h_{22}(t) = - f^2_2\int_0^t d\tau\,\, \frac{1}{2}\left[ d_1(\tau)\sin(\omega_0\tau) - d_2(\tau) \cos(\omega_0\tau) \right]\\
&h_{13}(t) = \left( h_{31}(t) \right)^* = \sqrt{2}\left(f_1 f_2\right) \int_0^t d\tau\, e^{-i\frac{\omega_0\tau}{2}} \,\,\frac{1}{2}\left[ \tilde{d}_1(\tau)\sin\left(\frac{\omega_0\tau}{2}\right) - \tilde{d}_2(\tau) \cos\left(\frac{\omega_0\tau}{2}\right) \right]
\\
&h_{23}(t) = \left( h_{32}(t) \right)^* = \sqrt{2}\left(f_1 f_2\right) \int_0^t d\tau\, e^{i\frac{\omega_0\tau}{2}} \,\,\frac{1}{2}\left[ -d_1(\tau)\sin\left(\frac{\omega_0\tau}{2}\right) + d_2(\tau) \cos\left(\frac{\omega_0\tau}{2}\right) \right],
\end{align*}
and
\begin{align*}
&a_{11}(t) = f^2_2 \int_0^t d\tau\,\, \left[d_1(\tau)\cos(\omega_0\tau) + d_2(\tau) \sin(\omega_0\tau)\right]\\
&a_{22}(t) =   f^2_2 \int_0^t d\tau\,\, \left[\tilde{d}_1(\tau)\cos(\omega_0\tau) + \tilde{d}_2(\tau) \sin(\omega_0\tau)\right]\\
&a_{33}(t) =  f^2_1\int_0^t d\tau\, 2 \left( d_1(\tau) + \tilde{d}_1(\tau) \right)\\
&a_{12}(t) =  \left(a_{21}(t)\right)^* = 0 \\
&a_{13}(t) =  \left(a_{31}(t)\right)^* = \sqrt{2}\left(f_1 f_2\right) \int_0^t d\tau\, e^{-i\frac{\omega_0\tau}{2}}\left[d_1(\tau)\cos\left(\frac{\omega_0\tau}{2}\right) + d_2(\tau) \sin\left(\frac{\omega_0\tau}{2}\right)\right]\\
&a_{23}(t) = \left(a_{32}(t)\right)^* = \sqrt{2}\left(f_1 f_2\right) \int_0^t d\tau\, e^{i\frac{\omega_0\tau}{2}}\left[\tilde{d}_1(\tau)\cos\left(\frac{\omega_0\tau}{2}\right) + \tilde{d}_2(\tau) \sin\left(\frac{\omega_0\tau}{2}\right)\right].
\end{align*}

\subsection{B.2 --- From the master equation to the Bloch differential equations}
\label{Subsec:Bloch2}

\subsubsection{Equation of motion for $v_3(t)$}

Let us then start with the differential equation ruling over the evolution of $v_3(t)$. A simple calculation yields to
\begin{align}\label{v3.1Mod2}
\frac{d}{dt}v_3(t) &= -\left[a_{11}(t) + a_{22}(t)\right] v_3(t) + \left[ a_{11}(t) - a_{22}(t)\right] + \frac{1}{\sqrt{2}} v_1(t) \mathrm{Re}\left[ f(t) \right] -  \frac{1}{\sqrt{2}} v_2(t) \mathrm{Im}\left[ f(t) \right]\notag\\
&= \sum_{j=1}^3 G_{3j}(t) v_j(t) + b_3(t)
\end{align}
where
\begin{equation}
f(t) \equiv a_{13}(t) + a_{32}(t) + 2i \left[ h_{13}(t) - h_{32}(t) \right].
\end{equation}
In order to gain some insight on this expression, we firstly substitute the expressions of the entries of the Lamb-Shift Hamiltonian and of the Kossakowski matrix derived above.
Moreover, we will exploit relation \eqref{ReIm}.
Simple calculations lead to
\begin{align*}
G_{13}(t) \equiv \frac{1}{\sqrt{2}} \mathrm{Re}\left[ f(t) \right] &= f_1f_2\int_0^t d\tau\,\, \left[ d_1(\tau) + \tilde{d}_1(\tau) \right] \notag\\
&= 2 f_1f_2\int_0^{+\infty}\,d\omega J_{\mathrm{eff}}(\omega,T) \int_0^t\,d\tau\cos(\omega\tau)
\end{align*}
with $J_{\mathrm{eff}}(\omega,T) \equiv J(\omega) \coth\left(\frac{\omega}{2T}\right)$.
Similarly we get
\begin{equation}
G_{23}(t) \equiv \frac{1}{\sqrt{2}} \mathrm{Im}\left[ f(t) \right] = 0.
\end{equation}
as well as
\begin{align*}
& G_{33} (t) \equiv -\left[ a_{11}(t) + a_{22}(t)\right] = -2 f^2_2 \int_0^{+\infty} d\omega\, J_{\mathrm{eff}}(\omega,T) \int_0^t\,d\tau\,\cos\left[\left(\omega-\omega_0\right)\tau\right] \\
& b_3(t) \equiv a_{11}(t) - a_{22}(t) = -2 f^2_2 \int_0^{+\infty} d\omega\, J(\omega) \int_0^t\,d\tau\,\cos\left[\left(\omega-\omega_0\right)\tau\right].
\end{align*}

\subsubsection{Equation of motion for $v_1(t)$}

We now consider the differential equation that governs the evolution of $v_1(t)$.
While the third component rules over the population (inversion) dynamics, the first and second component takes into account for the coherence dynamics. 
We have that
\begin{align}\label{v1.1Mod2}
\frac{d}{dt}v_1(t) &= -\frac{1}{2}\mathrm{Re}\left[l(t)\right] v_1(t) + \frac{1}{2}\mathrm{Im}\left[l(t)\right] v_2(t) + \frac{1}{\sqrt{2}} m(t) v_3(t) + \frac{2}{\sqrt{2}} \left( \mathrm{Re}\left[a_{23}(t)\right]  - \mathrm{Re}\left[a_{13}(t)\right]  \right)
\notag\\
&= \sum_{j=1}^3 G_{1j}(t) v_j(t) + b_1(t)
\end{align}
with
\begin{align*}
&l(t) \equiv a_{11}(t) + a_{22}(t) + 2a_{33}(t) - 2i \left( h_{11}(t) - h_{22}(t) + \omega_0 \right)\\
&m(t) \equiv \mathrm{Re}\left[a_{13}(t)\right]  + \mathrm{Re}\left[a_{23}(t)\right] + 2\mathrm{Im}\left[h_{13}(t)\right] + 2\mathrm{Im}\left[h_{23}(t)\right].
\end{align*}
Making use of \eqref{ReIm} and of the fact that $a_{ii}(t), h_{ii}(t) \in \mathbb{R}, \,\,\forall i=1,2,3$, we have that
\begin{align}
&G_{11}(t) \equiv -\frac{1}{2}\mathrm{Re}\left[l(t)\right] = -\frac{1}{2}\left[ a_{11}(t) + a_{22}(t) + 2a_{33}(t)\right] \notag\\
&\qquad\qquad= \frac{1}{2}\left[ G_{33}(t) - 2 a_{33}(t) \right]\\
&G_{12}(t) \equiv \frac{1}{2}\mathrm{Im}\left[l(t)\right] =  - \left( h_{11}(t) - h_{22}(t) + \omega_0 \right)\notag\\
&\qquad\qquad = -\omega_0 + f^2_2\int_0^{+\infty}\,d\omega\, J_{\mathrm{eff}}(\omega,T) \int_0^t\,d\tau\, \sin\left[\left(\omega-\omega_0\right)\tau\right]\\
&G_{13}(t) \equiv \frac{1}{\sqrt{2}} m(t) = 2f_1f_2 \int_0^{+\infty}\,d\omega\, J_{\mathrm{eff}}(\omega,T) \int_0^t\,d\tau\, \cos\left[\left(\omega-\omega_0\right)\tau\right]
\end{align}
and
\begin{align}
b_1(t) &\equiv \frac{2}{\sqrt{2}} \left( \mathrm{Re}\left[a_{23}(t)\right]  - \mathrm{Re}\left[a_{13}(t)\right]  \right) \notag\\
&= 2f_1f_2 \int_0^{+\infty}\,d\omega\, J(\omega) \int_0^t\,d\tau\, \left(\cos(\omega\tau) + \cos\left[\left(\omega-\omega_0\right)\tau\right]\right)
\end{align}

\subsubsection{Equation of motion for $v_2(t)$}

We finally derive the proper differential equation for $v_2(t)$. We have that
\begin{align}\label{v2.1Mod2}
\frac{d}{dt}v_2(t) &= \frac{1}{2}\mathrm{Re}\left[r(t)\right] v_1(t) - \frac{1}{2}\mathrm{Im}\left[r(t)\right] v_2(t) + \frac{1}{\sqrt{2}}  n(t) v_3(t) + \frac{2}{\sqrt{2}}  \mathrm{Im}\left[a_{23}(t)\right]  + \frac{2}{\sqrt{2}}  \mathrm{Im}\left[a_{13}(t)\right] notag\\
&= \sum_{j=1}^3 G_{2j}(t) v_j(t) + b_2(t)
\end{align}
with
\begin{align*}
&r(t) \equiv i a_{11}(t) + i a_{22}(t) + 2i a_{33}(t) + 2 \left( h_{11}(t) - h_{22}(t) + \omega_0 \right)\\
&n(t) \equiv \mathrm{Im}\left[a_{23}(t)\right]  - \mathrm{Im}\left[a_{13}(t)\right] + 2\mathrm{Re}\left[h_{13}(t)\right] - 2\mathrm{Re}\left[h_{23}(t)\right].
\end{align*}
Making use of \eqref{ReIm} we have that
\begin{align}
&G_{21}(t) \equiv  \frac{1}{2}\mathrm{Re}\left[r(t)\right] = 2\left( h_{11}(t) - h_{22}(t) + \omega_0 \right) \notag\\
&\qquad\qquad = -G_{12}(t) = \omega_0 + f^2_2\int_0^{+\infty}\,d\omega\, J_{\mathrm{eff}}(\omega,T) \int_0^t\,d\tau\, \sin\left[\left(\omega-\omega_0\right)\tau\right]\\
&G_{22}(t) \equiv - \frac{1}{2}\mathrm{Im}\left[r(t)\right] =  a_{11}(t) + a_{22}(t) + 2a_{33}(t) \notag\\
&\qquad\qquad = G_{11}(t) =  \frac{1}{2}\left[ G_{33}(t) - 2 a_{33}(t) \right]\\
&G_{23}(t) \equiv \frac{1}{\sqrt{2}} n(t) = -2f_1f_2 \int_0^{+\infty}\,d\omega\, J_{\mathrm{eff}}(\omega,T) \int_0^t\,d\tau\, \sin\left[\left(\omega-\omega_0\right)\tau\right]
\end{align}
and finally
\begin{align}
b_2(t) &\equiv \frac{2}{\sqrt{2}}  \left(\mathrm{Im}\left[a_{23}(t)\right]  +   \mathrm{Im}\left[a_{13}(t)\right]\right) \notag\\
&= 2f_1f_2 \int_0^{+\infty}\,d\omega\, J(\omega) \int_0^t\,d\tau\, \left(\sin(\omega\tau) - \sin\left[\left(\omega-\omega_0\right)\tau\right]\right)
\end{align}

\subsection{B.3 --- The steady-state solution}
\label{Subsec:LTL2}

In the present Section we give and discuss the long-time limit version of the Bloch differential equations for the model presented in Section 2.

\subsubsection{The long-time limit version of the coefficients}

Here we compute the long-time limit version of the time-dependent coefficients of the Bloch differential equations.
\begin{itemize}
\item 
\begin{align*}
G_{11}(+\infty) &= \frac{1}{2} \left[ G_{33}(+\infty) - 2a_{33}(+\infty) \right] \notag\\
&= - f^2_2 \int_0^{+\infty} d\omega\, J_{\mathrm{eff}}(\omega,T) \int_0^{+\infty}\,d\tau\,\cos\left[\left(\omega-\omega_0\right)\tau\right] - 2 f^2_1\int_0^{+\infty} d\omega\, J_{\mathrm{eff}}(\omega,T) \int_0^{+\infty}\,d\tau\,\cos(\omega\tau)\notag\\
&= - \int_0^{+\infty} d\omega\, J_{\mathrm{eff}}(\omega,T) \left( f^2_2 \mathrm{Re}\left[\int_0^{+\infty}\,d\tau\,e^{i(\omega-\omega_0)\tau}\right] + 2f^2_1 \mathrm{Re}\left[\int_0^{+\infty}\,d\tau\,e^{i\omega\tau}\right]\right) \notag\\
&= -\pi \left( f^2_2 J_{\mathrm{eff}}(\omega_0,T) + 2f^2_1 J_{\mathrm{eff}}(0,T) \right) = -\pi f^2_2 J_{\mathrm{eff}}(\omega_0,T) 
\end{align*}
\item 
\begin{align*}
G_{12}(+\infty) &= -\omega_0 + f^2_2\int_0^{+\infty}\,d\omega\, J_{\mathrm{eff}}(\omega,T) \int_0^t\,d\tau\, \sin\left[\left(\omega-\omega_0\right)\tau\right] \notag\\
&= -\omega_0 + f^2_2\int_0^{+\infty}\,d\omega\, J_{\mathrm{eff}}(\omega,T) \mathrm{Im}\left[ \int_0^t\,d\tau\, e^{i(\omega-\omega_0)\tau}\right] \notag\\
&= -\omega_0 - f^2_2 \delta_1(\Omega,T)
\end{align*}
where 
\begin{equation}
\delta_1 (\Omega,T)\equiv \mathcal{P}\int_0^{+\infty}\,d\omega\, \frac{J_{\mathrm{eff}}(\omega,T)}{\omega-\omega_0}
\end{equation}
with $\mathcal{P}$ denoting the Cauchy principal value integral.
\item
\begin{align*}
G_{13}(+\infty) &=2f_1f_2 \int_0^{+\infty}\,d\omega\, J_{\mathrm{eff}}(\omega,T) \int_0^{+\infty}\,d\tau\, \cos\left[\left(\omega-\omega_0\right)\tau\right]\notag\\
&= 2f_1f_2 \int_0^{+\infty}\,d\omega\, J_{\mathrm{eff}}(\omega,T) \mathrm{Re}\left[ \int_0^t\,d\tau\, e^{i\left(\omega-\omega_0\right)\tau}\right] = 2\pi f_1f_2 J_{\mathrm{eff}}(\omega_0,T)
\end{align*}
\item
\begin{align*}
b_1(+\infty) &= 2f_1f_2 \int_0^{+\infty}\,d\omega\, J(\omega) \int_0^{+\infty} \,d\tau\, \left(\cos(\omega\tau) + \cos\left[\left(\omega-\omega_0\right)\tau\right]\right)\notag\\
&= 2f_1f_2 \int_0^{+\infty}\,d\omega\, J(\omega) \left(\mathrm{Re}\left[\int_0^{+\infty}\,d\tau\, e^{i\omega\tau}\right] + \mathrm{Re}\left[\int_0^{+\infty}\,d\tau\, e^{i(\omega-\omega_0)\tau}\right] \right) \notag\\
&= 2\pi f_1f_2 \left[ J(0) + J(\omega_0) \right] = 2\pi f_1f_2 J(\omega_0)
\end{align*}
\item
\begin{equation*}
G_{21}(+\infty) = -G_{12}(+\infty) = \omega_0 + f^2_2 \delta_1(\Omega,T)
\end{equation*}
\item
\begin{equation*}
G_{22}(+\infty) = G_{11}(+\infty) = -\pi f^2_2 J_{\mathrm{eff}}(\omega_0,T) 
\end{equation*}
\item
\begin{align*}
G_{23}(+\infty) &= - 2f_1f_2 \int_0^{+\infty}\,d\omega\, J_{\mathrm{eff}}(\omega,T) \int_0^{+\infty}\,d\tau\, \sin\left[\left(\omega-\omega_0\right)\tau\right]\notag\\
&= -2f_1f_2 \int_0^{+\infty}\,d\omega\, J_{\mathrm{eff}}(\omega,T) \mathrm{Im}\left[ \int_0^{+\infty}\,d\tau\, e^{i\left(\omega-\omega_0\right)\tau}\right] = 2 f_1f_2 \delta_1(\Omega,T)
\end{align*}
\item
\begin{align*}
b_2(+\infty) &= 2f_1f_2 \int_0^{+\infty}\,d\omega\, J(\omega) \int_0^{+\infty} \,d\tau\, \left(\sin(\omega\tau) - \sin\left[\left(\omega-\omega_0\right)\tau\right]\right)\notag\\
&= 2f_1f_2 \int_0^{+\infty}\,d\omega\, J(\omega) \left(\mathrm{Im}\left[\int_0^{+\infty}\,d\tau\, e^{i\omega\tau}\right] - \mathrm{Im}\left[\int_0^{+\infty}\,d\tau\, e^{i(\omega-\omega_0)\tau}\right] \right) \notag\\
&= -2f_1f_2 \left( \delta_2 (\Omega,T)+ \lambda\Omega\Gamma(s) \right) 
\end{align*}
where
\begin{equation}
\delta_2(\Omega,T) \equiv \mathcal{P}\int_0^{+\infty}\,d\omega\, \frac{J(\omega)}{\omega-\omega_0}
\end{equation}
and where we used the fact that
\begin{equation}
\mathcal{P} \int_0^{+\infty}\,d\omega \frac{J(\omega)}{\omega} = \lambda \int_0^{+\infty}\,d\omega\left(\frac{\omega}{\Omega}\right)^{s-1} e^{-\omega/\Omega} = \lambda\Omega\Gamma(s).
\end{equation}
\item
\begin{align*}
G_{31}(+\infty) &= 2 f_1f_2\int_0^{+\infty}\,d\omega J_{\mathrm{eff}}(\omega,T) \int_0^{+\infty}\,d\tau\cos(\omega\tau)\notag\\
&= 2f_1f_2 \int_0^{+\infty}\,d\omega\, J_{\mathrm{eff}}(\omega,T) \mathrm{Re}\left[ \int_0^{+\infty}\,d\tau\, e^{i\left(\omega-\omega_0\right)\tau}\right] = 2\pi f_1f_2 J_{\mathrm{eff}}(0,T) = 0
\end{align*}
\item
\begin{equation}
G_{32}(+\infty) = 0
\end{equation}
\item
\begin{align*}
G_{33}(+\infty) &= - 2 f^2_2 \int_0^{+\infty} d\omega\, J_{\mathrm{eff}}(\omega,T) \int_0^{+\infty}\,d\tau\,\cos\left[\left(\omega-\omega_0\right)\tau\right] \notag\\
&= - 2 f^2_2 \int_0^{+\infty} d\omega\, J_{\mathrm{eff}}(\omega,T) \mathrm{Re}\left[\int_0^{+\infty}\,d\tau\,e^{i(\omega-\omega_0)\tau}\right] \notag\\
&= -2\pi f^2_2 J_{\mathrm{eff}}(\omega_0,T) 
\end{align*}
\item
\begin{align*}
b_3(+\infty) &= - 2 f^2_2 \int_0^{+\infty} d\omega\, J(\omega) \int_0^{+\infty}\,d\tau\,\cos\left[\left(\omega-\omega_0\right)\tau\right] \notag\\
&= - 2 f^2_2 \int_0^{+\infty} d\omega\, J(\omega) \mathrm{Re}\left[\int_0^{+\infty}\,d\tau\,e^{i(\omega-\omega_0)\tau}\right] \notag\\
&= -2\pi f^2_2 J(\omega_0) 
\end{align*}
\end{itemize}

\subsubsection{The steady-state Bloch equations and their solution}
\label{Subsubsec:SSol2}

Exploiting the above results, we find that the long-time limit version of the differential equations for the Bloch vector components have the following form
\begin{align}\label{eq:2ModelBlochEq}
&\frac{d v_1(t)}{dt} = -\pi f^2_2 J_{\mathrm{eff}}(\omega_0,T) v_1(t) - \left[\omega_0 + f^2_2\delta_1\right] v_2(t) + 2\pi f_1f_2 J(\omega_0) \left[\coth\left(\frac{\omega_0}{2T}\right)v_3(t) + 1\right]\notag\\
&\frac{d v_2(t)}{dt} = \left[\omega_0 + f^2_2\delta_1\right] v_1(t) - \pi f^2_2 J_{\mathrm{eff}}(\omega_0,T) v_2(t) + 2\pi f_1f_2 \left[ \delta_1 v_3(t) - \delta_2 + \lambda\Omega\Gamma(s)\right]\notag\\
& \frac{d v_3(t)}{dt} = - 2\pi f^2_2 J(\omega_0) \left[\coth\left(\frac{\omega_0}{2T}\right)v_3(t) + 1\right],
\end{align}
where $\delta_{1,2} \equiv \delta_{1,2}(\Omega,T)$ for brevity of notation.
The structure of these Bloch equations is manifestly more symmetrical, concerning $v_{1,2}(t)$, than Eq. \eqref{eq:LTBVeq}. More specifically, if we denote by
\begin{equation}
\frac{d v_i(t)}{dt} = \sum_{j=1}^3 B_{ij} v_j(t) + b_i
\end{equation}
a generic set of Bloch equations, then one can immediately notice that in the first model considered Section A (see in particular Eq. \eqref{eq:LTBVeq}) we had that $B_{12} \neq B_{21}$ and $B_{11} = 0 \neq B_{22}$, alongside with $b_1 = 0 \neq b_2$. In this second model instead, the structure of the Bloch equations \eqref{eq:2ModelBlochEq} shows that $B_{12} = B_{21}$ and $B_{11} = B_{22}$, despite $b_1 \neq b_2$. As we will see in a moment, this difference reflects in the fact that both first two components of the Bloch vector, i.e. $\overline{v}_{1,2}$, reach a non-zero steady-state value.

First however, let us stress that the differential equation for $v_3(t)$ (as in the first model, see Section A.3) does not depend on $v_{1,2}(t)$ components and therefore the stationary solution is readily found once again to be given by
\begin{equation}
\overline{v}_3 = - \tanh\left(\frac{\omega_0}{2T}\right),
\end{equation}
which represents the usual thermal distribution \cite{Breuer2002} also found in the previous model \eqref{eq:v3s}. Same considerations concerning this solution as those done in the previous model still hold true.
Inserting this solution in the differential equations for the other two components allows to find the stationary solutions $\overline{v}_1$ and $\overline{v}_2$ which read
\begin{align}\label{eq:SSCMod2v2}
\overline{v}_1 &= 2 f_1f_2\alpha(\omega_0,T,s) \mathrm{Re}\left[\frac{1}{\left(\omega_0 + f^2_2 \delta_1\right) - i \left(\pi f^2_2 J_{\mathrm{eff}}(\omega_0,T)\right)}\right],\\
\overline{v}_2 &= 2 f_1f_2\alpha(\omega_0,T,s) \mathrm{Im}\left[\frac{1}{\left(\omega_0 + f^2_2 \delta_1\right) - i \left(\pi f^2_2 J_{\mathrm{eff}}(\omega_0,T)\right)}\right],
\end{align}
where
\begin{equation}
\alpha(\omega_0,T,s) \equiv \delta_1\tanh\left(\frac{\omega_0}{2T}\right) + \delta_2 + \lambda\Omega\Gamma(s)
\end{equation}
which very naturally retraces the role of $v_{1,2}$ as real and imaginary part of the coherences, respectively.

\subsection{B.4 --- Additional details on the steady-state coherences and their measure}
\label{Subsec:SSC2}

\begin{figure}[!htbp]
\includegraphics[width=.3\columnwidth]{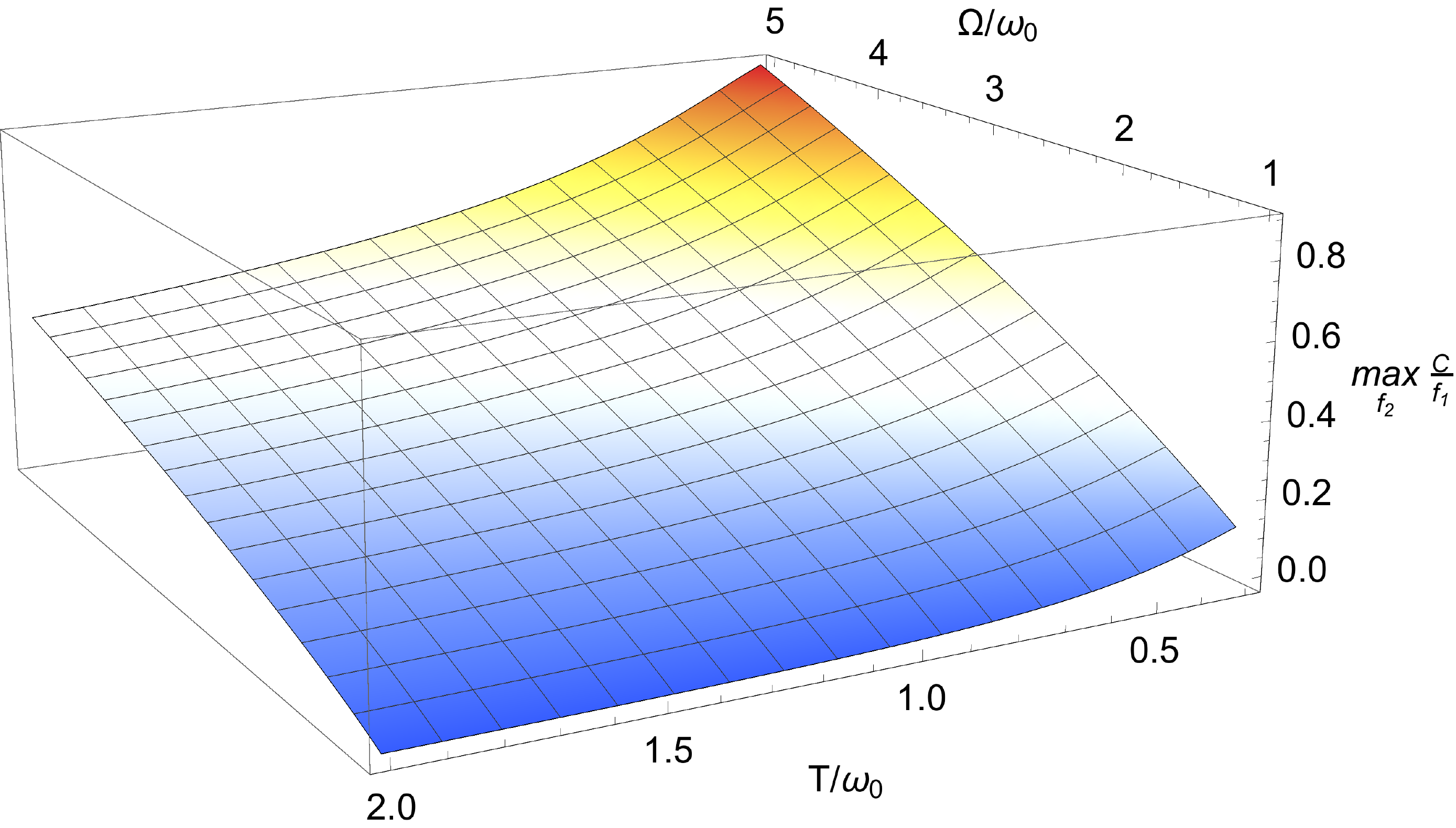}
\includegraphics[width=.3\columnwidth]{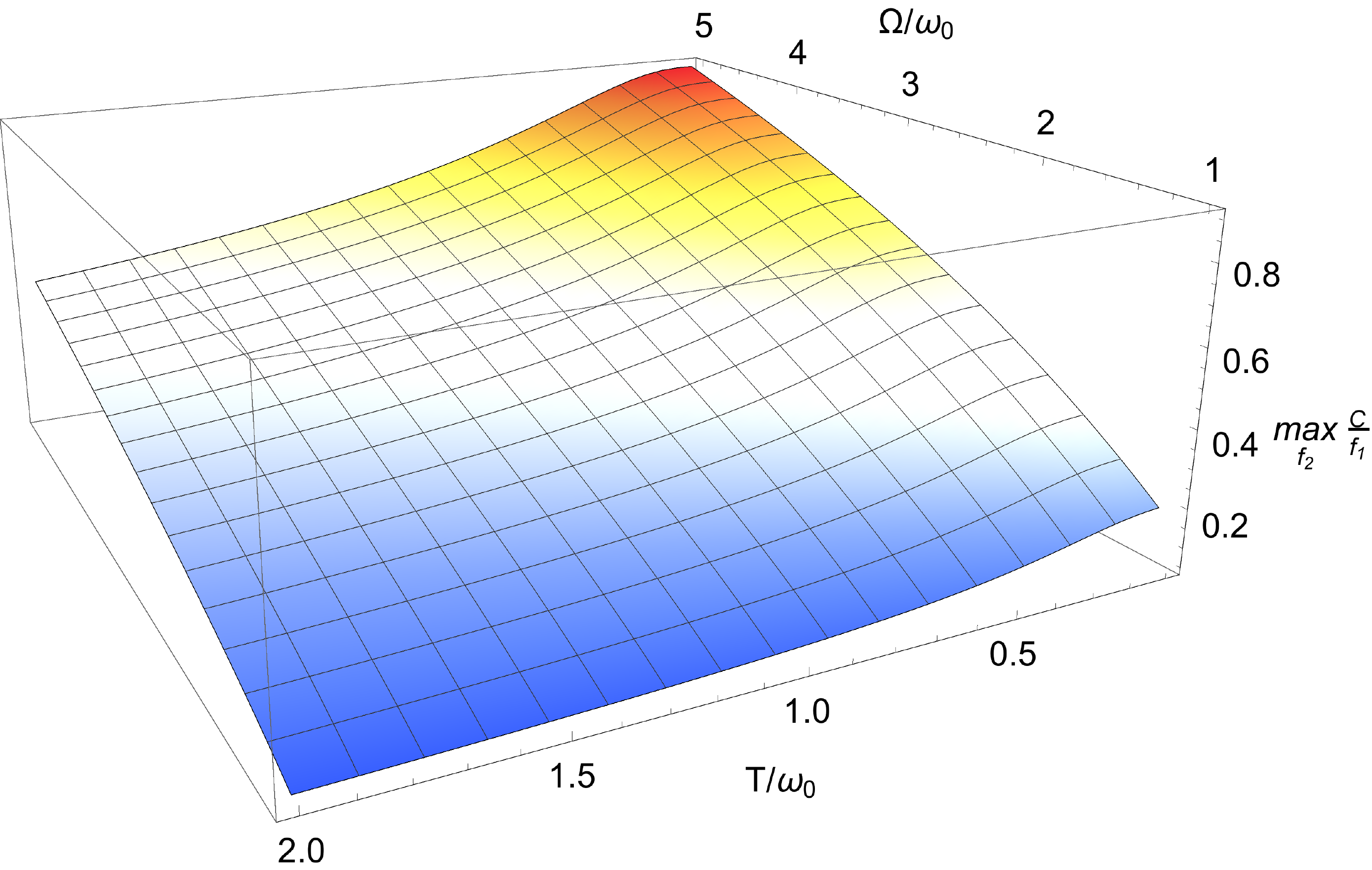}
\includegraphics[width=.3\columnwidth]{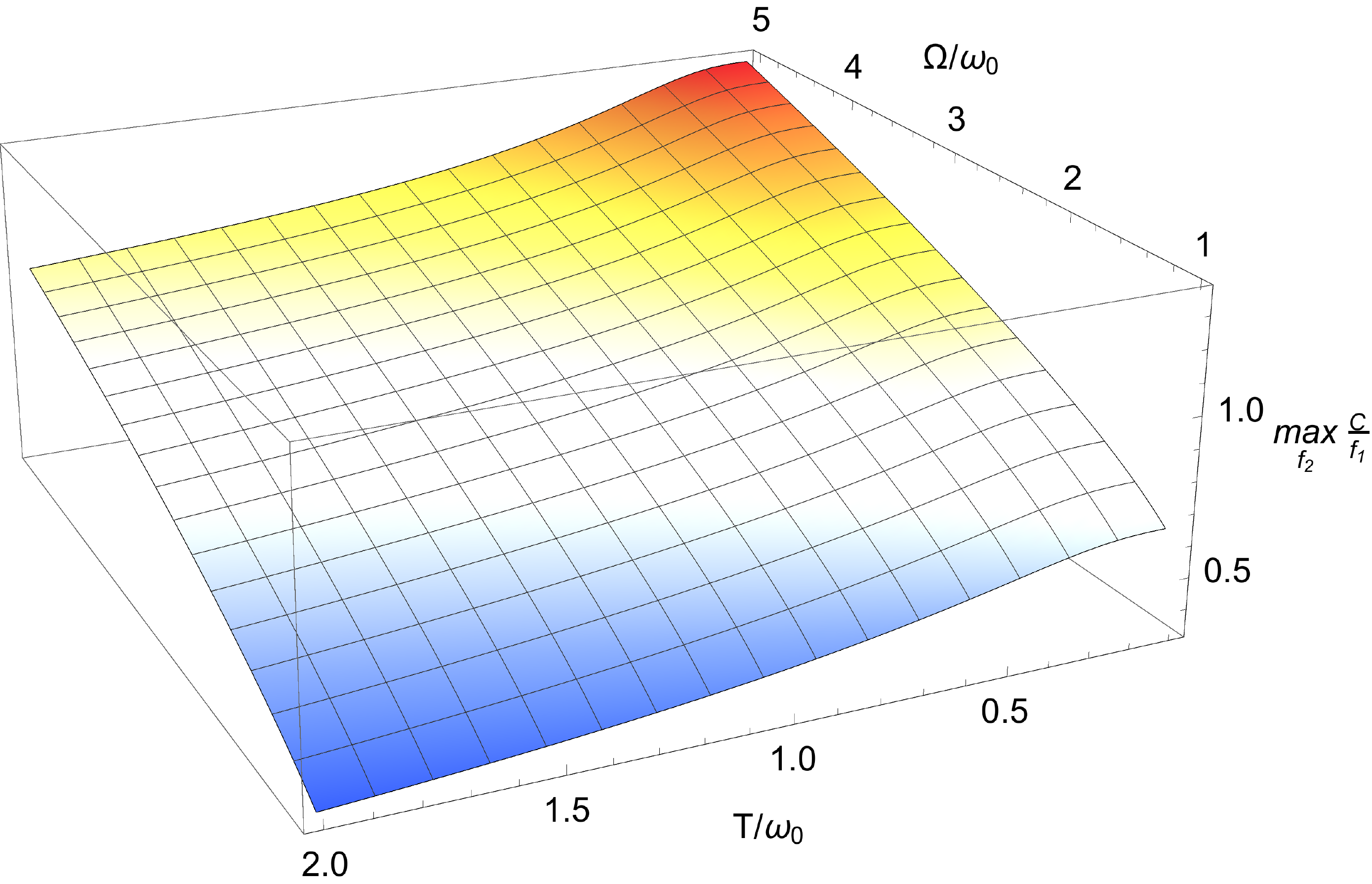}\\
\vspace*{0.2cm}
{\bf{(a)}}\hspace*{5.5cm}{\bf{(b)}}\hspace*{5.5cm}{\bf{(c)}}
\caption{(Color online) Plot of the maximum over $f_2 $ of Eq. \eqref{eq:CohMeasMod2} as a function of the bath temperature $T$ and of the cutoff frequency $\Omega$ for {\bf{(a)}} a sub-Ohmic ($s=0.5$), {\bf{(b)}} an Ohmic ($s=1$) and {\bf{(c)}} a super-Ohmic ($s=3$) environment. The coupling constant $\lambda$ has been fixed to $10^{-2}\omega_0$.}
\label{fig3SM}
\end{figure}

Exploiting the result \eqref{eq:SSCMod2v2}, we can easily compute the coherence measure \eqref{eq:CohMeas} which, after few simplifications, reads
\begin{equation}\label{eq:CohMeasMod2}
\mathcal{C} = \frac{|2 f_1f_2 \alpha(\omega_0,T,s)|}{\sqrt{\left(\omega_0 + f^2_2 \delta_1\right)^2 + \left(\pi f^2_2 J_{\mathrm{eff}}(\omega_0,T)\right)^2}}.
\end{equation}
As in the first model considered in Section A, the dependence of \eqref{eq:CohMeasMod2} on $f_1$ is linear and therefore all the quantities will be plotted in units of such coupling. A maximization over $f_2$ is instead performed analytically (the result is not shown here due to its complexity) and plotted in Fig. \ref{fig3SM}, in the weak coupling limit $\lambda=10^{-2}\omega_0$, as function of the temperature $T$ and cutoff frequency $\Omega$ for {\bf{(a)}} a sub-Ohmic spectral density ($s=0.5$), {\bf{(b)}} and Ohmic ($s=1$) and {\bf{(c)}} a super-Ohmic one ($s=3$).

Due to the different form of the interaction considered, the two quantities $\delta_{1,2}$ determining the coherence measure do not show the same singular behavior as $\Delta_{1,2}$ of the first model (see Subsection A.4), this resulting in the absence of enhancing peaks, for Ohmic and sub-Ohmic regimes, of the coherence measure. 

\begin{figure}[htbp!]
\begin{tikzpicture} 
  \node (img1)  {\includegraphics[width=0.48\columnwidth]{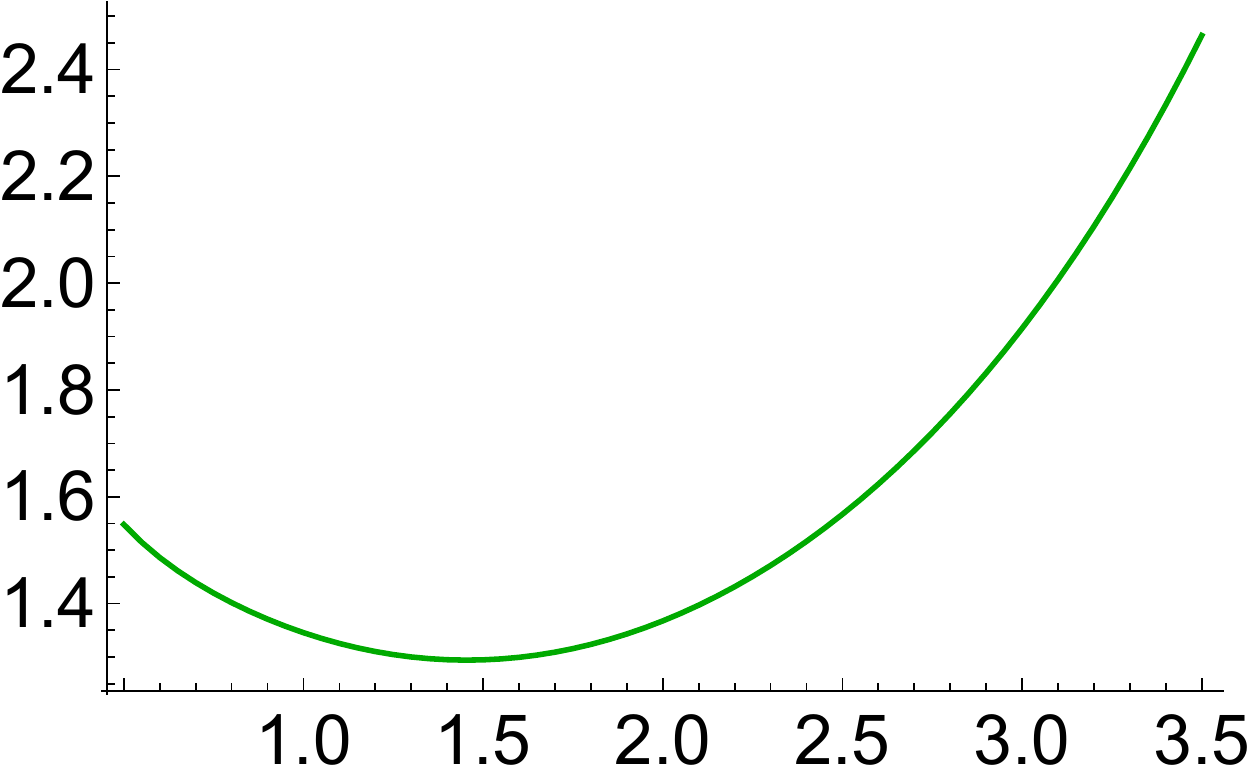}};
  \node[left=of img1, node distance=0cm, rotate=90, anchor=center, yshift=-.4cm,xshift=.3cm] {{\fontsize{14}{18}$\max_{f_2}{C/f_1}$}};
       \node[above=of img1, node distance=0cm, yshift=-7cm,xshift=4.5cm] {{\fontsize{14}{18}$s$}};
\end{tikzpicture}
\caption{Plot of the coherence measure Eq. \eqref{eq:maxCoherences} as function of the Ohmicity parameter $s$ for fixed values of the remaining parameters, i.e. $\Omega=10\omega_0$, $T=10^{-3}\omega_0$ and $\lambda=10^{-2}\omega_0$.}
\label{figCohVSsMod2}
\end{figure}

We close this Section by plotting in Fig. \ref{figCohVSsMod2} the dependence of the maximized coherence measure as function of $s$ for the same values of the parameters chosen in Fig. \ref{figCohVSsMod1}, in order to allow for a comparison between them, once again showing that the SSC are present for every value of the Ohmicity parameter.

\section{C --- Splitting of the interaction to two independent baths}
\label{Sec:Split}

The present Section is devoted to show what happens to the SSC when the two projections of the interaction Hamiltonian, e.g., considered in Eq. \eqref{eq:Ham1} along the parallel direction and orthogonal subspace with respect to the free Hamiltonian $\Ham_S$ are separately attributed to two independent baths.
Let us thus take into account the situation where the two-level system is coupled to two bosonic baths, denoted with $E_1$ and $E_2$ such that the Hamiltonian generating the dynamics of the overall system is given by
\begin{align}\label{eq:Ham2Baths}
\Ham &= \Ham_S + \sum_{i=1}^2 \Ham_{E_i} + \Ham_{SE_1} + \Ham_{SE_2}\notag \\
&= \frac{\omega_0}{2}\sigma_z + \sum_{i=1}^2 \sum_{k_i} \omega_{k_i} b^{\dagger}_{k_i} b_{k_i} + f_1\sigma_z \otimes B_{E_1} + f_2\sigma_x \otimes B_{E_2},
\end{align}
with $B_{E_i} = \sum_{k_i} g_{k_i} \left(b_{k_i} + b^{\dagger}_{k_i}\right)$.
For consistency we will assume to start in a product state of the form $\rho(0) = \rho_S(0)\otimes\rho_{\beta_1}\otimes\rho_{\beta_2}$, where $\rho_S(0)$ is a completely generic state of the two-level reduced system while $\rho_{\beta_i} = Z_i^{-1} e^{-\beta_i\Ham_{E_i}}$ with $Z_i \equiv \mathrm{Tr}_{E_i}\left[e^{-\beta_i\Ham_{E_i}}\right]$ are two Gibbs states relative to generic inverse temperatures $\beta_{1,2} = \left(k_B T_{1,2}\right)^{-1}$. 
Note that the two baths are considered to be independent and consequently no interaction term between them appears in Eq. \eqref{eq:Ham2Baths}. In a sense, therefore, the system serves the role of a junction, a framework widely employed in many open quantum systems and especially thermodynamical scenario.

The crucial property in what follows is that the odd moments of the interaction Hamiltonian with respect to the reference thermal states vanish \cite{Breuer2002}, i.e.
\begin{equation}\label{oddmeanval0}
\mathrm{Tr}_{E_i} \left[ \Ham_{SE_i}(t_1)  \Ham_{SE_i}(t_2)\ldots  \Ham_{SE_i}(t_{2n+1}) \right] = 0, \qquad \forall n,\,\, i=1,2.
\end{equation}
By employing the same approach used above, namely assuming weak coupling between the two-level system and the two baths in order to truncate the expansion of the dynamical generator to second order in the coupling strength, one gets once again to the master equation in time-convolutionless form Eq. \eqref{eq:MEgeneral} which in this case reads
\begin{align}\label{eq:2oTCL2Env}
&\frac{d}{dt}\rho_S(t) = -i \left[ \Ham_S,\rho_S(t) \right] \notag\\
&\quad- \sum_{i=1}^2 \int_0^t d\tau \mathrm{Tr}_{E_i} \left\{ \left[ \tilde{\Ham}_{SE_i},\, \left[ \tilde{\Ham}_{SE_i}(-\tau),\,\rho_S(t) \right]\right] \right\}.
\end{align}
where the particular structure of the interaction Hamiltonian $\Ham_{SE_1E_2} = \Ham_{SE_1} + \Ham_{SE_2}$ and the linearity of the trace have been used and where $\tilde{\Ham}_{SE_i} (t)$ once again denotes the time-evolved version of the interaction in the interaction picture and thus, explicitly, 
\begin{align}
\tilde{\Ham}_{SE_1} (t) &= f_1 \sigma_z \otimes \sum_k g_k \left( b_k e^{-i\omega_k t} +  b^{\dagger}_k e^{i\omega_k t}\right),\notag\\
\tilde{\Ham}_{SE_2} (t) &=  f_2 \left( \sigma_+e^{i\omega_0 t} + \sigma_-e^{-i\omega_0 t} \right) \otimes \sum_k g_k \left( b_k e^{-i\omega_k t} +  b^{\dagger}_k e^{i\omega_k t}\right).
\end{align}
It is now immediate to see that, by virtue of Eq. \eqref{oddmeanval0}, the resulting master equation for this model will be just the sum of the two channels, namely the purely dephasing one induced by the first bath $E_1$ and the one induced by the second bath $E_2$, each with the respective temperature and spectral density leading to the introduction of two environmental correlation functions.
The main point here is however that all the terms responsible for the formation of the SSC, which in the master equation were proportional to $\sigma_z\sigma_{\pm}$ or its adjoint, are no more present in this structure as the partial trace together with property Eq. \eqref{oddmeanval0} wash them out. It is therefore superfluous to discuss further such example since it is well known from the literature of open quantum systems \cite{Breuer2002} that both such dynamics taken singularly inevitably lead to a steady-state solution which show no coherence and so does also their linear combination.

This analysis shows the impossibility to simulate or mimic the effect of a bath possessing an interaction Hamiltonian with the composite structure put under evidence in this work by means of any sequence of alternate interaction with different baths each one being either parallel or perpendicular to the free system's Hamiltonian.

\section{D --- Analysis of the robustness of SSC formation with respect to the presence of a second unwanted bath}
\label{Sec:Robust}

It is almost always the case, in optomechanical or quantum optical experimental platforms, that the system of interest not only interacts with the surrounding controlled environment, e.g. the cavity, but also with a second unwanted external bath which typically induces dephasing. This is mainly due to the fact that perfect insulation of the 'system + cavity' is almost impossible and there will usually be some leakage towards the environment which is usually represented by an additional dephasing channel.
Since one of the aims of the present work is to provide a benchmark result for possible experimental implementations, it is then important to verify that the formation of SSC by means of an interaction of the form, e.g., Eq. \eqref{eq:Ham1} is not completely and immediately spoiled by the presence of a second bath inducing dephasing. 
For this reason, let us reconsider the first model analyzed in Section A on top of which we add a second bosonic bath, also starting in a thermal state relative to a different temperature $\beta_2 = \left(k_B T_2\right)^{-1}$, that interacts with the qubit through a $\sigma_z$ type of interaction, namely
\begin{align}\label{eq:HamRobust}
\Ham &= \Ham_S + \sum_{i=1}^2 \Ham_{E_i} + \Ham_{SE_1} + \Ham_{SE_2}\notag \\
&=  \frac{\omega_0}{2}\sigma_z + \sum_k \omega_k b^{\dagger}_k b_k + \left( f_1\sigma_z + f_2\sigma_x\right)\otimes B_E + \sum_{k_2} \omega_{k_2} c^{\dagger}_{k_2} c_{k_2} + f_3\sigma_z \otimes C_{E_1},
\end{align}
with $B_E = \sum_k g_k \left(b_k + b^{\dagger}_k\right)$, $C_E = \sum_{k_2} g_{k_2} \left(c_{k_2} + c^{\dagger}_{k_2}\right)$ denotes the multimode position operator of the additional bath ($\lbrace c_j, c^{\dagger}_j\rbrace_j$ are usual bosonic annihilation and creation operators).
Furthermore, also in this case we will assume to start in a product state $\rho(0) = \rho_S(0)\otimes\rho_{\beta_1}\otimes\rho_{\beta_2}$, where $\rho_{\beta_i} = Z_i^{-1} e^{-\beta_i\Ham_{E_i}}$ with $Z_i \equiv \mathrm{Tr}_{E_i}\left[e^{-\beta_i\Ham_{E_i}}\right]$ are two Gibbs states relative to generic inverse temperatures $\beta_{1,2} = \left(k_B T_{1,2}\right)^{-1}$. 

The same technique of second-order time-convolutionless expansion of the dynamical generator can be used also in this case to show easily that the resulting master equation has the form shown in \eqref{eq:2oTCL2Env} where now
\begin{align}
\tilde{\Ham}_{SE_1} (t) &= f\left[f_1 \sigma_z + f_2 \left( \sigma_+e^{i\omega_0 t} + \sigma_-e^{-i\omega_0 t} \right)\right]\otimes \sum_k g_k \left( b_k e^{-i\omega_k t} +  b^{\dagger}_k e^{i\omega_k t}\right),\notag\\
\tilde{\Ham}_{SE_2} (t) &=  f_3 \left( \sigma_+e^{i\omega_0 t} + \sigma_-e^{-i\omega_0 t} \right) \otimes \sum_{k_2} g_{k_2} \left( c_{k_2} e^{-i\omega_{k_2} t} + c^{\dagger}_{k_2} e^{i\omega_{k_2} t}\right).
\end{align}
It is straightforward now to realize that, due to Eq. \eqref{oddmeanval0}, the resulting master equation will be given by the sum of the pure dephasing channel induced by the interaction $\Ham_{SE_2}$  with the second environment and the complex dynamics generated by $\Ham_{SE_1}$ discussed in detail in Section A. This implies that the only dynamical coefficient modified by the presence of the second environment will be $a_{33}(t)$ which becomes 
\begin{equation}
a_{33}(t) = f^2_1 \int_0^t\,d\tau 2D_1(\tau) + f^2_3 \int_0^t\,d\tau 2\tilde{D}_1(\tau),
\end{equation}
with $\tilde{D}_1(t) = 2\int_0^{+\infty} d\omega J_{\mathrm{eff}}(\omega,\Omega_2,T_{E_2}) \cos(\omega\tau)$ being the noise kernel of the second bath, assumed to be once again described by a spectral density of the Ohmic-dependent form. Importantly, none of the coefficients of the master equation guaranteeing the formation of SSC, namely $a_{13}(t)$ and $a_{23}(t)$ making up the affine terms $b_{1,2}(t)$ of the Bloch differential equations, are affected by the second dephasing channel in this limit. Finally, the coefficient $a_{33}(t)$ vanishes in the long-time limit $t\to +\infty$, independently of the two bath temperatures. 

This allows therefore to conclude that the results presented in Section A are still valid, and thus robust, even when a second independent external bath inducing dephasing is taken into account, this being a very strong statement in view of possible experimental applications. 

\section{E --- Alternative approach to the characterization of SSC: Equilibration theory}
\label{Sec:Equilibration}

We conclude this Supplementary Material by discussing in more detail the alternative approach, namely the one based on \textit{equilibration theory}, to show the formation of SSC by means of the composite system-bath interaction Hamiltonian put in evidence in the present work.
As stated in the main body of the paper, the two formal approaches show consistency with each other thus adding one more layer of robustness to the results. However, there are both conceptual and small quantitative differences in the two approaches which make them both worthy to consider.
Last but not least, the equilibration theoretical approach, at the price of being able to only provide information about the steady state solution of the dynamics, has the advantage of being much more easily applicable from a numerical point of view, thus opening the door also to fast simulations of the strong coupling dynamical regime and/or to more complex systems.

This alternative approach has its roots in the strong suggestion that a quantum system, when coupled to a large thermal bath, reaches thermal equilibrium.
This means that, if $\Ham = \Ham_S + \Ham_E + \Ham_{SE}$ is the total Hamiltonian generating the unitary dynamics of the composite system through the Liouville-von Neumann equation
\begin{equation}
\frac{\partial \rho_{SE}(t)}{\partial t} = -i \left[\Ham\,,\, \rho_{SE}(t) \right],\qquad (\hbar = 1)
\end{equation}
the steady-state solution satisfying $\partial_t\rho_{SE}^{eq} = 0$ is given by
\begin{equation}
\rho_{SE}^{eq} = \frac{e^{-\beta\Ham}}{Z},\qquad Z_{SE} = \mathrm{Tr}_{SE}\left[e^{-\beta\Ham}\right].
\end{equation}
As a direct consequence of this, it is very natural to conclude that the stationary solution for the reduced system will be simply given by
\begin{equation}\label{equilibriumstate}
\rho_S^{eq} = \mathrm{Tr}_E\left[\rho_{SE}^{eq} \right],
\end{equation}
which however differs from the local Gibbs state $Z_S^{-1} e^{-\beta\Ham_S}$, $Z_S = \mathrm{Tr}_S\left[ e^{-\beta\Ham_S}\right]$, due to the presence of the interaction Hamiltonian $\Ham_{SE}$ \cite{Subasi2012,FordPRL1985}.
This difference will therefore be of course enhanced in the strong coupling regime, but even in the weak coupling the corrections can become significant \cite{Mori2008a, SeifertPRL2016,kolar}.

In order to quantitatively see this, one can start from Eq. \eqref{equilibriumstate} and make use of the following operator expansion, valid for any $A, B$
\begin{equation}
e^{A + \alpha B} = e^{A} \left[ 1 + \alpha\int_0^1 dx \,e^{-Ax} B e^{Ax} + \alpha^2\int_0^1 dx \int_0^x dy\, e^{-Ay} B e^{Ay} e^{-Ax} B e^{Ax} \right] + o\left(\lambda^3\right),
\end{equation}
to immediately see (using the identification $\alpha = -\beta$) that the steady-state solution for the reduced system is, up to second order in the interaction Hamiltonian $\Ham_{SE}$, given by
\begin{align}\label{redeqstate}
\rho_S^{eq} &= \frac{e^{-\beta\Ham_S }}{Z'_S} \left[(1 -\beta\int_0^1 dx \,\mathrm{Tr}_E\left[ e^{x\beta (\Ham_S+\Ham_E)} \Ham_{SE} e^{- x\beta (\Ham_S+\Ham_E)}\rho_{\beta}\right]\right.\notag\\
& \qquad+ \left.\beta^2\int_0^1 dx \int_0^x \,dy \mathrm{Tr}_E\left[e^{x\beta (\Ham_S+\Ham_E)} \Ham_{SE} e^{- x\beta (\Ham_S+\Ham_E)}e^{y\beta (\Ham_S+\Ham_E)} \Ham_{SE} e^{- y\beta (\Ham_S+\Ham_E)}\rho_{\beta}\right]\right),
\end{align}
with $\rho_{\beta}$ denoting the thermal state of the bath and  $Z'_S = Z_S -\beta a + \beta^2 b$, where
\begin{align}
a &= \mathrm{Tr}_{SE}\left[\int_0^1 dx\,e^{x\beta (\Ham_S+\Ham_E)} \Ham_{SE} e^{- x\beta (\Ham_S+\Ham_E)}\right]\\
b &= \mathrm{Tr}_{SE}\left[\int_0^1 dx\int_0^x dy\, e^{x\beta (\Ham_S+\Ham_E)} \Ham_{SE} e^{- x\beta (\Ham_S+\Ham_E)}e^{y\beta (\Ham_S+\Ham_E)} \Ham_{SE} e^{- y\beta (\Ham_S+\Ham_E)}\right].
\end{align}
In line with Eq. \eqref{oddmeanval0}, since the bath is initially in a thermal state, we can assume that the mean value of any odd power of the interaction Hamiltonian vanishes, this having the consequence that $a=0$ as well as the second term in the expression of $\rho_S^{eq}$. 
The expression of \eqref{redeqstate} therefore becomes
\begin{equation}\label{newredeqstate}
\rho_S^{eq} = \frac{e^{-\beta\Ham_S }}{Z_S} \left( 1 -\beta^2 b + \beta^2\int_0^1 dx \int_0^x dy \,\mathrm{Tr}_E\left[e^{x\beta (\Ham_S+\Ham_E)} \Ham_{SE} e^{- x\beta (\Ham_S+\Ham_E)}e^{y\beta (\Ham_S+\Ham_E)} \Ham_{SE} e^{- y\beta (\Ham_S+\Ham_E)}\rho_{\beta}\right] \right).
\end{equation}
In order to easily calculate the quantities inside the above integrals, one can notice that, for any operator $B$
\begin{equation}\label{this}
e^{-\beta (\Ham_S+\Ham_E))} B e^{\beta (\Ham_S+\Ham_E))} = B(-i\beta),
\end{equation}
namely the l.h.s. or \eqref{this} corresponds to the evolved version in Heisenberg picture of the operator $B$ up to the imaginary time $-i\beta$. This allows to rewrite \eqref{newredeqstate} in the following compact form
\begin{equation}\label{eq:thermalstateapproach}
\rho_S^{eq} = \tau_S \left( 1 -\beta^2 b + \beta^2\int_0^1 dx \int_0^x dy \langle\Ham_{SE}(-i\beta x) \Ham_{SE}(-i\beta y) \rangle_{\beta} \right),
\end{equation}
where, for simplicity of notation, $\tau_S = \frac{e^{-\beta\Ham_S }}{Z_S} $ denotes the Gibbs state of the bare system and 
\begin{equation}
b = \mathrm{Tr}_{SE}\left[\int_0^1dx\int_0^xdy\, \Ham_{SE}(-i\beta x)\Ham_{SE}(-i\beta y)\right].
\end{equation}
Let us now apply this formalism to the first model put forward in Section A  in order to derive the steady-state values for the three Bloch-vector components $\overline{v}_{1,2,3}$.
In light of Eq. \eqref{eq:interactionHamevol} we have explicitly that 
\begin{equation}
\Ham (-i\beta x) = \left[f_1 \sigma_z + f_2 \left( \sigma_+e^{\omega_0 \beta x} + \sigma_-e^{-\omega_0 \beta x} \right)\right]\otimes \sum_k g_k \left( b_k e^{-\omega_k \beta x} +  b^{\dagger}_k e^{\omega_k \beta x}\right).
\end{equation}
We will start from evaluating the steady-state coherences.
Exploiting the algebraic properties of the Pauli matrices, one obtains straightforwardly that
\begin{align}\label{eq-2nd-sx}
\overline{v}_1 &\equiv \mathrm{Tr}_S\left[ \sigma_x \rho_S^{eq} \right] \notag\\
&= \mathrm{Tr}_S\left[ \sigma_x \tau_S \left(1 - \beta^2 b +\beta^2 \int_0^1dx\int_0^xdy\, \mathrm{Tr}_{E}\left[\Ham_{SE}(-i\beta x)\Ham_{SE}(-i\beta y)\right]\right) \right]\notag\\
&= \beta^2  f_1f_2\int_0^1dx\int_0^xdy\, \mean{B_E(-i\beta x)B_E(-i\beta y)}_{\beta} \left(\mathrm{Tr}_S\left[\sigma_x\tau_S \sigma_z\sigma_x(-i\beta y)\right] \mathrm{Tr}_S\left[\sigma_x\tau_S \sigma_x(-i\beta x) \sigma_z\right] \right)\notag\\
&= \beta^2  f_1f_2\int_0^1dx\int_0^xdy\, \mean{B_E(-i\beta x)B_E(-i\beta y)}_{\beta} \left( e^{\omega_0\beta y} \mathrm{Tr}_S\left[\sigma_x\tau_S \sigma_z\sigma_+\right]  + e^{-\omega_0\beta y} \mathrm{Tr}_S\left[\sigma_x\tau_S \sigma_z\sigma_-\right]\right)  \notag\\
& +\beta^2  f_1f_2\int_0^1dx\int_0^xdy\, \mean{B_E(-i\beta x)B_E(-i\beta y)}_{\beta} \left( e^{\omega_0\beta x} \mathrm{Tr}_S\left[\sigma_x\tau_S \sigma_+\sigma_z\right]  + e^{-\omega_0\beta x} \mathrm{Tr}_S\left[\sigma_x\tau_S \sigma_-\sigma_z\right]\right)  \notag\\
&= \beta^2  f_1f_2\int_0^1dx\int_0^xdy\, \mean{B_E(-i\beta x)B_E(-i\beta y)}_{\beta} \left( e^{\omega_0\beta y} \overline{n}_F(\beta) +  e^{-\omega_0\beta y} \left(\overline{n}_F(\beta)-1\right)\right)  \notag\\
& -\beta^2  f_1f_2\int_0^1dx\int_0^xdy\, \mean{B_E(-i\beta x)B_E(-i\beta y)}_{\beta} \left( e^{\omega_0\beta x} \overline{n}_F(\beta) + e^{-\omega_0\beta x} \left(\overline{n}_F(\beta) - 1\right)\right),
\end{align}
where $\overline{n}_F(\beta) = \left( e^{\beta\omega_0} + 1 \right)^{-1}$ denotes the Fermi-Dirac mean occupation number of the spin system.
Using the fact that the environmental correlation function is time-homogeneous and comparing with Eq. \eqref{eq:envcorrD1D2} one can easily recognize that 
\begin{align}
\mean{B_E(-i\beta x)B_E(-i\beta y)}_{\beta} = \mean{B_E B_E(i\beta (x-y))}_{\beta} = \frac{1}{2}\left[ D_1(-i\beta (x-y)) - i D_2(-i\beta (x-y)) \right],
\end{align}
where
\begin{align}\label{eq:D1D2new}
&D_1(-i\beta (x-y)) = 2\int_0^{+\infty} d\omega J_{\mathrm{eff}}(\omega,\Omega,T_E) \cosh(\omega\beta (x-y)) \quad\quad \left( \text{with} \,\, J_{\mathrm{eff}}(\omega,\Omega,T_E) \equiv J(\omega) \coth\left(\frac{\omega}{2T_E}\right)\right)\\
&D_2(-i\beta (x-y)) = -2 i \int_0^{+\infty} d\omega J(\omega) \sinh(\omega\beta (x-y)).
\end{align}
Using then the relation $2 \overline{n}_F(\beta) - 1 = \tanh(\beta\omega_0/2)$, one finally obtains
\begin{align}
\overline{v}_1 &= \beta^2 f_1f_2\int_0^1dx\int_0^xdy\, \frac{1}{2}\left[D_1(-i\beta (x-y) - i D_2(-i\beta (x-y)) \right]\notag\\
&\qquad\qquad\qquad\qquad\left[-\tanh\left(\frac{\beta\omega_0}{2}\right)\left(\cosh\left(\beta\omega_0 y\right) - \cosh\left(\beta\omega_0 x\right)\right) -\sinh\left(\beta\omega_0 y\right) + \sinh\left(\beta\omega_0 x\right)\right].
\end{align}
Even though one can go further in the above expression for $\overline{v}_1$ by exchanging the order of integration and performing first the integrals over the variables $x$ and $y$, we will stop here and simply note that this expression gives a non-zero value for the SSC which again depends on the product $f_1f_2$ and thus it is in agreement with the predictions based on the master-equation approach put forward in Section A. Moreover, it is even more clear in this expansion that when the bath temperature is lowered, i.e. $\beta \to +\infty$, the SSC gets enhanced.

Analogue calculations and considerations, which we will omit for brevity, show that the second component of the Bloch vector $\overline{v}_2$ is equal to zero also according to this approach and the expansion of $\rho^{eq}_S$ in terms of Eq. \eqref{eq:thermalstateapproach}.

\begin{figure}[htbp!]
\begin{tikzpicture} 
  \node (img1)  {\includegraphics[width=.44\columnwidth]{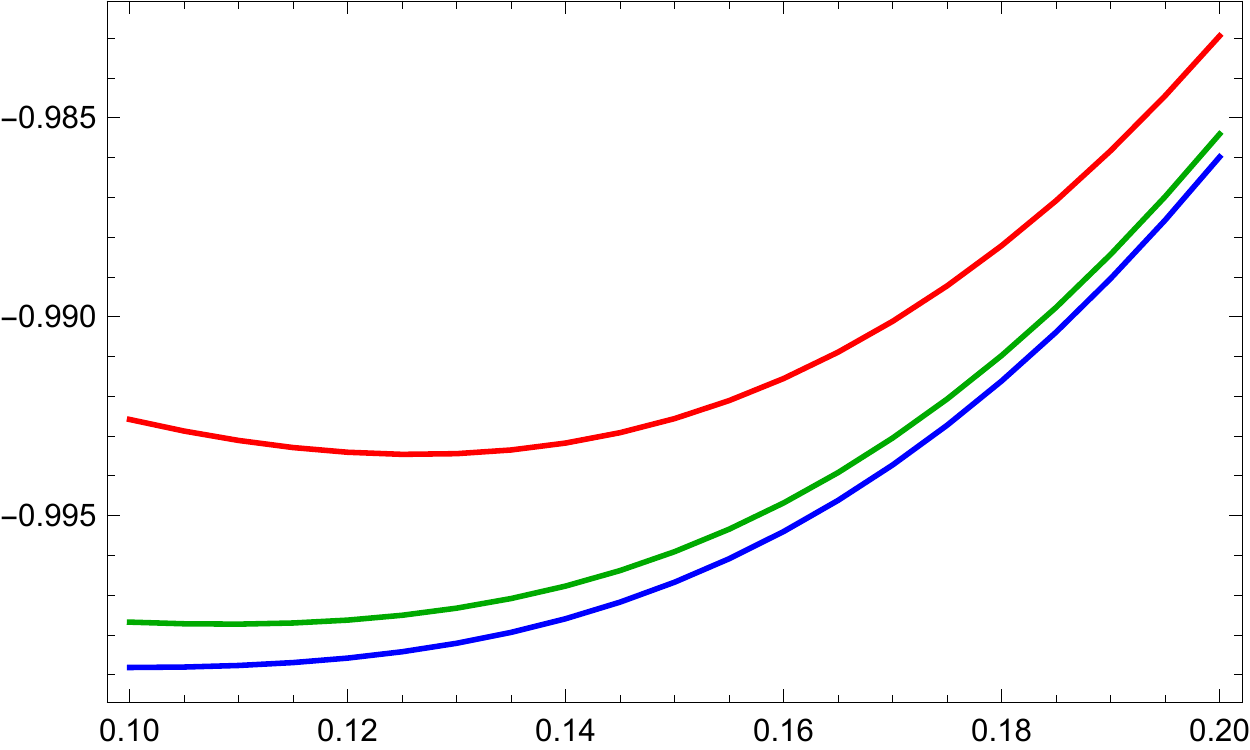}};
     \node[above=of img1, node distance=0cm, yshift=-4cm,xshift=-1.5cm] {{\color{red}$s=0.5$}};
        \node[above=of img1, node distance=0cm, yshift=-5.5cm,xshift=3.2cm] {{\color{black}{\bf{(a)}}}};
     \node[above=of img1, node distance=0cm, yshift=-4.7cm,xshift=-0.3cm] {{\color{black!40!green}$s=1$}};
   \node[above=of img1, node distance=0cm, yshift=-5.3cm,xshift=1cm] {{\color{blue}$s=3$}};
      \node[left=of img1, node distance=0cm, rotate=0, anchor=center, yshift=0cm,xshift=0.8cm] {{$\overline{v}_3$}};
       \node[above=of img1, node distance=0cm, yshift=-6.2cm,xshift=3.2cm] {{$T/\omega_0$}};
\end{tikzpicture}
\begin{tikzpicture} 
  \node (img1)  {\includegraphics[width=.44\columnwidth]{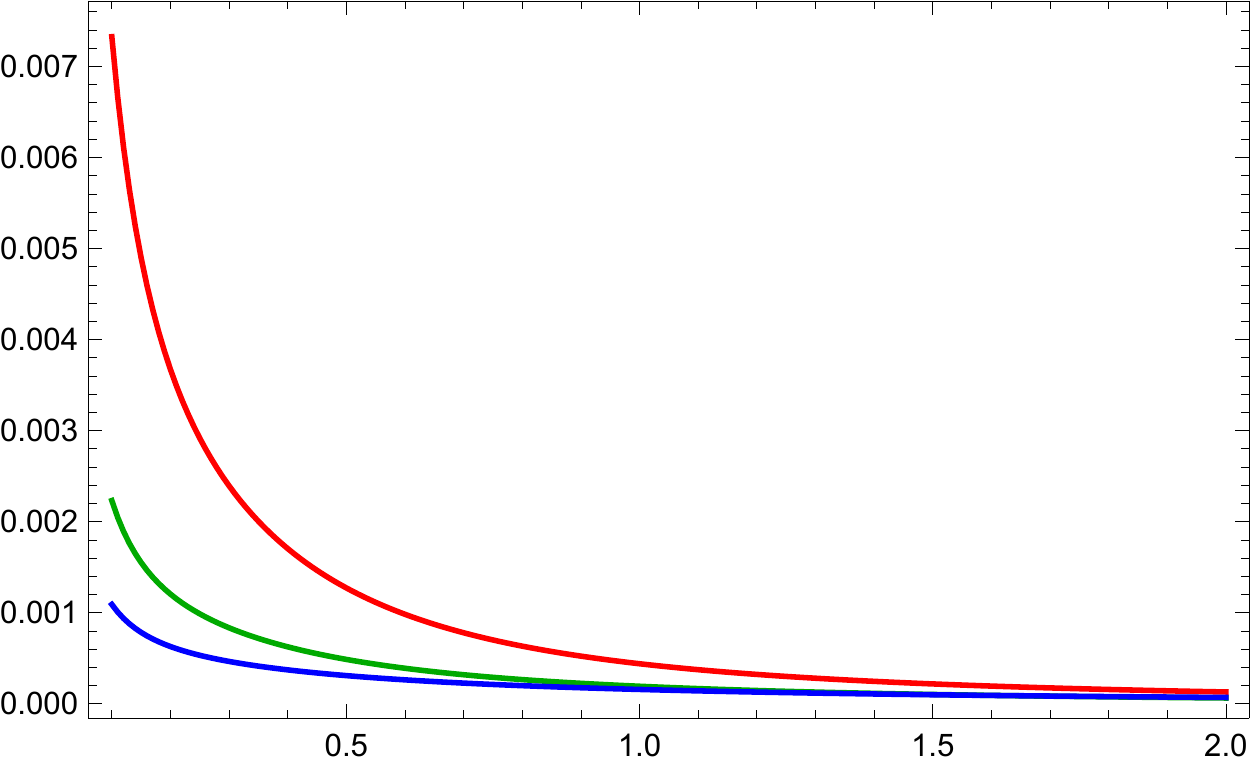}};
\node[above=of img1, node distance=0cm, yshift=-4cm,xshift=-1.5cm] {{\color{red}$s=0.5$}};
        \node[above=of img1, node distance=0cm, yshift=-5.5cm,xshift=3.2cm] {{\color{black}{\bf{(b)}}}};
     \node[above=of img1, node distance=0cm, yshift=-4.95cm,xshift=-2.45cm] {{\color{black!40!green}$s=1$}};
   \node[above=of img1, node distance=0cm, yshift=-5.65cm,xshift=-2.8cm] {{\color{blue}$s=3$}};
      \node[left=of img1, node distance=0cm, rotate=90, anchor=center, yshift=-0.8cm,xshift=0cm] {{$\overline{v}_3 + \tanh\left(\frac{\beta\omega_0}{2}\right) $}};
       \node[above=of img1, node distance=0cm, yshift=-6.2cm,xshift=3.2cm] {{$T/\omega_0$}};
\end{tikzpicture}\\
\caption{(Color online) Temperature dependence of {\bf{(a)}} $\overline{v}_3$ and of {\bf{(b)}} $\overline{v}_3 + \tanh\left(\frac{\beta\omega_0}{2}\right) $, the latter giving the correction to the Boltzmann factor $\tanh\left(\frac{\beta\omega_0}{2}\right)$ according to Eq. \eqref{eq:ResV3}.  The other parameters have been taken as $\lambda=10^{-2}\omega_0$, $\Omega=10\omega_0$ and finally  $f_1 = f_2 = 10^{-1}$. In each panel, the three curves refer to the sub-Ohmic $s=0.5$, Ohmic $s=1$ and super-Ohmic $s=3$ as indicated.}
\label{figv3Model1}
\end{figure}

Let us then finally move to the steady-state solution for the $\overline{v}_3$ component of the Bloch vector, which gives the stationary value of the population imbalance of the qubit and thus of its energy.
We remind the reader that we had postponed the discussion on this quantity until this moment because, as already stressed above, the equilibration approach allows, already at second-order, to have access to the corrections induced on $\overline{v}_3$ by the formation of SSC $\overline{v}_{1,2}$, at variance with what happens using the master-equation dynamical approach where a fourth-order expansion would be required. This difference relies on the fact that the second-order expansion is not performed on the dynamical generator as in the master equation approach, but directly at the level of the steady-state solution.
By applying the same calculations as for $\overline{v}_1$ and exploiting the fact that $\mathrm{Tr}_S\left[ \sigma_z \tau_S \right] = (2\overline{n}_F(\beta) - 1) = -\tanh\left(\beta\omega_0/2\right)$, one gets
\begin{align}
\overline{v}_3 &\equiv \mathrm{Tr}_S\left[ \sigma_z\rho_S^{eq} \right] \notag\\
&= \mathrm{Tr}_S\left[ \sigma_z \tau_S \left(1 - \beta^2 b +\beta^2 \int_0^1dx\int_0^xdy\, \mathrm{Tr}_{E}\left[\Ham_{SE}(-i\beta x)\Ham_{SE}(-i\beta y)\right]\right) \right]\notag\\
&= -\tanh\left(\frac{\beta\omega_0}{2}\right) \left[1 - \beta^2 b\right] + \beta^2 \int_0^1dx\int_0^xdy\, \mean{B_E(-i\beta x)B_E(-i\beta y)}_{\beta} \left[ f^2_1  \mathrm{Tr}_S\left[\sigma_z\tau_S \sigma_z\sigma_z\right] \right.\notag\\
&\left.\qquad\qquad\qquad\qquad\qquad\qquad\qquad\qquad+ f^2_2 e^{\omega_0\beta(x-y)}  \mathrm{Tr}_S\left[\sigma_z\tau_S \sigma_+\sigma_-\right] + f^2_2 e^{-\omega_0\beta(x-y)}  \mathrm{Tr}_S\left[\sigma_z\tau_S \sigma_-\sigma_+\right] \right]\notag\\
&= -\tanh\left(\frac{\beta\omega_0}{2}\right) \left[1 - \beta^2 b\right] + \beta^2 \int_0^1dx\int_0^xdy\, \mean{B_E(-i\beta x)B_E(-i\beta y)}_{\beta} \left[ -f^2_1 \tanh\left(\frac{\beta\omega_0}{2}\right)  \right.\notag\\
&\left.\qquad\qquad\qquad\qquad\qquad\qquad\qquad\qquad\qquad\qquad\qquad\qquad+f^2_2 e^{\omega_0\beta(x-y)}  \overline{n}_F(\beta) + f^2_2 e^{-\omega_0\beta(x-y)}  ( \overline{n}_F(\beta)-1)\right]\notag\\
&=  -\tanh\left(\frac{\beta\omega_0}{2}\right) \left[1 - \beta^2 b + f^2_1 \beta^2 \int_0^1dx\int_0^xdy\, \mean{B_E(-i\beta x)B_E(-i\beta y)}_{\beta} \right]\notag\\
& \qquad +\beta^2f^2_2\int_0^1dx\int_0^xdy\, \mean{B_E(-i\beta x)B_E(-i\beta y)}_{\beta} \left[\cosh\left(\omega_0\beta (x-y)\right) \left(2\overline{n}_F(\beta) -1\right) + \sinh\left(\omega_0\beta (x-y)\right)\right]\notag\\
&= -\tanh\left(\frac{\beta\omega_0}{2}\right) \left[1 - \beta^2 b +\beta^2 \int_0^1dx\int_0^xdy\, \mean{B_E(-i\beta x)B_E(-i\beta y)}_{\beta} \left(f^2_1 + f^2_2 \cosh\left(\omega_0\beta (x-y)\right)\right)\right]\notag\\
&\qquad\qquad\qquad\qquad\qquad\qquad\qquad\qquad+\beta^2f^2_2\int_0^1dx\int_0^xdy\, \mean{B_E(-i\beta x)B_E(-i\beta y)}_{\beta} \sinh\left(\omega_0\beta (x-y)\right),
\end{align}
where
\begin{align}
b &= \mathrm{Tr}_{SE}\left[\int_0^1dx\int_0^xdy\, \Ham_{SE}(-i\beta x)\Ham_{SE}(-i\beta y)\right] \notag\\
&= \int_0^1dx\int_0^xdy\, \mean{B_E(-i\beta x)B_E(-i\beta y)}_{\beta} \left[ f^2_1 \mathrm{Tr}_S\left[\sigma_z\sigma_z\right] + f^2_2 e^{\omega_0\beta (x-y)}\mathrm{Tr}_S\left[ \sigma_+\sigma_- \right] + f^2_2 e^{-\omega_0\beta (x-y)}\mathrm{Tr}_S\left[ \sigma_-\sigma_+ \right] \right]\notag\\
&= 2 \int_0^1dx\int_0^xdy\, \mean{B_E(-i\beta x)B_E(-i\beta y)}_{\beta}  \left[f^2_1 + f^2_2\cosh\left( \beta\omega_0(x-y) \right) \right].
\end{align}
Substitution of this term into the expression for $\overline{v}_3$ leads after simple algebra to
\begin{equation}\label{eq:ResV3}
\overline{v}_3 = -\tanh\left(\frac{\beta\omega_0}{2}\right) \left[1 - \beta^2 \left( f^2_1 \zeta - f^2_2 \zeta_C\right) \right] + \beta^2f^2_2\zeta_S,
\end{equation}
with 
\begin{align}
\zeta &\equiv \int_0^1dx\int_0^xdy\, \mean{B_E(-i\beta x)B_E(-i\beta y)}_{\beta}\notag\\
\zeta_C &\equiv \int_0^1dx\int_0^xdy\, \mean{B_E(-i\beta x)B_E(-i\beta y)}_{\beta}\cosh\left( \beta\omega_0(x-y) \right) \notag\\
\zeta_S &\equiv \int_0^1dx\int_0^xdy\, \mean{B_E(-i\beta x)B_E(-i\beta y)}_{\beta}\sinh\left( \beta\omega_0(x-y) \right).
\end{align}
It is immediate to see that the solution for the steady-state population this way obtained is given by the usual Boltzmann factor $2\overline{n}_F(\beta)-1 = -\tanh\left(\beta\omega_0/2\right)$ plus corrections induced by the presence of SSC which depend both on $f^2_1$ and $f^2_2$ (and so having the same order of magnitude of the SSC as well).
The behavior of both the steady-state population $\overline{v}_3$ and this correction term $\overline{v}_3 + \tanh\left(\frac{\beta\omega_0}{2}\right) $ are given in Fig. \ref{figv3Model1} for the same choices of Ohmicity parameters ($s = 0.5$ for sub-Ohmic, $s=1$ for Ohmic and finally $s=3$ for super-Ohmic spectral density) as in the thorough analysis carried out in Section A.4.
From the analysis of Fig. \ref{figv3Model1} {\bf{(b)}}, one can deduce that the corrections to the thermal value are increased monotonously with the inverse temperature $\beta$, thus becoming more pronounced as the SSC build up.

\begin{figure}[htbp!]
\vspace*{-1cm}
\begin{tikzpicture} 
  \node (img1)  {\includegraphics[width=.48\columnwidth]{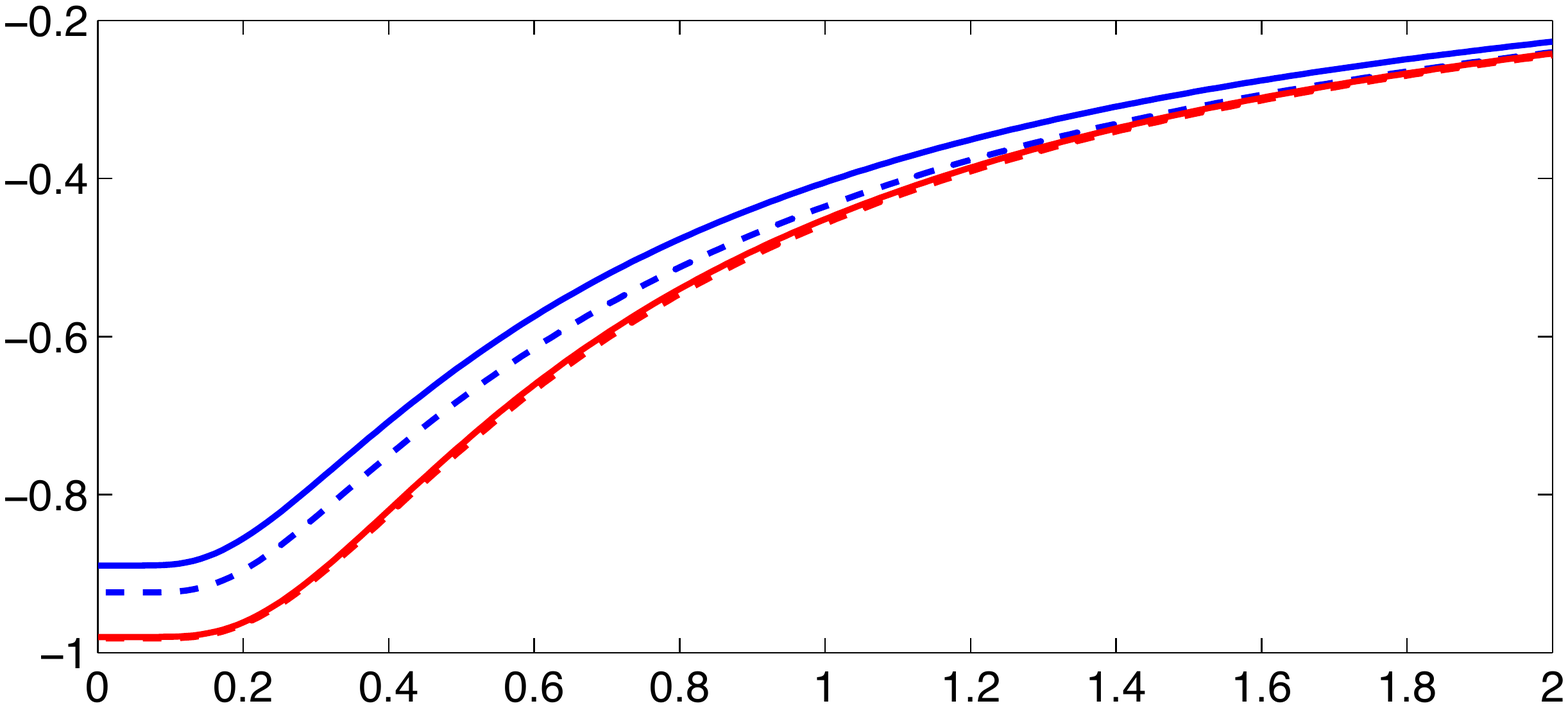}};
    \node[above=of img1, node distance=0cm, yshift=-4.7cm,xshift=0cm] {{\color{red}$\kappa=0.2$}};
   \node[above=of img1, node distance=0cm, yshift=-3.5cm,xshift=-0.8cm] {{\color{blue}$\kappa=0.5$}};
      \node[left=of img1, node distance=0cm, rotate=90, anchor=center, yshift=-1.6cm,xshift=0cm] {{$\overline{v}_3(T)$}};
       \node[above=of img1, node distance=0cm, yshift=-6.5cm,xshift=2.9cm] {{$T/\omega_0$}};
\end{tikzpicture}
\begin{tikzpicture} 
  \node (img1)  {\includegraphics[width=.48\linewidth]{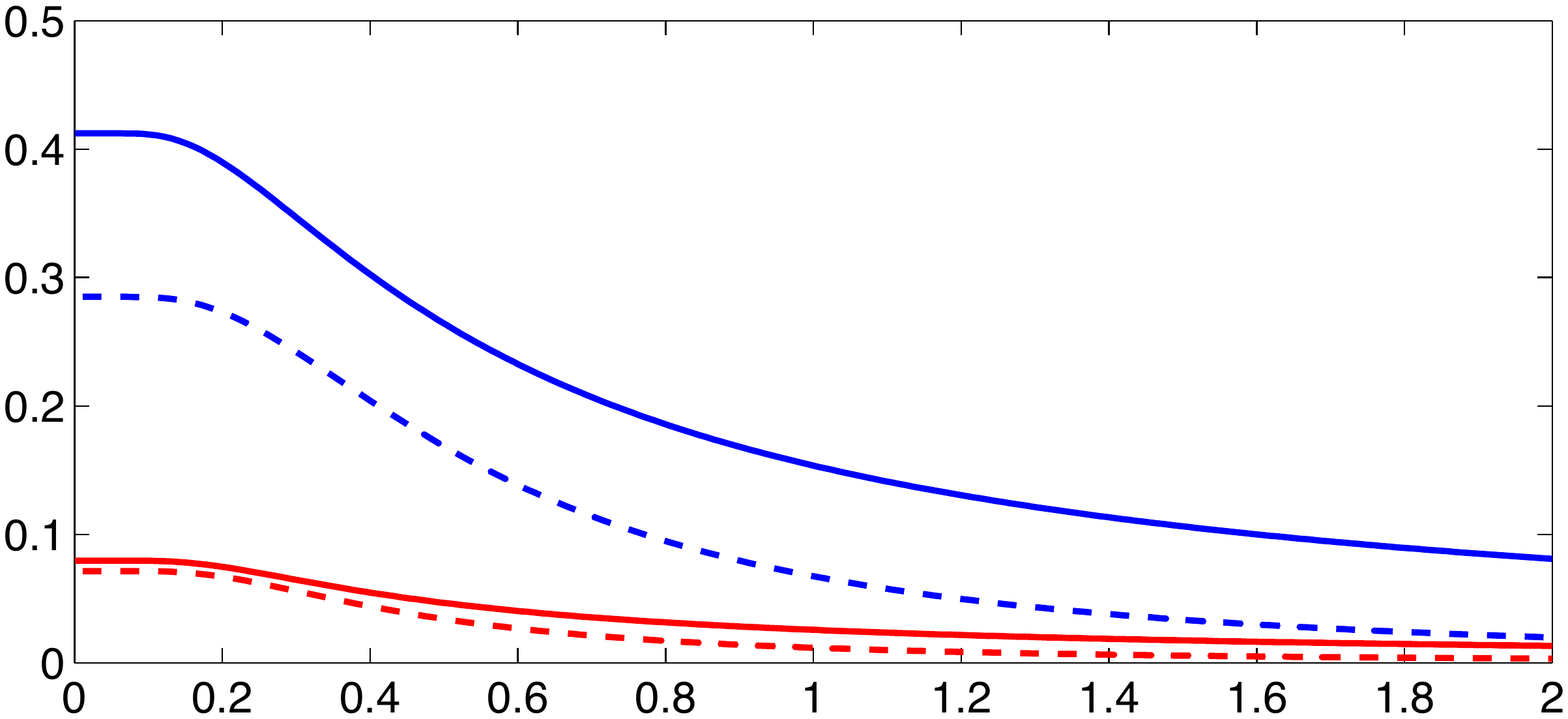}};
  \node[above=of img1, node distance=0cm, yshift=-5cm,xshift=-2.4cm] {{\color{red}$\kappa=0.2$}};
   \node[above=of img1, node distance=0cm, yshift=-3.4cm,xshift=-1.5cm] {{\color{blue}$\kappa=0.5$}};
 \node[left=of img1, node distance=0cm, rotate=90, anchor=center, yshift=-1.6cm,xshift=0cm] {{$\theta(T)$}};
       \node[above=of img1, node distance=0cm, yshift=-6.5cm,xshift=2.9cm] {{$T/\omega_0$}};
\end{tikzpicture}\\
\vspace*{-1cm}
{\bf{(a)}}\hspace*{9cm}{\bf{(b)}}
\caption{(Color online) Temperature dependence of {\bf{(a)}} $\overline{v}_3$ and of {\bf{(b)}} $\theta$ for different coupling $\kappa$ between the qubit and its effective bath, namely an harmonic mode in turn coupled to a thermal bath (solid lines) and a spin-chain in turn coupled to a thermal bath (dashed lines).}
\label{figV3ToyModels}
\end{figure}

Finally we have considered two different models than those presented in Sections 1 and 2.
The first model consists of a qubit, which is the subsystem of interest, coupled to an harmonic oscillator which is in turn equilibrated by means of a thermal reservoir. The interaction between the spin and the harmonic oscillator has the same structure as in Eq. \eqref{eq:Ham1} so that the total Hamiltonian is of the form
\begin{equation}\label{eq:HamToy1SM}
\Ham = \frac{\omega_0}{2}\sigma_z + \omega_1\left( a^{\dagger} a+1/2\right) + \left(\kappa_1 \sigma_x+\kappa_2\sigma_z\right)\otimes\left(a + a^{\dagger}\right) + \sum_k \omega_k b^{\dagger}_k b_k + \left( a+a^{\dagger}\right)\otimes B_E,
\end{equation}
with$B_E = \sum_k g_k \left(b_k + b^{\dagger}_k\right)$.
The second model studied has another two-level system in place of the harmonic oscillator, such that the overall dynamics is generated by
\begin{equation}\label{eq:HamToy2SM}
\Ham = \frac{\omega_0}{2}\sigma^{(1)}_z + \frac{\omega_1}{2}\sigma^{(2)}_z  + \left(\kappa_1 \sigma^{(1)}_x+\kappa_2\sigma^{(1)}_z\right)\otimes\sigma^{(2)}_x + \sum_k \omega_k b^{\dagger}_k b_k + \sigma^{(2)}_x\otimes B_E.
\end{equation}
These two open quantum systems have been investigated numerically by relying on the equilibration-picture approach and no weak-coupling approximation has been assumed.
The purpose of this last analysis was to prove the generality of the results above obtained and their robustness even outside the weak-coupling regime (which was invoked in order to employ the master-equation formalism).
Here in the SM we complement and complete the plot displaying the temperature dependence of the SSC in these two models (see Fig. 3 of the main body of the paper) by showing in Fig. \ref{figV3ToyModels} {\bf{(a)}}  the temperature dependence of the steady-state population imbalance $\overline{v}_3(T)$ and {\bf{(b)}} of the angle $\theta \equiv \arctan\left(\mathcal{C}/|\overline{v}_3|\right)$ of deviation from the $z-$axis in the Bloch sphere representation of the qubit system (see also the Schematics in the main text).
Both quantities are plotted for the same choices of parameters, i.e.  $\kappa_1 =\kappa_2 =\kappa$  and $\omega_0 = \omega_1$, where the solid curves refer to Eq. \eqref{eq:HamToy1SM} while the dashed ones to Eq. \eqref{eq:HamToy2SM}.
As one can immediately see, the steady-state population once does not go to the ground state when $T\to 0$, as would happen if the solution would be $-\tanh\left(\beta\omega_0/2\right)$, in accordance with what happens in Fig. \ref{figv3Model1} {\bf{(a)}}.
The final value of $\overline{v}_3$ acquires in fact a correction to the Boltzmann factor which gets more pronounced for decreasing temperature, i.e. for increasing values of SSC, as can be also seen from the enhancement of the deviation angle $\theta$ for $T\to 0$ in Fig. \ref{figv3Model1} {\bf{(b)}}. 

%
%

\end{document}